\documentclass[fleqn,usenatbib]{mnras}

\usepackage{newtxtext,newtxmath}

\usepackage[T1]{fontenc}
\usepackage{ae,aecompl}

\usepackage{graphicx}	
\usepackage{amsmath}	
\usepackage{amssymb}	
\usepackage{rotating}
\usepackage{url}
\usepackage{float,lscape}
\usepackage{rotating,fp}
\usepackage{array}
\usepackage{caption}
\usepackage{subcaption}
\usepackage{cleveref}

\newcommand{\cosmolt}{Hayden et al.\ (in prep)}
\newcommand{\cosmolp}{(Hayden et al.\ in prep)}

\title[See Change: VLT spectroscopy of host galaxies]{See Change: VLT spectroscopy of a sample of high-redshift Type Ia supernova host galaxies}

\author[S.~C. Williams et al.]{
S.~C. Williams,$^{1, 2, 3}$\thanks{E-mail: steven.williams@utu.fi}
I.~M. Hook,$^{1}$
B. Hayden,$^{4, 5, 6}$
J. Nordin,$^{7}$ 
G. Aldering,$^{4}$ \newauthor 
K. Boone,$^{4, 5, 8}$
A. Goobar,$^{9}$
C.~E. Lidman,$^{10}$
S. Perlmutter,$^{4, 5}$
D. Rubin,$^{4, 6}$ \newauthor
P. Ruiz-Lapuente,$^{11, 12}$
C. Saunders,$^{4, 13, 14}$
\newauthor (The Supernova Cosmology Project)\medskip\\
\\
$^{1}$Physics Department, Lancaster University, Lancaster, LA1 4YB, UK\\
$^{2}$Finnish Centre for Astronomy with ESO (FINCA), Quantum, Vesilinnantie 5, University of Turku, 20014 Turku, Finland\\
$^{3}$Department of Physics and Astronomy, University of Turku, 20014 Turku, Finland\\
$^{4}$E.~O.~Lawrence Berkeley National Laboratory, Berkeley, CA 94720, USA\\
$^{5}$Department of Physics, University of California Berkeley, Berkeley, CA 94720, USA\\
$^{6}$Space Telescope Science Institute, 3700 San Martin Drive, Baltimore, MD 21218, USA\\
$^{7}$Humboldt-Universit{\"a}t zu Berlin, Institut f{\"u}r Physik, Newtonstrasse 15, 12589, Berlin, Germany\\
$^{8}$DIRAC Institute, Department of Astronomy, University of Washington, 3910 15th Ave NE, Seattle, WA 98195, USA\\
$^{9}$The Oskar Klein Centre, Department of Physics, AlbaNova, Stockholm University, SE-106 91 Stockholm, Sweden\\
$^{10}$The Research School of Astronomy and Astrophysics, Australian National University, ACT 2601, Australia\\
$^{11}$Instituto de F\'isica Fundamental, Consejo Superior de Investigaciones Cient\'ificas, c/. Serrano 121, E-28006, Madrid, Spain\\
$^{12}$Institut de Ci\`encies del Cosmos (UB-IEEC), c/. Mart\'i i Franques 1, E-08028 Barcelona, Spain\\
$^{13}$Sorbonne Universit\'e, Universit\'e Paris Diderot, CNRS/IN2P3, Laboratoire de Physique Nuc\'eaire et de Hautes\\ \'Energies, 4 Place Jussieu, Paris, France\\
$^{14}$Sorbonne Universi\'es, Institut Lagrange de Paris (ILP), 98 bis Boulevard Arago, 75014 Paris, France\\
}

\date{Accepted XXX. Received YYY; in original form ZZZ}

\pubyear{2020}

\begin{document}
\label{firstpage}
\pagerange{\pageref{firstpage}--\pageref{lastpage}}
\maketitle

\begin{abstract}
The Supernova Cosmology Project has conducted the `See Change' programme, aimed at discovering and observing high-redshift ($1.13 \leq z \leq 1.75$) Type Ia supernovae (SNe~Ia). We used multi-filter {\it Hubble Space Telescope} (\textit{HST}) observations of massive galaxy clusters with sufficient cadence to make the observed SN~Ia light curves suitable for a cosmological probe of dark energy at $z>0.5$. This See Change sample of SNe~Ia with multi-colour light curves will be the largest to date at these redshifts. As part of the See Change programme, we obtained ground-based spectroscopy of each discovered transient and/or its host galaxy. Here we present Very Large Telescope (VLT) spectra of See Change transient host galaxies, deriving their redshifts, and host parameters such as stellar mass and star formation rate. Of the 39 See Change transients/hosts that were observed with the VLT, we successfully determined the redshift for 26, including 15 SNe~Ia at $z>0.97$. We show that even in passive environments, it is possible to recover secure redshifts for the majority of SN hosts out to $z=1.5$. We find that with typical exposure times of $3-4$\,hrs on an 8m-class telescope we can recover $\sim$75\% of SN~Ia redshifts in the range of $0.97<z<1.5$. Furthermore, we show that the combination of \textit{HST} photometry and VLT spectroscopy is able to provide estimates of host galaxy stellar mass that are sufficiently accurate for use in a mass-step correction in the cosmological analysis.
\end{abstract}




\section{Introduction}

Observations of Type Ia supernovae (SNe~Ia) at different redshifts by \citet{1998AJ....116.1009R} and \citet{1999ApJ...517..565P} provided the first evidence that the expansion of the Universe is accelerating. Since those works, the number of SNe~Ia with high-quality, multi-colour light curves has increased enormously at redshifts $z < 1$, from surveys such as ESSENCE \citep{2007ApJ...666..674M}, Supernova Legacy Survey \citep{2010A&A...523A...7G}, Carnegie Supernova Project \citep{2011AJ....142..156S}, Hubble Space Telescope Cluster Supernova Survey \citep{2012ApJ...746...85S}, Lick Observatory Supernova Search \citep{2013MNRAS.433.2240G}, Sloan Digital Sky Survey-II \citep{2018PASP..130f4002S}, Pan-STARRS 1 \citep[PS1;][]{2018ApJ...859..101S} and the
Dark Energy Survey \citep{2019ApJ...872L..30A}. 

\citet{2018ApJ...859..101S} combined PS1 SNe~Ia with those from several other surveys to produce the `Pantheon Sample' of 1048 SNe~Ia at $0.01<z<2.3$. When combined with Cosmic Microwave Background constraints, the sample gave a measurement of a constant dark energy equation of state parameter, $w=-1.026\pm0.041$, consistent with the cosmological constant value of $w=-1$. At $z \geq 1$, the SN~Ia Hubble diagram is still relatively poorly populated, meaning that for SNe~Ia, the dark energy equation of state is effectively unconstrained at $z>0.5$. As the accelerating expansion is not understood, it is important to measure the equation of state of the dark energy component of the Universe to better understand its nature. Measuring \textit{w} at high-\textit{z} is important because any deviation from $-1$ or variation with time would imply that dark energy cannot be in the form of the cosmological constant.

The peak optical luminosities of Type~Ia supernovae (SNe~Ia) are strongly correlated with the width of their light curves, with more luminous SNe~Ia having broader light curves. This allows the width of the light curve to be used to standardise the peak absolute brightness of individual SNe~Ia \citep{1993ApJ...413L.105P}. A correction for the SN colour further reduces the scatter in the standardised brightnesses \citep{1996ApJ...473...88R,1998A&A...331..815T}. After corrections for light curve width (or `stretch') and colour, there is typically an intrinsic scatter in the \citet{1929PNAS...15..168H} diagram of $\lesssim$\,0.15\,mag. This, in conjunction with their high luminosities (typically peaking at around $M_B=-19$), is what makes SNe~Ia good distance indicators and probes of dark energy.

The observed properties of SNe~Ia vary with the nature of their host galaxies. SNe~Ia occurring in star forming disk galaxies tend to be slower declining and more luminous than those in passive early-type galaxies \citep{1995AJ....109....1H,2006ApJ...648..868S}. Even after standardisation of the SN~Ia lightcurves using stretch and colour, a correlation between Hubble residual and host galaxy type remains. The Hubble residual of a SN~Ia is the offset of its standardised brightness from the expected standardised brightness for SNe~Ia (i.e.\ improved standardisation of the SNe would yield smaller Hubble residuals). Standardised SNe~Ia in passive galaxies have Hubble residuals that are slightly brighter than those in younger environments \citep{2009ApJ...700.1097H,2009ApJ...707.1449N}. A similar relationship is observed with host galaxy stellar mass, where the standardised brighnesses of SNe~Ia occurring in massive galaxies tend to be brighter than those occuring in galaxies with a lower stellar mass \citep{2010MNRAS.406..782S}. Hubble residuals have also shown to be correlated with the local environment of SNe~Ia. After stretch and colour standisation, \citet{2013A&A...560A..66R} found that SNe~Ia with local (within 1\,kpc of the SN position) H$\alpha$ emission were $0.094\pm0.031$\,mag fainter than those from passive regions. Correlations have been found between local specific star formation rate (sSFR) and Hubble residuals \citep{2018arXiv180603849R}, and between local colour and Hubble residuals \citep{2018A&A...615A..68R}. For example, \citet{2018A&A...615A..68R} found a systematic difference of $0.091\pm0.013$\,mag between the standardised peak magnitudes of SNe~Ia that have a bluer local $U-V$ host colour and those with a redder $U-V$ colour. Hubble residuals may correlate more strongly with local host galaxy information than they do with global host galaxy properties \citep{2018A&A...615A..68R}. \citet{2018ApJ...867..108J} found that local properties did tend to correlate more strongly with distance residuals, compared to random locations in the SN host galaxy, however the significance of this difference was relatively low. These works show that host galaxy (and ideally maybe even local) information is an important consideration when doing precision cosmology.

\subsection{The See Change survey}

The Supernova Cosmology Project has conducted the `See Change' survey, an extensive space and ground-based observing programme to efficiently discover very high-redshift SNe~Ia at $z\geq1$, with the aim of obtaining a measurement of the dark energy density at 0.5 < z < 1 that is independent of any assumption regarding the redshift continuity (e.g.\ via an equation of state parameterisation) of dark energy. A detailed overview of the See Change survey will be given by \citet[][submitted]{see-change-intro}. Observing these very high-redshift SNe~Ia probes dark energy at the time the Universe is thought to have transitioned from being matter dominated to dark-energy dominated. The See Change survey was based on a two-year, 174-orbit {\it Hubble Space Telescope} ({\it HST}) programme (PI: Perlmutter; Program IDs: 13677 and 14327) to observe large galaxy clusters at $1.13\leq z\leq1.75$, with the aim of discovering cluster-SNe~Ia, and obtaining the high-cadence, multi-colour light curves needed for cosmology. Four orbits were also used from \textit{HST} Program 13747 (PI: Tracy Webb). Targeting large galaxy clusters gives many more galaxies in the relatively small field of view (FoV) of {\it HST} WFC3 than can be achieved in the general field. It also takes advantage of the potentially higher SN~Ia rate in clusters at higher redshift, compared to clusters in the local universe \citep{2012ApJ...745...32B}. One of the cosmological aims of See Change was also to make weak lensing measurements (e.g.\ \citealp{2017ApJ...847..117J}). See Change has also yielded, or contributed to, a number of galaxy cluster studies \citep[e.g.][]{2015ApJ...812L..40G,2015ApJ...809..173W,2017ApJ...843..126D,2017ApJ...842L..21N,2018ApJ...866..136F,2018ApJ...852...96W,2018arXiv180906820G}.

The {\it HST} component of the See Change survey has been supported by several large ground-based spectroscopic follow-up programmes on the Very Large Telescope (VLT), W.~M.\ Keck Observatory (Keck), Gemini Telescopes, Gran Telescopio Canarias (GTC) and Subaru Telescope. This spectroscopy is required to determine precise redshifts for the host galaxies of the cosmological SN~Ia sample. Obtaining spectroscopic redshifts for these SNe~Ia is vital if we are to optimise the power that the See Change sample of SNe~Ia has on constraining the cosmology. It is also important to help determine the SN type. In the majority of cases, live SN spectroscopy to determine type was not feasible due to the faintness of the SNe or difficulties in obtaining spectra when the SNe were near maximum light. Most of the transients therefore had to be classified photometrically. Photometric classification of SNe uses the light curves in all available filters to constrain the possible type of a given SN (see e.g.\ \citealp{2002PASP..114..833P}). In this method of classification, determination of the precise redshift of the SN removes an important free parameter from the fitting model. This can often break degeneracies arising from the photometry alone, for example, between a lower redshift SN~CC and a higher redshift SN~Ia.

\begin{figure*}
\begin{center}
\includegraphics[height=9.98cm]{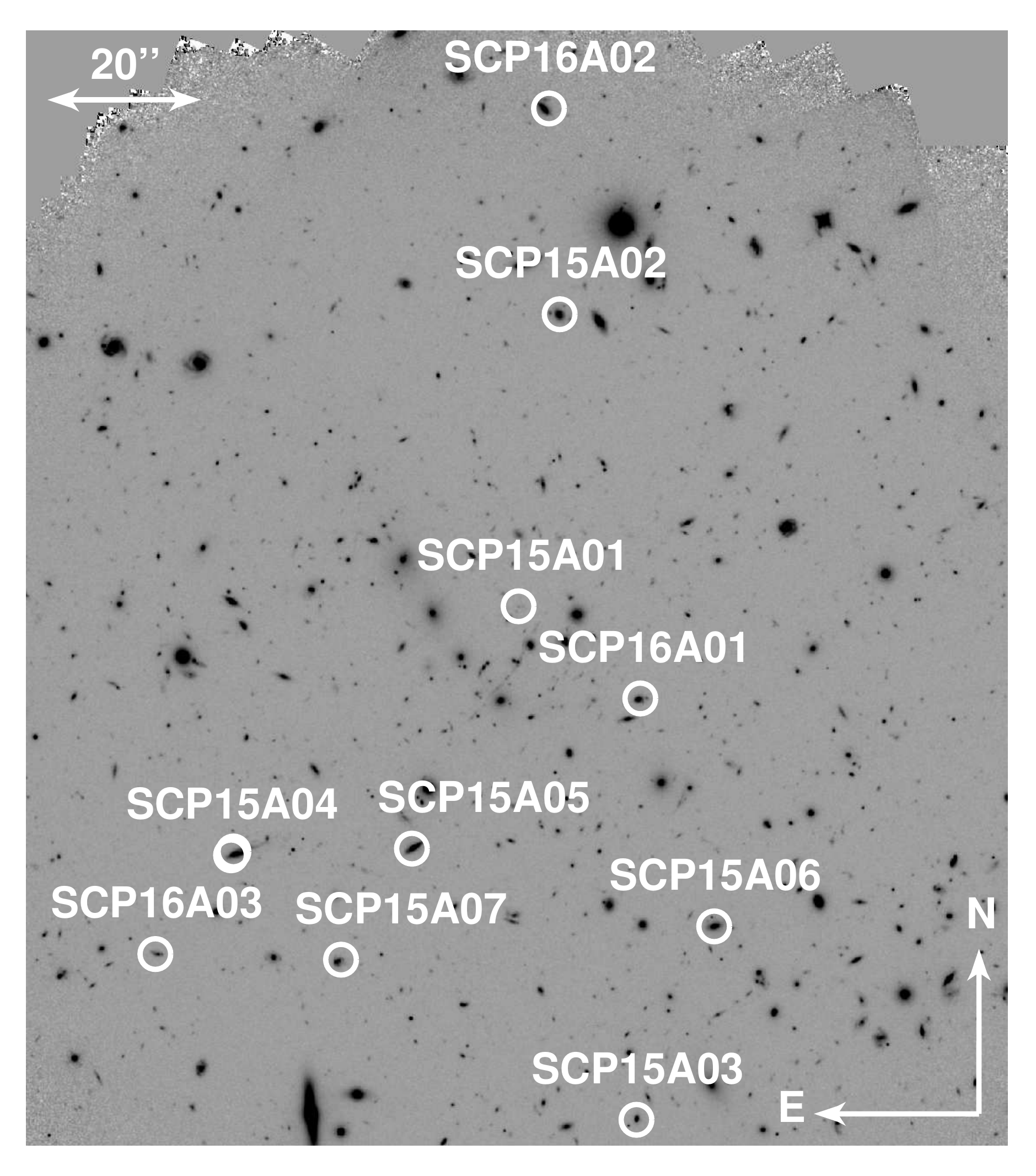} 
\includegraphics[height=9.98cm]{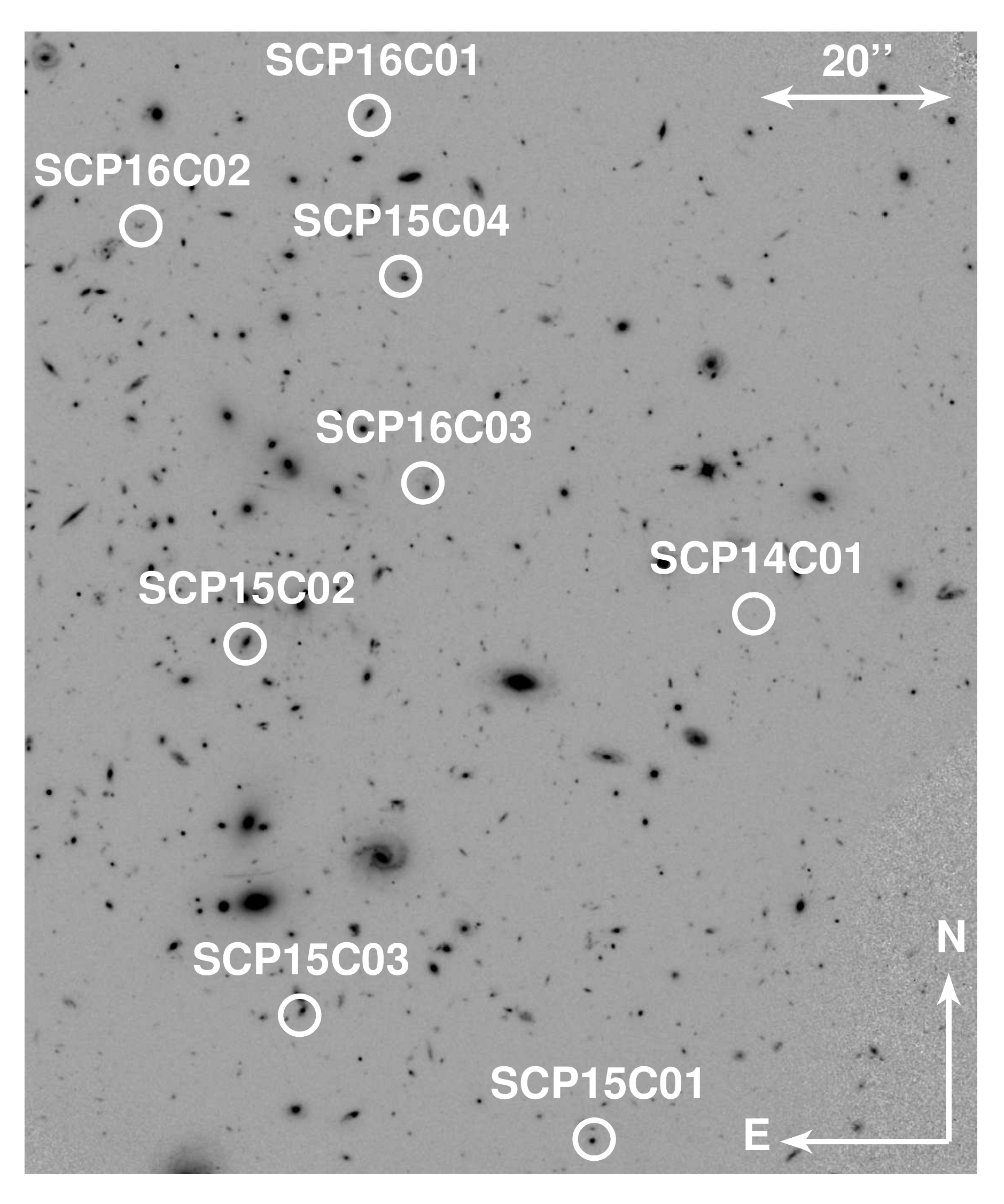} 
\caption{Stacked {\it HST} WFC3/IR F105W See Change image of clusters SPT0205 (left) and MOO1014 (right), with the positions of all transients discovered through the See Change programme indicated. A square-root flux scale is used to aid the visualisation of the faint objects. The galaxy that hosted SCP15A04, also produced SCP16A04, hence there are two circles appearing close together in the figure.\label{0205}}
\end{center}
\end{figure*}

\begin{figure*} 
\includegraphics[height=9.98cm]{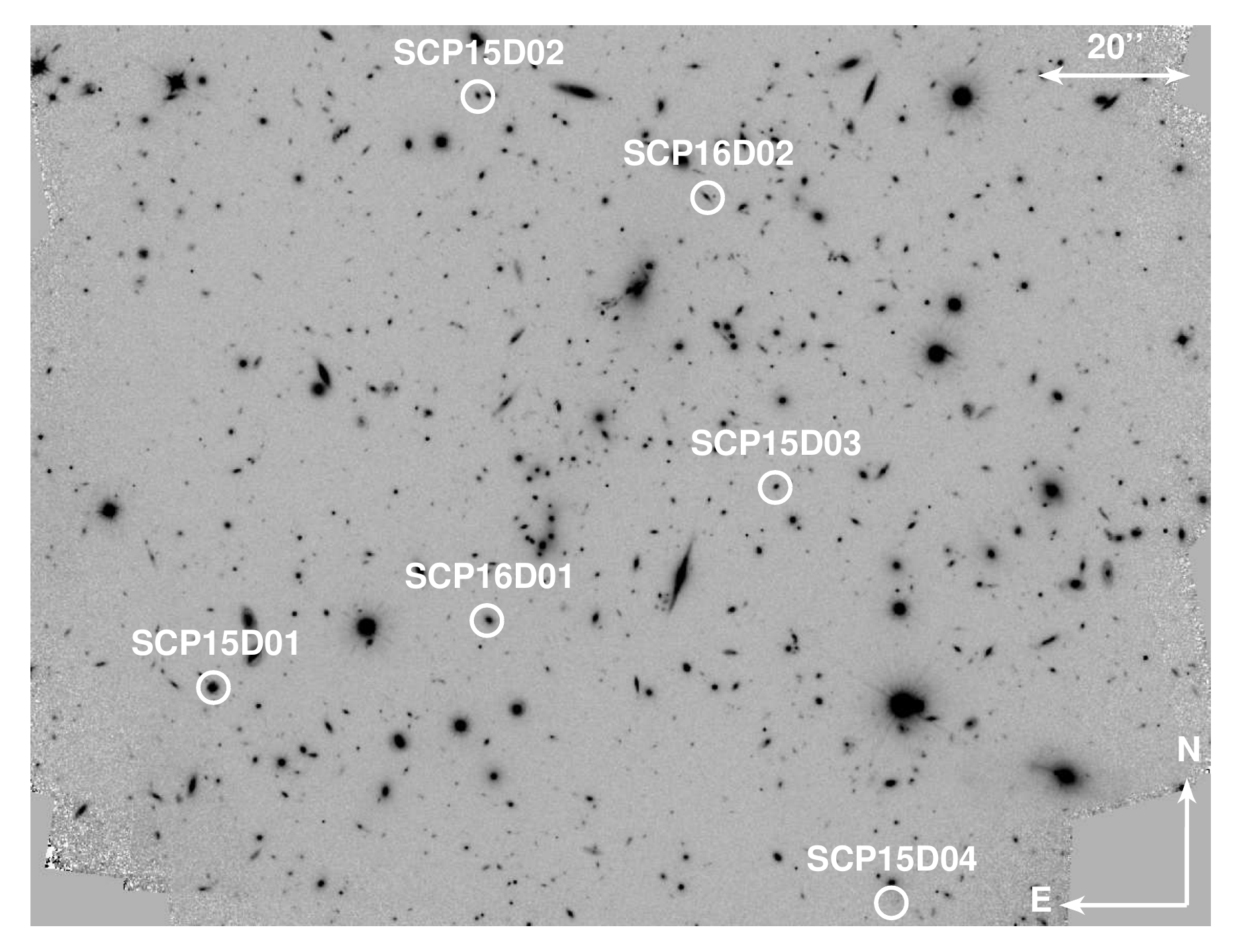}
\includegraphics[height=9.98cm]{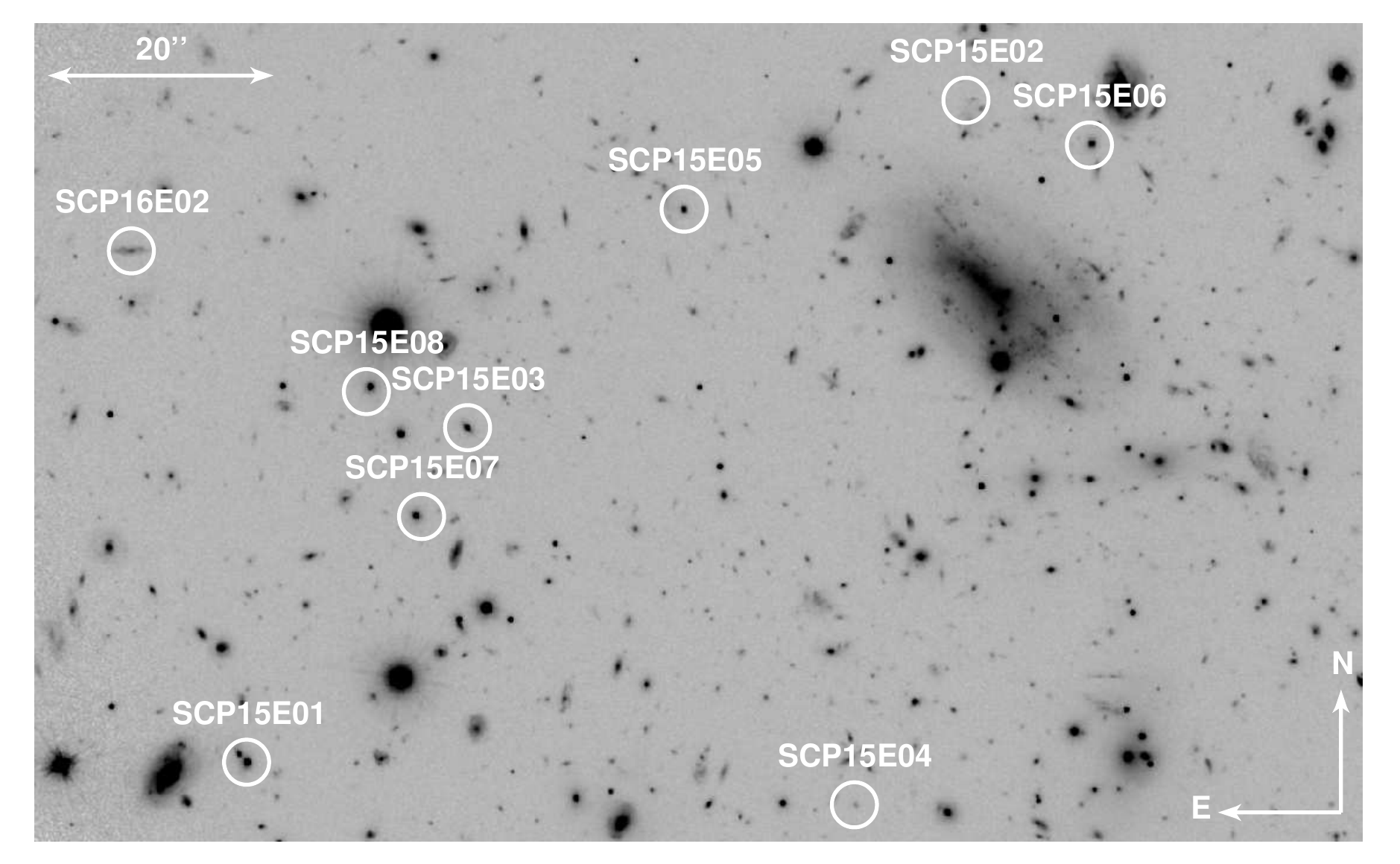} 
\caption{Stacked {\it HST} WFC3/IR F105W See Change image of clusters SPT2106 (top) and SPT2040 (bottom), with the positions of all transients discovered through the See Change programme indicated. A square-root flux scale is used to aid the visualisation of the faint objects.\label{2106}}
\end{figure*}

On overview of the See Change survey will be presented by \citet[][submitted]{see-change-intro}. \citet{2017arXiv170704606R} presented the See Change discovery and observations of a lensed SN~Ia at $z=2.22$. The See Change cosmological analysis will be presented by Hayden et al.\ (in prep).

\subsection{This work}

Here we present ground-based VLT spectroscopy of the host-galaxies of SN candidates discovered with {\it HST} during the See Change survey. We also obtained a live spectrum for an active transient with no detected host galaxy, found in the field of the MOO~J1014+0038 cluster. In this work, we use the host-galaxy spectra to obtain a precise redshift for each individual transient, which will be used in the cosmological analysis to be published by \cosmolt. The redshifts have also assisted the photometric classification of each transient, as they were often able to break degeneracies in the typing that still remain when considering the (relatively uncertain) photometric redshifts only. The See Change redshift range of $1.13\leq z \leq1.75$ prohibits detailed analysis of the local environments of the SNe~Ia, even when using \textit{HST}. Using a combination of our VLT spectra and \textit{HST} photometry of the SN~Ia host galaxies, we can however look at the global host properties, including stellar mass, star formation rate and stellar age. The main aim of this part of the work was to obtain the best possible constraints for the stellar mass of each SN~Ia host galaxy. This stellar mass measurement can then be used to `correct' the standardised peak SN~Ia brightnesses for the Hubble residual--stellar mass correlation discussed earlier. Given See Change is observing galaxy clusters, the average host galaxy properties are likely to be different from those of field SN~Ia surveys for example, and therefore the results of this work will be important to consider for the cosmological analysis.

This paper is structured as follows: we give a brief overview of the galaxy clusters for which we obtained VLT data as part of See Change in Section~\ref{sec:clusters}. The identification of VLT targets and transient classification is discussed in Section~\ref{sec:id}. The VLT data and the data reduction are described in Sections~\ref{sec:spec} and~\ref{sec:red} respectively. We describe how we determined the redshifts and galaxy parameters in Sections~\ref{sec:cross} and~\ref{sec:galpar}, before presenting the spectra themselves in Section~\ref{sec:res}. We then discuss our results, how the host galaxies compare to other galaxy cluster members, and how well our results constrain the galaxy parameters in Section~\ref{sec:dis}, before summarising the work in Section~\ref{sec:sum}.

\section{Target Clusters}\label{sec:clusters}
The See Change programme targeted twelve of the most massive known clusters in the redshift range of $1.13\leq z\leq1.75$. An overview of the cluster selection will be presented in \citet[][submitted]{see-change-intro}. The seven clusters targeted with the VLT observations are briefly discussed below.

The South Pole Telescope Survey (SPT) is a survey to discover galaxy clusters through the Sunyaev-Zel'dovich effect covering 2500 deg$^2$ of the southern sky \citep{2015ApJS..216...27B}. Three clusters discovered by this survey were observed as part of See Change: SPT-CL J0205--5829 (hereafter SPT0205) is a massive cluster at $z=1.32$ identified by \citet{2013ApJ...763..127R}. It has properties similar to lower-redshift clusters of similar mass and \citet{2013ApJ...763...93S} conclude that the majority of star formation in the cluster occurred at $z > 2.5$. The stacked \textit{HST} image of this cluster, along with the positions of the transients discovered during See Change are shown in Fig.~\ref{0205}. SPT-CL J2106--5844 (hereafter SPT2106) is the most massive cluster known at $z > 1$ \citep{2011ApJ...731...86F}. The \textit{HST} image of this cluster, along with the positions of the transients discovered during See Change are shown in Fig.~\ref{2106}. SPT-CL J2040--4451 (hereafter SPT2040) was discovered by the SPT-SZ survey \citep{2013ApJ...763..127R} at $z=1.48$. From spectra of 15 cluster member galaxies, \citet{2014ApJ...794...12B} found an elevated occurrence of star formation in the galaxies of this cluster. All 15 members had star formation rates $\geq1.5$~M$_{\odot}$~yr$^{-1}$. The \textit{HST} image of this cluster, along with the positions of the transients discovered during See Change are also shown in Fig.~\ref{2106}.

The \textit{Spitzer} Adaptation of the Red-sequence Cluster Survey (SpARCS) was a {\it z}$^{\prime}$-band imaging programme of the SWIRE {\it Spitzer} fields (which have been observed with all seven {\it Spitzer} passbands). It used the $z- 3.6\,{\mathrm{\mu{m}}}$ colour to find candidate clusters (by flagging large overdensities in a narrow colour range; see e.g.\ \citealp{2009ApJ...698.1943W}, \citealp{2009ApJ...698.1934M}). The stellar bump sequence method was also used to discover clusters (e.g.\ \citealp{2013ApJ...767...39M}). Three clusters discovered in the SpARCS survey were monitored by See Change: SpARCS J022427--032354 (hereafter SpARCS0224) at $z=1.63$, SpARCS J033056--284300 (hereafter SpARCS0330) at $z=1.63$ and SpARCS J003550--431224 (hereafter SpARCS0035) at $z=1.34$ \citep{2012MNRAS.427..550L}.

XDCP J0044.0-2033 (hereafter XDCP0044) is an X-ray luminous massive cluster at $z = 1.58$ \citep{2011A&A...531L..15S}. The cluster shows a reversal of the star-formation density relation, where the star formation rate in the projected cluster core is four times higher that in the outskirts \citep{2015MNRAS.447L..65S}. See also \citet{2014A&A...568A...5F} for detailed observations of XDCP0044. MOO J1014+0038 (hereafter MOO1014) was found by the Massive and Distant Clusters of WISE Survey \citep{2015ApJ...806...26B} and is at a redshift of $z=1.231$ \citep{2019ApJ...878...72D}.

\begin{table}
	\centering
	\caption{The letter used in the naming of each transient discovered in the field of a given cluster.}
	\label{tab:letter}
	\begin{tabular}{lcc} 
\hline
	Cluster &Redshift &SN Designation\\
\hline
    SPT0205 &1.32 &A \\
	MOO1014 &1.23 &C \\
	SPT2106 &1.13 &D \\
	SPT2040 &1.48 &E \\
    SpARCS0035 &1.34 &F\\
    XDCP0044 &1.58  &G\\
    SpARCS0330 &1.63 &H\\
    SpARCS0224 &1.63 &I\\
\hline
\end{tabular}
\end{table}

\section{Identification of Transients} \label{sec:id}
The See Change programme observed these $z>1.1$ clusters with \textit{HST}, using a five week cadence, for one orbit split between WFC3/UVIS F814W, and the WFC3/IR F105W and F140W filters at each epoch. Additional target of opportunity (ToO) \textit{HST} orbits were assigned to clusters with live SNe to ensure high signal-to-noise (S/N) observations in at least three bands, allowing the applied colour standardisation to be tested, as well as increasing the sampling of the light curve. The 50\% completeness for a simulated SN detection was AB mag 26.6 in F105W + F140W filters \citep[][submitted]{see-change-intro}. The transients were discovered by subtracting \textit{HST} reference images (built up as part of this survey) from the most recent image, and then performing photometry on any excess flux that was consistent with the expected PSF of a point source.

The highest priority targets for the ground-based spectroscopic programmes were candidate $z>1$ SNe~Ia. For the purpose of triggering spectroscopic follow-up, an object was determined to be a potential SN~Ia if the magnitude and colours of the transient were consistent with a SN~Ia and the photometric redshift of the host galaxy gave the possibility that it was at $z>1$. At the point of obtaining spectroscopy, we therefore did not have to be confident that a given transient was a $z>1$ SN~Ia, just that there was a reasonable \textit{possibility} that it was.

The See Change naming system indicates the sequential order of the transients discovered in the field of a given cluster in a given year, where each cluster has an associated letter, summarised in Table~\ref{tab:letter}. For example, SCP16A04 was the fourth transient to be discovered in the field of SPT0205 during the year 2016.

Due to the lack of live SN spectra, See Change relies on photometric classification of each transient. The photometric typing employed for the See Change survey combines the light curve classification procedure discussed in detail by \citet{2017arXiv170704606R}, the redshift information, and a single-epoch typing. This single-epoch typing will be presented by \citet[][submitted]{see-change-intro} and is based on comparing the observed flux in each \textit{HST} filter, when the SN is near peak brightness, to Monte Carlo simulations of different SN types at different redshifts. This single-epoch typing yields a type probability at any given redshift (in the range $0.5\leq{z}\leq2.0$), which can be combined with the spectroscopic redshift to give a classification. The light curve classifier fits the photometry with various templates of different SN types, both light curve shape in a band, and the colour. A sample of SN~Ia templates was constructed from SALT2 \citep{2007A&A...466...11G,2014ApJ...793...16M}, to which CC templates from SNANA \citep{2009PASP..121.1028K} and SN~Ia subtypes were added. The flux in each \textit{HST} band was then compared to the templates in a Monte Carlo simulation.

A transient is classified as a SN~Ia if its probability of being a SN~Ia, P(Ia)~$\geq0.9$. A likely SN~Ia has $0.75\leq$~P(Ia)~$<0.9$, with a possible SN~Ia having $0.25\leq$~P(Ia)~$<0.7$ and non-SN~Ia P(Ia)~$<0.25$. These classifications use all available photometry and redshift information (including the spectra presented here), and are thus different from the initial classifications of a transients as candidate~SN~Ia when deciding on follow-up observations.

\section{Spectroscopic Observations} \label{sec:spec}
The VLT data were obtained through programmes 294.A-5025, 095.A-0830, 096.A-0926, 097.A-0442 and 0100.A-0851. We used  X-Shooter \citep{2011A&A...536A.105V} and the MOS mode of the FOcal Reducer/low dispersion Spectrograph 2 (FORS2; \citealp{1998Msngr..94....1A}). If suitable, FORS2 was generally preferred over X-Shooter, due to the MOS giving us the capability to obtain spectra of multiple host galaxies simultaneously, and therefore being much more time-efficient. The objects that were deemed to potentially be $z>1$ SNe~Ia were the highest priority for VLT spectroscopy. These were the only targets observed with X-Shooter, and had priority over other objects when assigning the FORS2 slit positions. Non-SN~Ia See Change transients were the second priority when assigning the FORS2 slits. The VLT observations of See Change transient host galaxies are summarised in Table~\ref{tab:hostobs}. The average brightness of our target SN~Ia host galaxies was $F814W=23.7$\,mag (AB), ranging from the brightest SCP15D01 at $F814W=21.4$\,mag to SCP15A01 at mag~$\sim$28.

\subsection{FORS2}
The positions of the FORS2 slitlets were assigned using the FORS Instrumental Mask Simulator (\texttt{FIMS}). The FoV of the {\it HST} WFC3/IR channel is $123^{\prime\prime}\times136^{\prime\prime}$ and we targeted the centre of each cluster with a single {\it HST} field repeatedly. The FoV of the FORS2 MOS is approximately $6.8^{\prime}\times6.8^{\prime}$, so the host galaxies of the See Change transients never filled all 19 slitlets on the MOS. Therefore, if possible we placed remaining slitlets on cluster-galaxy candidates. Typically the bluest feature we targeted for redshift determination was [O~{\sc ii}] 3727\,\AA, which at $z=1.5$ is at an observed wavelength of 9318\,\AA. The optical FORS2 spectroscopy is therefore only realistically feasible for the $z < 1.5$ targets. At the lower redshift end of our SN~Ia target sample, $z=1$, the line is at 7454\,\AA. We therefore selected the GRIS\_300I+11 grism, which gives a wavelength coverage of 6000 -- 11000\,\AA\ and a resolution of $R=660$ at the central wavelength of 8600\,\AA. We used a slit width of 1$^{\prime\prime}$ and split each observation into two exposures, with a small spatial offset applied, typically $\sim3^{\prime\prime}$. While our top priority targets when observing with FORS2 were always the See Change candidates deemed likely to be high redshift SNe~Ia, all See Change transients discovered in any of the clusters targeted with FORS2 were observed at least once. As will be described below, we were able to obtain secure redshifts for the majority of both the SNe~Ia and non-SNe~Ia that we targeted with FORS2.

\subsection{X-Shooter}
The X-Shooter time was split between ToO and non-ToO observations. The non-ToO observations were primarily focused on obtaining redshifts of the host galaxies of SN~Ia candidates that were likely at $z > 1.4$, meaning the [O~{\sc ii}] 3727\,\AA\ and Ca~{\sc ii} H\,\&\,K lines start to become redshifted too far out of the optical, making a redshift difficult to obtain from FORS2 data. The ToO time was primarily used when a redshift was vital for deciding whether to trigger {\it HST} ToO orbits on a particular candidate, and therefore obtaining a redshift was time critical. Our X-Shooter observing strategy was optimised for the NIR observations. The total integration time in the NIR-arm of each Observation Block (OB) was 1\,hr (this was split into several individual exposures), with the UVB and VIS-arms exposures then set at the maximum possible without adding to the total observation time (typically $\sim0.758$ and $\sim0.872$\,hrs respectively). To aid the NIR sky-subtraction we employed nodding and a 1$^{\prime\prime}$ dither. The spatial distance of the nodding was sometimes restricted by the presence of other sources that would be on the slit at the requested position angle, and was therefore set on a case-by-case basis. Slits widths of 1.0$^{\prime\prime}$, 0.9$^{\prime\prime}$ and 0.9$^{\prime\prime}$ were used in the UVB, VIS and NIR arms respectively.

\section{Data Reduction} \label{sec:red}
\subsection{FORS2 Spectra} \label{fors}
The FORS2 spectra were bias-corrected, flat-fielded and wavelength calibrated using the \texttt{ESOReflex} pipeline \citep{2013A&A...559A..96F}. As each OB consisted of two exposures with a spatial offset, the sky subtraction is performed by subtracting one exposure from the other, thus sampling the same background as the traces in the two positions along the slit. This sky-subtraction method will generally reduce the S/N by $\sqrt{2}$. However, we found that the systematics associated with the sky-subtraction were significantly reduced. Given that the majority of the features we were trying to detect were at observed wavelengths of $>$\,8000\,\AA, where for spectroscopy of such faint sources there are many sky lines, we therefore determined that on balance utilising the offsets was the better method for recovering reliable redshifts for these faint host galaxies.

\subsubsection{Extraction of 1d spectra}
The path of the trace was defined using a low-order polynomial fit in the \texttt{aptrace} algorithm in \texttt{IRAF}\footnote{IRAF is distributed by the National Optical Astronomy Observatory, which is operated by the Association of Universities for Research in Astronomy (AURA) under a cooperative agreement with the National Science Foundation.} \citep{1986SPIE..627..733T}. The traces were fit by a Gaussian in the spatial direction (after binning in wavelength due to the low S/N per pixel). As we were observing extended sources, the width of the traces would be influenced by both the dimensions of the target and the atmospheric seeing. Therefore the FWHM of the traces were measured independently for each object. The extracted pixels were then optimally combined using the methodology outlined by \citet{1986PASP...98..609H}, where each spatial pixel at a given wavelength is assigned a weight according to its position compared to the centre of the trace and the width of the trace itself. Each pair of 1d spectra were then summed. This completes the sky subtraction by removing sky variations that are dependent on the position along the slit. We find that this produces a good sky subtraction, with relatively little systematic error associated with the sky lines.

\subsubsection{Flux Calibration} \label{fluxcal}
Flux calibration was performed using the ESO standards and calculated using data products from the \texttt{ESOReflex} pipeline. This was then applied to each optimally combined 1d spectrum. The individual flux-calibrated 1d spectra of each host galaxy were then stacked using a weighted mean to produce the final spectra presented here.

\subsubsection{Error spectra}
The error spectra of the wavelength-calibrated 2d spectra were obtained from the \texttt{ESOReflex} pipeline. The errors were then propagated through the extraction, combining and flux calibration process described above. These final spectra, with accurate errors, allowed us to compute realistic uncertainties on the emission line fluxes. This was particularly important for the apparently passive galaxies, as the errors were used to compute the upper limits on the emission lines. The  error spectra were also used in the cross-correlation analysis (see Section~\ref{sec:cross}).

\subsection{X-Shooter Spectra}
The X-Shooter data were reduced to 2d, wavelength-calibrated, sky-subtracted, flux-calibrated spectra using the \texttt{ESOReflex} pipeline. The utilisation of the nod and dither gave a very good sky subtraction. The extraction and optimal combining were then performed using a modified (due to the differing sky-subtraction method and resolution) version of the FORS2 reduction described in Section~\ref{fors}.

\begin{table*}
	\centering
	\caption{\textit{HST} photometry of galaxies that hosted a SN~Ia, likely SN~Ia or possible SN~Ia, and for which we were able to obtain a spectroscopic redshift in this work. This photometry was used in the SED fitting.
	}
	\label{tab:phot}
	\begin{tabular}{lcccccc}
\hline
	Transient &\multicolumn{6}{c}{Host galaxy photometry ($\mu$J)}\\
	Name &F606W &F814W &F105W &F125W &F140W &F160W\\
\hline
SCP15A03 &...	&$1.00\pm0.09$	&$3.68\pm0.02$	&...	&$6.24\pm0.02$	&$7.38\pm0.03$\\
SCP15A04 &$1.87\pm0.14$	&$3.93\pm0.34$	&$11.79\pm0.19$	&...	&$16.85\pm0.26$	&$19.46\pm0.40$\\
SCP15A05 &$1.55\pm0.19$	&$3.58\pm0.38$	&$10.65\pm0.11$	&...	&$14.18\pm0.24$	&$15.59\pm0.18$\\
SCP15A06 &...	&$2.37\pm0.17$	&$10.07\pm0.13$	&...	&$19.89\pm0.29$	&$23.72\pm0.26$\\
SCP15C01 &...	&$1.37\pm0.14$	&$2.15\pm0.04$	&...	&$2.81\pm0.06$	&...\\
SCP15C03 &...	&$1.09\pm0.50$	&$3.09\pm1.15$	&$4.59\pm1.40$	&$5.64\pm1.63$	&$6.96\pm2.04$\\
SCP15C04 &...	&$1.50\pm0.05$	&$3.72\pm0.02$	&$4.87\pm0.03$	&$5.94\pm0.03$	&$6.86\pm0.04$\\
SCP15D01 &...   &$9.96\pm0.24$ &$13.22\pm0.24$ &... &$14.75\pm0.25$ &...\\
SCP15D02 &$0.92\pm0.05$	&$2.17\pm0.14$	&$3.61\pm0.07$	&...	&$5.64\pm0.11$	&...\\
SCP15D03 &...	&$1.25\pm0.14$	&$2.90\pm0.05$	&...	&$3.93\pm0.04$	&...\\
SCP16D01 &...	&$2.09\pm0.12$	&$7.20\pm0.07$	&...	&$12.25\pm0.14$	&...\\
SCP15E06 &$0.38\pm0.22$	&$0.79\pm0.25$	&$4.12\pm0.26$	&...	&$8.21\pm0.16$	&$10.01\pm0.25$\\
SCP15E07 &$0.28\pm0.08$	&$1.18\pm0.10$	&$5.44\pm0.07$	&...	&$11.43\pm0.11$	&$13.96\pm0.15$\\
SCP15E08 &$3.98\pm0.12$	&$7.98\pm0.17$	&$12.28\pm0.19$	&...	&$16.81\pm0.19$	&$18.67\pm0.31$\\
SCP15G01 &...	&$1.21\pm0.39$	&$4.57\pm0.22$	&...	&$10.65\pm0.36$	&$13.63\pm0.37$\\
\hline
\end{tabular}
\end{table*}

\subsection{Telluric correction} \label{sec:tel}
We corrected the spectra for telluric absorption using \texttt{Molecfit} \citep{2015A&A...576A..78K,2015A&A...576A..77S}. The choice of spectrum used for the telluric fitting was made on a case-by-case basis for the FORS2 data. Continuum fitting was required, which can prove problematic if the S/N is particularly poor, exacerbated in some cases by emission or absorption lines being coincident with regions near telluric absorption. Due to this, the final combined spectrum was only used to determine the telluric correction for the higher S/N cases. In the majority of cases, the brightest object that appears in any of the slits of a single MOS observation was extracted and the atmospheric absorption fitted to that. The resulting telluric correction was then applied to the (fainter) objects in the other slits. The individual telluric-corrected spectra of a particular object were then combined as outlined in Section~\ref{fors}.

\begin{figure*}
\includegraphics[width=\linewidth]{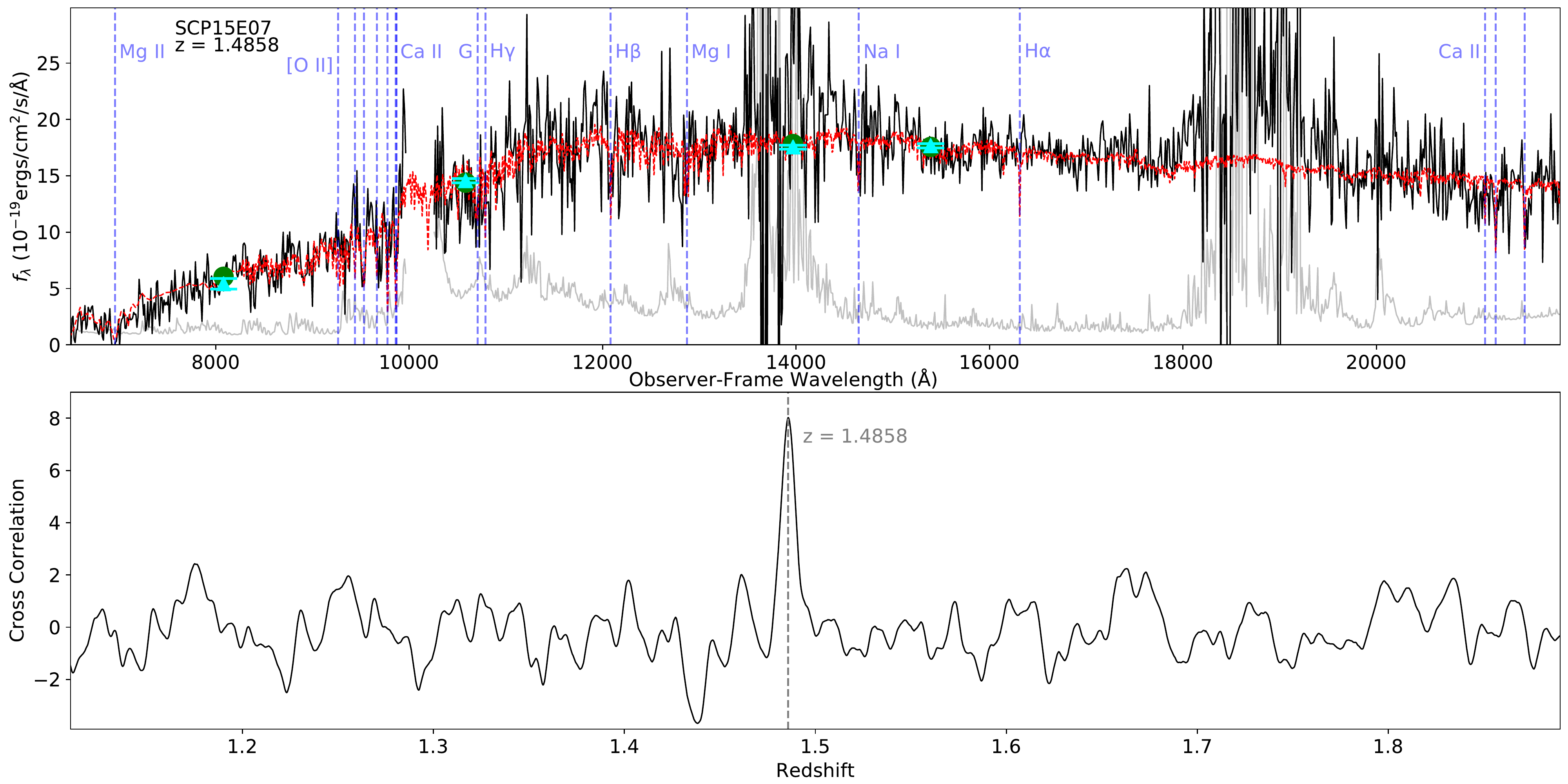}
\caption{\textit{Upper Panel:} Combined FORS2 and X-Shooter spectrum of the host galaxy of SN~Ia SCP15E07. The binned VLT data is the solid black line, with the best-fit \texttt{FAST} SED shown by the dashed red line. The solid grey line shows the errors on the spectrum. Common lines seen in galaxy spectra are also indicated. The green filled circles show synthetic photometry of the best-fit \texttt{FAST} SED, using the \textit{HST} filters, with the cyan triangular points showing the actual \textit{HST} photometry of the host galaxies (in the same filters). \textit{Lower Panel:} The only significant peak shown in the cross-correlation is centred at $z=1.4858$, with $\hat{r}>5$. The telluric correction was generally good, however in this particular case there appears some over-correction in the H$_2$O feature between $\sim$\,1.82--1.92\,$\mu$m. As noted in the text, regions of high atmospheric absorption are de-weighted in the cross-correlation. The NIR regions of very high atmospheric absorption (i.e. between $\sim$\,1.35--1.42\,$\mu$m, and between $\sim$\,1.82--1.92\,$\mu$m) are excluded from the \texttt{FAST} fitting of the stellar population, so systematic errors in the correction here will not affect galaxy parameters derived by \texttt{FAST}.} \label{xcorr}
\end{figure*}

The X-Shooter data were corrected for atmospheric absorption using standard stars observed shortly after or before the science observation, and at a similar airmass. The 1d spectrum of each standard star observation was extracted using the \texttt{ESOReflex} pipeline. We then used \texttt{Molecfit} to correct the standard observation for atmospheric absorption due to H$_2$O, O$_2$, CO$_2$, CO and CH$_4$. This correction was then applied to the corresponding science observation. Each corrected science observation was then combined as outlined above to produce a single spectrum for each object.

\subsection{Host galaxy photometry and absolute flux calibration}
One of the aims of this work was to obtain good constraints on the stellar mass of each SN~Ia host galaxy, which required absolute flux calibration of our spectra (see Section~\ref{sec:galpar} for details on how we determined the host galaxy parameters). For any galaxy that hosted a SN~Ia, likely SN~Ia or possible SN~Ia, we used the {\it HST} photometry and filter responses -- namely the F814W, F140W and F160W filters -- to make an absolute flux calibration. The {\it HST} photometry was performed using \texttt{LAMBDAR} \citep{2016MNRAS.460..765W}. Photometric apertures were defined using \texttt{SEP}\footnote{\url{https://sep.readthedocs.io/en/v1.0.x/}} \citep{1996A&AS..117..393B,2016JOSS....1...58B}. We detected objects greater than 2$\sigma$ above background, and determined their apertures using the Kron radius method, with a radius argument of 5 and an ellipse scaling factor of 2.5 (see \texttt{SEP} documentation for aperture photometry equivalent to FLUX\_AUTO in \texttt{Source Extractor}). Each aperture within 7\,arcsec of a SN was reviewed manually, and resized if the Kron radius was artificially enlarged by a nearby object. Possible contaminant galaxies overlapping the main apertures were added if they blended with the likely SN host. The units of the \textit{HST} stacked images were changed from electrons per second to electrons using the median exposure time of pixels in the drizzled image, so that \texttt{LAMBDAR} could accurately model the counting statistics of each pixel. The aperture was defined for the F105W stacked image, and the same aperture was then used in every available \textit{HST} filter. In addition to using the \textit{HST} photometry to perform an absolute flux calibration of our spectra, we also use it in the analysis when determining galaxy parameters. For this we use all available \textit{HST} photometry: $F606W$, $F814W$, $F105W$, $F125W$, $F140W$ and $F160W$ filters. All of our SN~Ia host galaxy \textit{HST} photometry is shown in Table~\ref{tab:phot}.

We assume that the galaxy light sampled in our spectra is representative of the overall galaxy light. It is possible this is not always the case. For example, in an edge-on spiral, it is possible that if the slit was placed parallel with the disk a larger emission line flux would be derived than if the slit were placed through the centre of the galaxy, perpendicular to the disk (thus sampling a lower proportion of the disk light). However, given our targets generally have a small angular size, which is convolved with the seeing, in most cases the assumption should be reasonable and we therefore assume this to be the case when deriving star-formation rates from emission line fluxes.

\begin{figure}
\includegraphics[width=\linewidth]{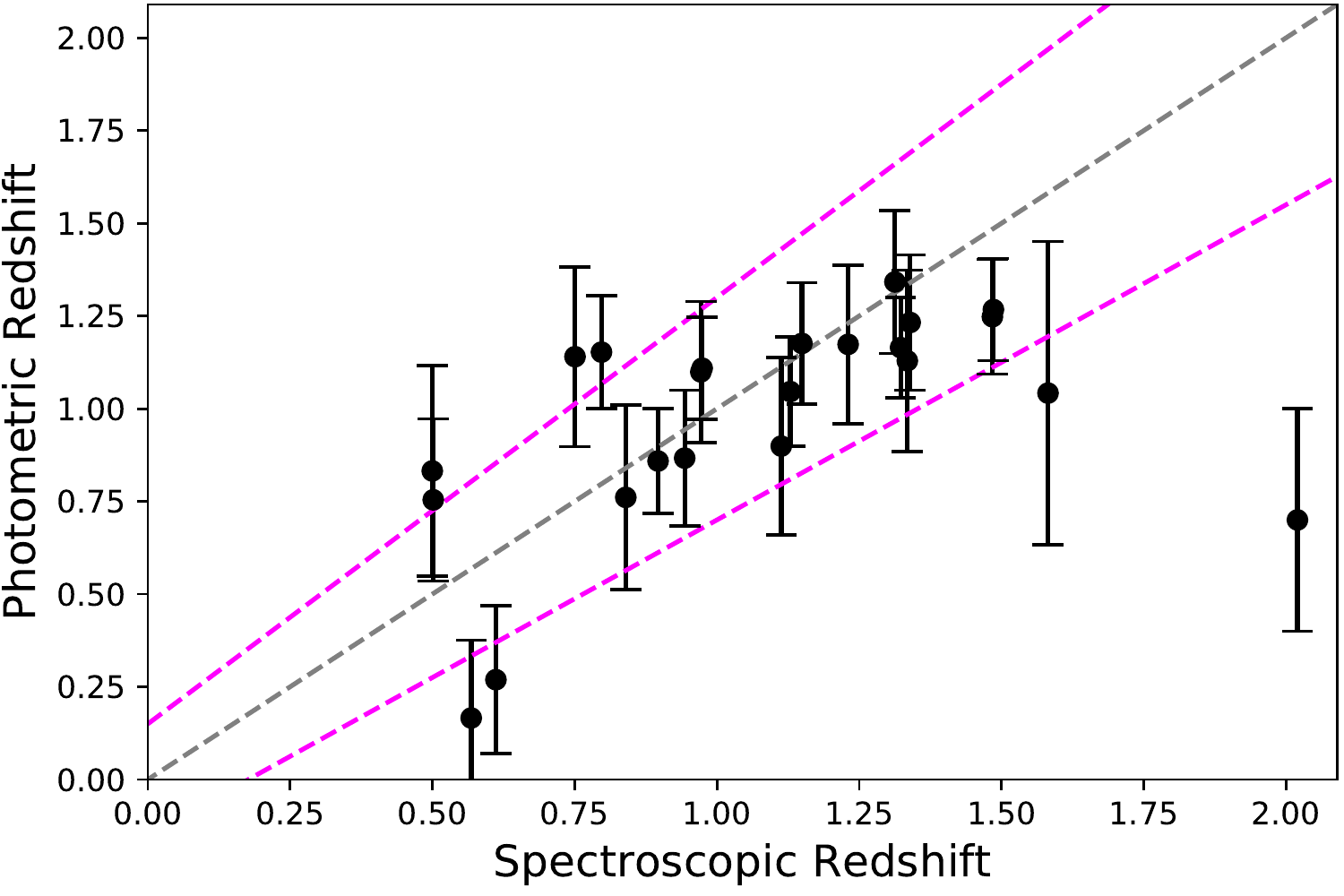}
\caption{Comparison between the photometric redshifts for the See Change host galaxies (both SN~Ia and non-Ia) and the spectroscopic redshifts derived from our spectra. The grey dashed line indicates $z_{\mathrm{phot}}=z_{\mathrm{spec}}$. A photometric reshift lying outside of the area encompassed by the two magenta dashed lines is considered an outlier, as defined in Section~\ref{sec:cross}. The outlier at $z=2.02$ is the host of SN~Ia SCP15E08. The light from this galaxy is likely dominated by the AGN at its centre.} \label{photoz}
\end{figure}

\section{Redshift Determination} \label{sec:cross}
Obtaining redshifts from passive galaxies can be difficult with low S/N spectra. This can be made even more difficult if, for example, the position of the Ca~{\sc ii} H\,\&\,K lines are in a low S/N region, either due to the throughput of the spectrograph, detector sensitivity, or atmospheric absorption or sky lines. We therefore employed a weighted cross-correlation to help derive secure redshifts, cross-correlating our spectra with the galaxy spectral cross-correlation templates from SDSS DR5 \citep{2007ApJS..172..634A}\footnote{\url{http://classic.sdss.org/dr5/algorithms/spectemplates/index.html}}. The weighted cross-correlation allows for the exclusion of bad data, with the remaining data then weighted by the inverse of the variance \citep{Kelson}. As our data are in the red/NIR, this is particularly useful as certain wavelength pixels will have significantly lower S/N than neighbouring pixels, due to sky lines and telluric features.

For the majority of objects where it was possible to derive a redshift with high confidence from the cross-correlation, we found that the features were already visible by visual inspection. This could be in part due to the relatively limited rest-frame wavelength coverage of the FORS2 data. The one example where this is not the case is SCP15E07 (with both FORS2 and X-Shooter data), which we therefore show as an example of the cross-correlation output in Fig.~\ref{xcorr}. The figure also illustrates the \textit{HST} photometry of the SCP15E07 host, and synthetic photometry of the best-fit SED in the same filters.

We compare all of our spectroscopic redshifts for the See Change host galaxies to the respective photometric redshifts in Fig.~\ref{photoz}. Following \citet{2010A&A...523A..31H}, we use the definition of a galaxy being a photometric redshift outlier if:

\begin{equation*}
\frac{|z_{\mathrm{phot}}-z_{\mathrm{spec}}|}{(1+z_{\mathrm{spec}})}>0.15.
\end{equation*}

The host galaxy of SCP15E08 (at $z_{\mathrm{spec}}=2.02$) is the only clear outlier. It is worth noting however that the light from this host galaxy is presumably dominated by the AGN at its core. We note that the relationship between the photometric redshifts and our spectroscopic redshifts appears `flatter' than the ideal scenario of $z_{\mathrm{phot}}=z_{\mathrm{spec}}$.

\section{Galaxy parameters} \label{sec:galpar}
The majority of our spectra are taken using FORS2, with a wavelength coverage of around $0.61-1$\,$\mu$m (the sensitivity drops off dramatically at $>$1\,$\mu$m). This, combined with the high redshifts of our targets, means that measurement of star formation rates (SFRs) from H$\alpha$ (or H$\beta$) is typically not possible, as the line is redshifted out of the observed wavelength range, and we have to rely on the [O~{\sc ii}] doublet at 3727\,\AA\ for our (emission-line) cluster galaxy SFR constraints. Any further mention of [O~{\sc ii}] in this paper refers to this 3727\,\AA\ doublet. The SFRs derived from [O~{\sc ii}] can vary significantly from those derived from H$\alpha$, due in large part to reddening and metallicity \citep{2004AJ....127.2002K}. At low SFRs the [O~{\sc ii}] emission from a galaxy can be contaminated by post-AGB stars \citep{2016MNRAS.461.3111B}. For these reasons, in addition to it being subject to less extinction, we use H$\alpha$ to estimate the SFR where we have a spectrum that includes both H$\alpha$ and [O~{\sc ii}]. For the SN~Ia host galaxies presented in this work, H$\alpha$ is always in the NIR, so it is only possible to use this line if we have an X-Shooter spectrum. For those where only [O~{\sc ii}] constraints are available, we assume all of the emission is associated with star formation and there is no contamination. Another potential source of emission line contamination is active galactic nuclei (AGN). Galaxies where the emission lines are dominated by the contribution from AGN are usually differentiated from those where the emission lines mainly arise from star formation using emission line ratios (e.g.\ [N~{\sc ii}]/H$\alpha$ and [O~{\sc iii}]/H$\beta$). However, these regions are typically not covered by our spectra, with only the [O~{\sc ii}] line well covered, so they cannot be conclusively filtered out.

The \texttt{FAST} \citep{2009ApJ...700..221K} code fits stellar population synthesis templates to both spectra and photometry. The code gives best-fit galaxy parameters, such as stellar age, stellar mass, extinction and star formation rate, with the errors on the derived parameters from Monte Carlo simulations. We utilised \texttt{FAST}, with the filters from \texttt{EAZY} \citep{2008ApJ...686.1503B}, on our SN~Ia host galaxy data. The use of \texttt{FAST} allowed us to simultaneously fit the \textit{HST} photometry presented in Table~\ref{tab:phot} and the calibrated ground-based spectra to constrain the stellar mass, SFR, internal extinction and stellar age of each galaxy. The inclusion of the spectra in the fitting tightens the constraints significantly, by breaking the degeneracy between reddening and age that arises from only fitting photometry. In some cases the uncertainties on the derived parameters are significantly influenced by the uncertainties on the \textit{HST} photometry, which can be relatively high in the F814W band.

\begin{figure}
\includegraphics[width=\columnwidth]{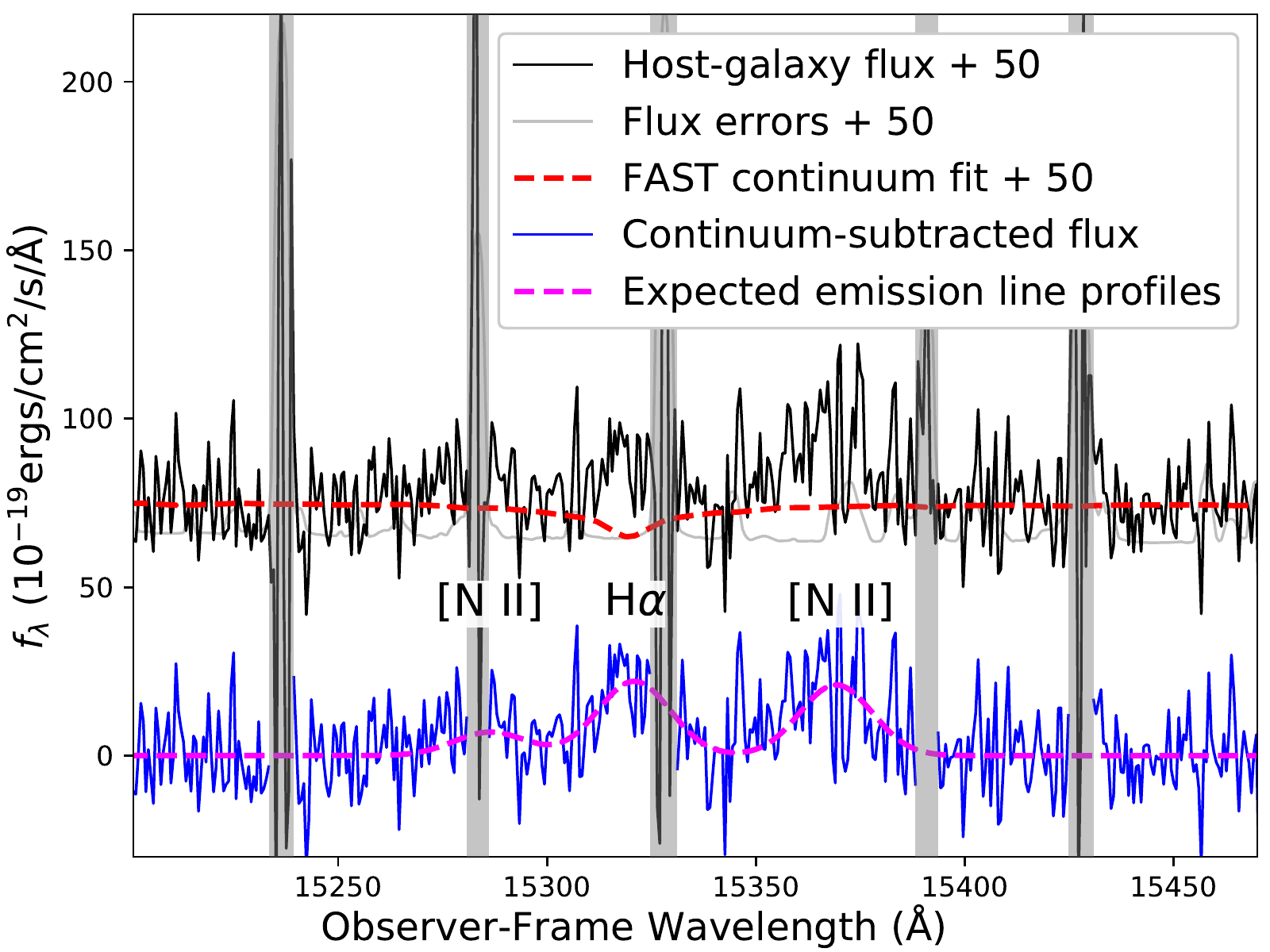}
\caption{Unbinned continuum-subtracted X-Shooter spectrum of the region around H$\alpha$ for the SCP15A04 and SCP16A04 host galaxy. The galaxy flux, errors and continuum fit are offset above. For visualisation, the expected (i.e.\ fixed [N~{\sc ii}] 6548/6583\,\AA\ ratio etc.) emission line profiles of H$\alpha$, along with [N~{\sc ii}] 6548 and 6583\,\AA\ are shown in magenta. When calculating the emission line fluxes in the X-Shooter data, regions affected by sky lines (indicated by the shaded grey regions) are not included in the measurement.}\label{fig:em}
\end{figure}

As \texttt{FAST} fits only the stellar population of a galaxy (and thus does not consider emission lines), we use the best-fit \texttt{FAST} SED of each galaxy as the continuum from which to measure any emission. Before the measurement is made, we normalise the flux of the \texttt{FAST} SED to the data in the region around the line of interest (but excluding the line itself). As \texttt{FAST} fits the entire spectrum and photometry, we do this in order to remove any systematic offset that may arise in specific regions of the spectrum from the fitting. If no emission line is obvious, we force an emission line measurement assuming a Gaussian with FWHM = $200\pm100$ km\,s$^{-1}$, which is convolved with the resolution of the spectra. The difference (positive or negative) in the spectrum flux from the \texttt{FAST} SED is then extracted using a weighted Gaussian. The uncertainty in the emission line measurement is then derived from a propagation of the errors on the spectra and the uncertainty in the \textit{HST} photometry used in the absolute flux calibration. Using the \texttt{FAST} SED in this way to aid in the emission line measurements has two advantages: 1) our spectra have relatively low S/N, so determining the continuum level in a small region near a given emission line can be difficult, and 2) it effectively corrects for stellar absorption, which is particularly important in our measurements of H$\alpha$ and H$\beta$ emission. An example of an emission line measurement is shown in Fig.~\ref{fig:em}. In the case of the (higher resolution) X-Shooter spectra, where the sky lines are narrow, the wavelengths that are badly affected by sky lines are masked from the emission line measurements. In such cases, the excess flux is extracted using a weighted Gaussian with those wavelengths excluded, with the total flux and errors then propagated through to account for the absent wavelength pixels.

The emission line measurements are converted into luminosities assuming a flat cosmology with $H_0=70$\,km\,s$^{-1}$\,Mpc$^{-1}$ and $\Omega_{\mathrm{M}}=0.3$. They are corrected for foreground Galactic extinction using extinction map values from \citealp{2011ApJ...737..103S}. To convert the emission-line luminosities into SFRs, we must also correct for host-galaxy extinction. The host-galaxy extinction constraints derived by \texttt{FAST} are for the stellar population. On average, the stellar population of a galaxy will be subject to lower dust extinction than the emission lines. In order to derive SFRs from our emission lines, we use $E(B-~V)_{\mathrm{stellar}}~=~0.44\times~E(B-~V)_{\mathrm{nebular}}$ from \citet{2000ApJ...533..682C}. We assume a \citet{1989ApJ...345..245C} reddening law with $R_V=3.1$ for both foreground and host galaxy extinction. These resulting dust-corrected line luminosities are then converted to SFRs using $\mathrm{SFR}=5.45\times 10^{-42}\, L(\textrm{H}\alpha)$~(M$_\odot$/yr)/(erg/s) from \citet{calzetti10} and $\mathrm{SFR}=2.65\times 10^{-41}\, L(\textrm{[O~{\sc ii}]})$~(M$_\odot$\,yr$^{-1}$)/(erg\,s$^{-1}$) from \citet{meyers12}. We also use H$\beta$ to determine SFR by assuming Case~B recombination (i.e.\ L(H$\alpha$)/L(H$\beta$)=2.86 at 10$^4$\,K) and the \citet{calzetti10} H$\alpha$ relationship, resulting in $\mathrm{SFR}=1.56\times 10^{-41}\, L(\textrm{H}\beta)$~(M$_\odot$/yr)/(erg/s). We caution here that these derived SFRs assume no emission line contamination (see discussion above).

\section{Results} \label{sec:res}
Here we present the See Change host-galaxy spectra and derived parameters. The host structure and where each SN resides with respect to the host can be seen in Appendix~\ref{appendix}.

\begin{table*}
	\centering
	\caption{Details of host galaxy spectra of See Change transients that are classified as non-SNe~Ia.
	}
	\label{tab:nonia}
	\begin{tabular}{lcc}
\hline
	SCP Transient &Redshift &Host features identified\\
\hline
	SCP15A02 &0.8966 &[O~{\sc ii}] 3727\,\AA, Ca~{\sc ii} H\,\&\,K \\
	SCP15A07 &0.5015 &H$\alpha$ (photo-z constrains to $z<1$, implying single emission line is H$\alpha$) \\
	SCP16A01 &0.4998 &H$\beta$, [O~{\sc iii}] 4959 \& 5007\,\AA, H$\alpha$ \\
	SCP16A02 &0.7971 &[O~{\sc ii}] 3727\,\AA, Ca~{\sc ii} H\,\&\,K, H$\beta$ \\
	SCP16A03 &1.335? &[O~{\sc ii}] 3727\,\AA? (Single strong emission line, alternatively H$\alpha$ at $z=0.326$) \\
    SCP15C02 &0.7506 &[O~{\sc ii}] 3727\,\AA, Ca~{\sc ii} H\,\&\,K, H$\beta$ \\
    SCP15C03 &-- &Emission line at $\sim$7490\,\AA\, possibly [O~{\sc ii}] 3727\,\AA\ \\
    SCP16C01 &0.9739 & [O~{\sc ii}], H$\epsilon$, H$\delta$, H$\gamma$, H$\beta$, [O~{\sc iii}] 4959 \& 5007\,\AA \\
    SCP16C02 &-- &{\it None} \\
	SCP16D02 &-- &\textit{None}\\
	SCP16D03 &0.6118 &[Ne~{\sc iii}] 3869\,\AA, H$\delta$, H$\gamma$, H$\beta$, [O~{\sc iii}] 4959 \& 5007\,\AA\ \\
	SCP15E01 &0.8398 &[O~{\sc ii}] 3727\,\AA, [Ne~{\sc iii}] 3869\,\AA, H$\gamma$, H$\beta$, [O~{\sc iii}] 4959 \& 5007\,\AA \\
    SCP15E02 &-- &{\it None} \\
	SCP15E05 &-- &\textit{None}\\
	SCP16E02 &0.9435 &[O~{\sc ii}] 3727\,\AA, [O~{\sc iii}] 4959 \& 5007\,\AA \\
\hline
\end{tabular}
\end{table*}

\begin{table}
	\centering
	\caption{Redshifts and emission line flux measurements for SN~Ia, likely SN~Ia and possible SN~Ia host-galaxies observed by the VLT. These measurements are not corrected for reddening.}
	\label{tab:z}
	\begin{tabular}{lcccc} 
\hline
	Transient &Type &Redshift &[O~{\sc ii}] flux &H$\alpha$ flux\\
    & & & \multicolumn{2}{c}{($\times10^{-18}$\,erg\,s$^{-1}$\,cm$^{-2}$)}\\
\hline
    SCP15A01 & SN~Ia &-- &N/A &N/A\\
	SCP15A03 & SN~Ia &1.3395 &$3.2\pm1.2$ &--\\
	SCP15A04 & SN~Ia &1.3345 &$22.9\pm3.9$ &$50.6\pm10.6$\\
	SCP15A05 & SN~Ia &1.3227 &$12.1\pm6.7$ &$5.2\pm3.8$\\
	SCP15A06 & SN~Ia &1.3128 &$-0.3\pm5.6$ &--\\
	SCP16A04$^{a}$ & SN~Ia &1.3345 &$22.9\pm3.9$ &$50.6\pm10.6$\\
    SCP14C01 & SN~Ia &1.23 &\multicolumn{2}{c}{{\it No detected host galaxy}}\\
    SCP15C01 & SN~Ia &0.9718 &$29.7\pm3.5$ &--\\
    SCP15C04 & SN~Ia &1.2305 &$-2.1\pm1.8$ &--\\
    SCP16C03 & SN~Ia &2.2216 &\multicolumn{2}{c}{{\it Background SN~Ia$^{b}$}}\\
    SCP15D01 &SN~Ia &0.5682 &-- &--\\
	SCP15D02 & poss~Ia &1.1493 &$18.7\pm2.6$ &--\\
	SCP15D03 & SN~Ia &1.1130 &$4.3\pm2.1$ &--\\
    SCP15D04 & SN~Ia &-- &N/A &N/A\\ 
	SCP16D01 & SN~Ia &1.1289 &$4.2\pm2.4$ &--\\ 
    SCP15E03 &poss~Ia &-- &N/A &N/A\\ 
    SCP15E04 &likely~Ia &-- &N/A &N/A\\
	SCP15E06  &SN~Ia &1.484 &$-0.6\pm1.1$ &--\\
	SCP15E07 &SN~Ia &1.4858  &$-0.2\pm1.8$ &$5.7\pm3.2$\\ 
    SCP15E08 & SN~Ia &2.02 &\multicolumn{2}{c}{{\it Broad emission at $\sim$8460\,\AA}}\\ 
	SCP15G01 & SN~Ia &1.5817$^{c}$ &-- &$5.5\pm3.8$\\
	SCP16H01 & SN~Ia &-- &N/A &N/A\\
  	SCP15I02 &poss~Ia &-- &N/A &N/A\\
\hline
\end{tabular}
\subcaption*{\textbf{Footnotes.} (a)~SCP16A04 had the same host galaxy as SCP15A04. (b)~A detailed analysis of SCP16C03 is presented in \citet{2017arXiv170704606R}. (c)~The redshift of the SCP15G01 host galaxy is mainly from Keck data, which will be presented elsewhere, however, the H$\alpha$ constraint is from our VLT data.}
\end{table}

\subsection{The spectroscopic redshift of MOO1014}
\begin{figure}
\includegraphics[width=\columnwidth]{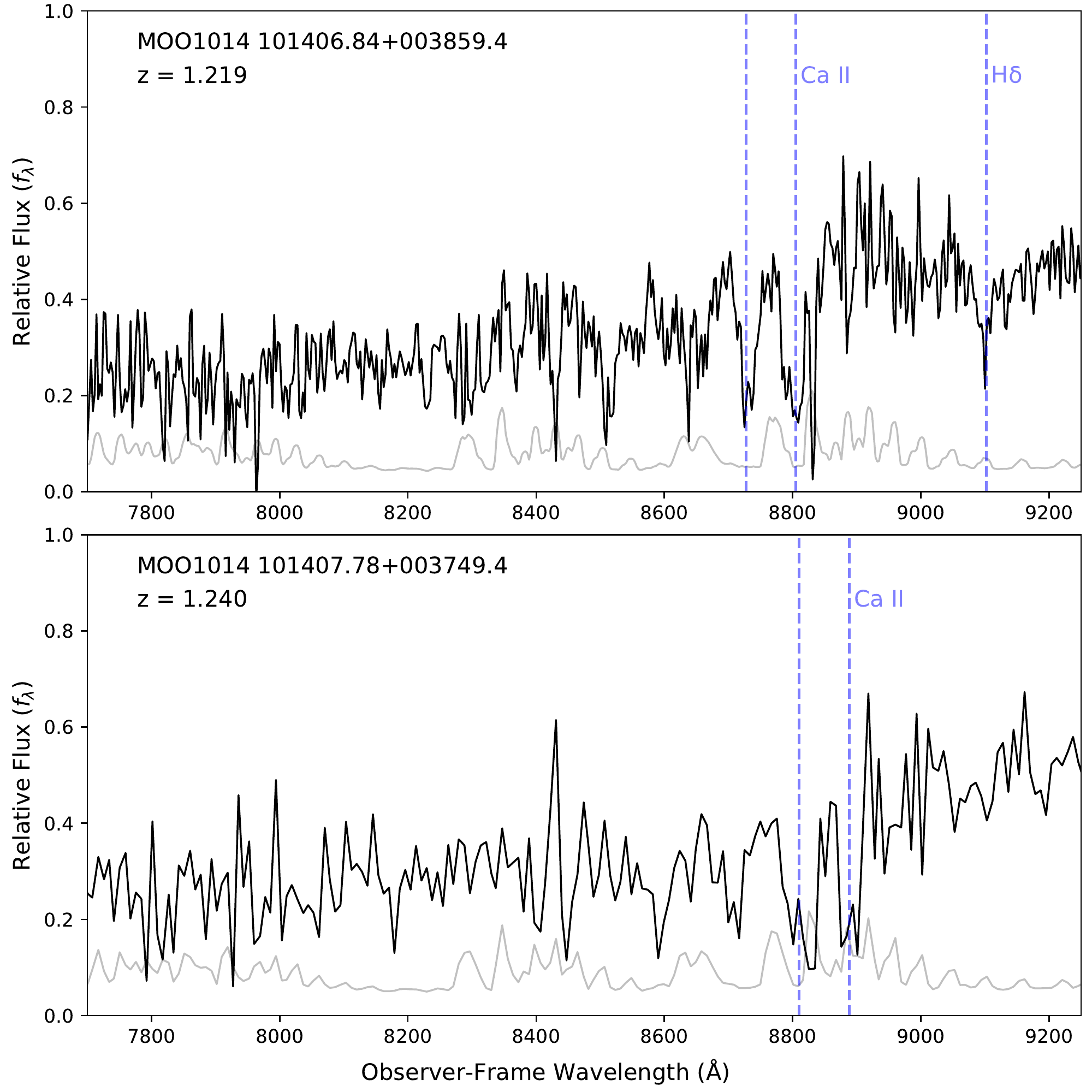}
\caption{Spectra of two galaxies in the field of cluster MOO1014 that did not host a transient during the See Change survey. These both show Ca~{\sc ii} H\,\&\,K absorption, enabling us to derive redshifts. The spectra are represented by black lines, with the grey lines showing the associated error spectra. The wavelengths of selected absorption lines are indicated.}\label{moocluster}
\end{figure}

At the beginning of the See Change survey, MOO1014 had a photometric redshift of $1.27\pm0.08$ \citep{2015ApJ...806...26B}, but no published spectroscopic redshift. We therefore used the spectra of galaxies observed with FORS2 to estimate the redshift of the cluster. Our spectra of two galaxies that did not host a See Change transient are displayed in Fig.~\ref{moocluster} and have redshifts of $z=1.219$ and $1.240$. Additionally, the host galaxy of SN~Ia SCP15C04 is at $z=1.232$ (see Section~\ref{sec:15c04} for further details on that host). These three galaxies are at a average redshift of $z=1.230$, essentially identical with the MOO1014 redshift of $z=1.231$ published by \citet{2019ApJ...878...72D}.

\subsection{Host galaxies of See Change transients that are not SNe~Ia}

The redshift of each host galaxy of a See Change transient that is not believed to be a SN~Ia is summarised in Table~\ref{tab:nonia}. The conclusion that these are not likely to be SNe~Ia is based on all of the available data, including the host spectroscopy itself, which in some cases directly helped break degeneracies in the photometric classification. The photometric classification of each See Change transient will be presented in \cosmolt. The individual spectra of galaxies that hosted non-SN~Ia transients during the See Change survey are shown in Fig.~\ref{non1}. Below we discuss specific cases where either the transient is worth noting, or the redshift determination requires further explanation.

\begin{figure*}
    \centering
    \begin{subfigure}[t]{\textwidth}
        \includegraphics[width=\linewidth]{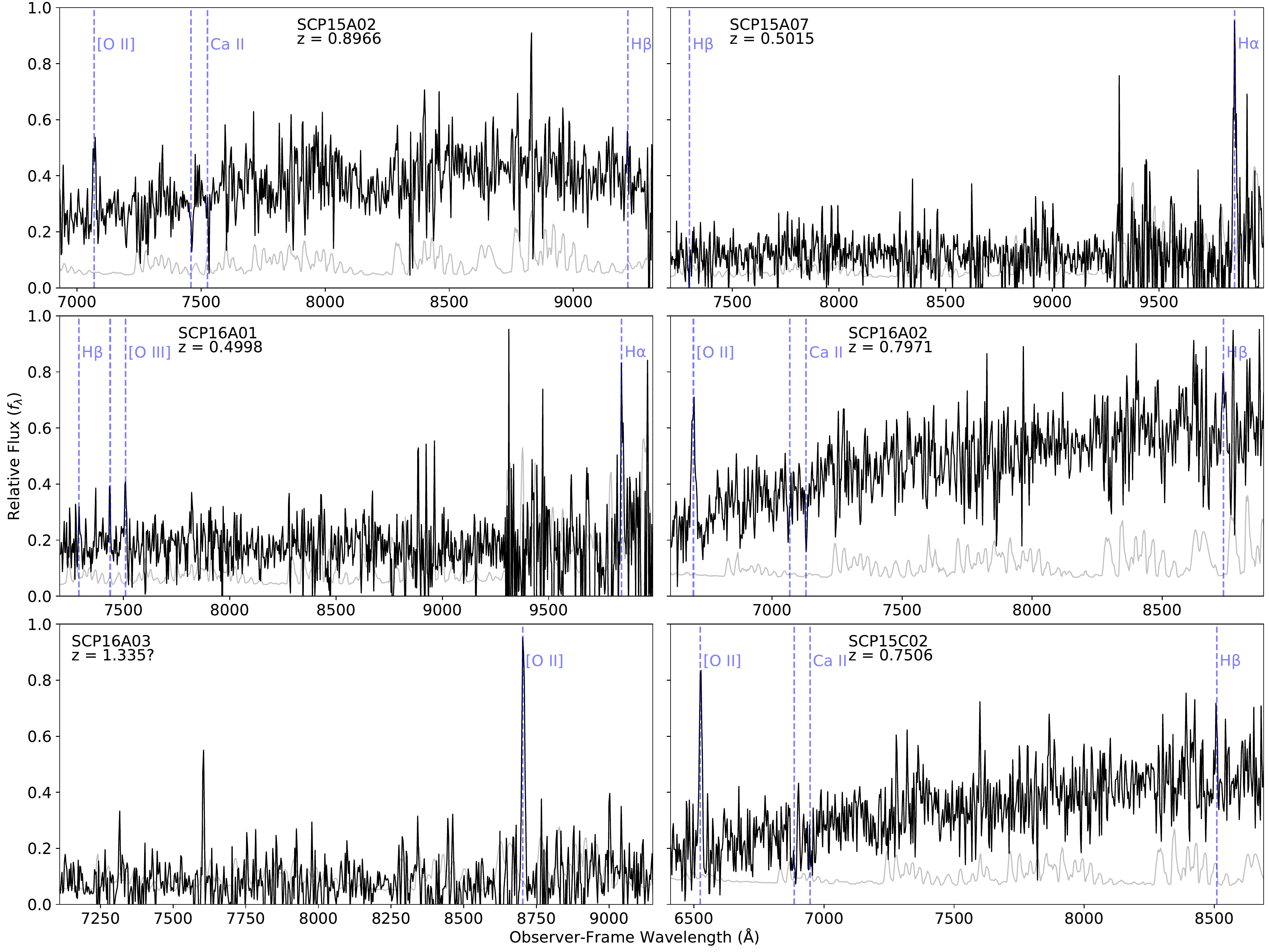}
    \end{subfigure}
\caption{The FORS2 spectra of the host galaxies of non-SN~Ia transients found during the See Change search. Key as in Fig.~\ref{moocluster}.}
\end{figure*}
\begin{figure*}\ContinuedFloat
    \begin{subfigure}[t]{\textwidth}
         \includegraphics[width=\linewidth]{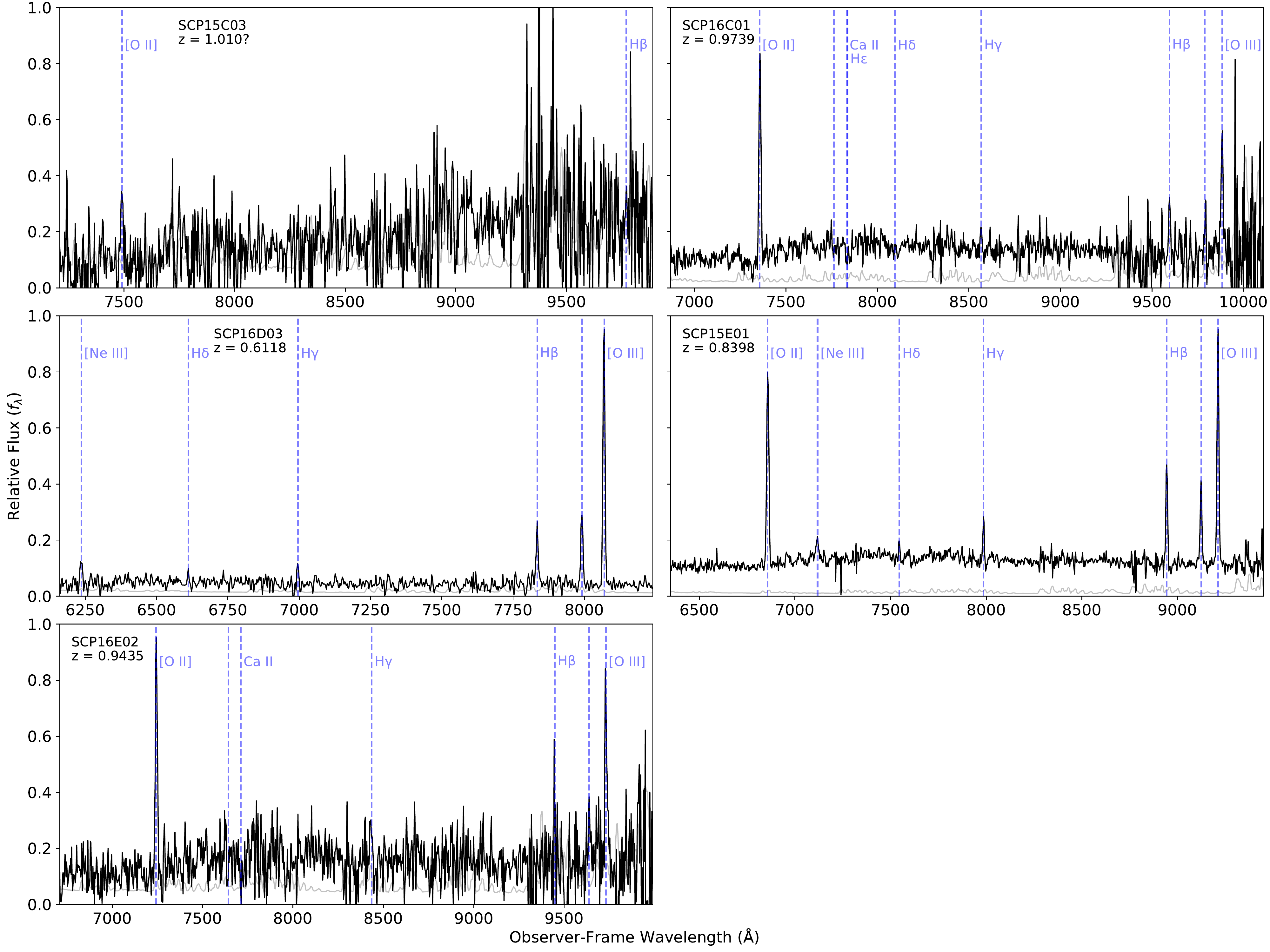}
    \end{subfigure}
\caption{\textbf{Continued.} The FORS2 spectra of the host galaxies of non-SN~Ia transients found during the See Change search. Key as in Fig.~\ref{moocluster}}
\label{non1}
\end{figure*}

The host galaxy of SCP15A07 has a single strong emission line, consistent with H$\alpha$ at $z=0.5015$. The line would also be consistent with [O~{\sc ii}] at $z=1.644$, but the photometric redshift strongly favours the lower redshift. The host galaxy of SCP16A03 shows a very strong emission line at 8705\,\AA. If this strong emission line is [O~{\sc ii}] 3727\,\AA, the galaxy is at $z=1.335$ and in the SPT0205 cluster. However, if the transient is in the cluster, it is too bright in the F814W {\it HST} observations (which would be rest-frame {\it U}-band) to be a normal SN~Ia, so may be a luminous SN~II. Alternatively the line may be H$\alpha$ at $z = 0.326$. As can be seen in Fig.~\ref{app1}, there is another galaxy very close to the host of SCP15C03, with the spiral arm extending to overlap the SN host galaxy. The emission line at $\sim$7490\,\AA\ could be associated with the spiral arm of this other galaxy and not the SN host. It is worth noting that at the cluster redshift of $z=1.23$, a 4000\,\AA\ break would be at $\sim$8900\,\AA. It can be seen from Figure\,\ref{non1} that there appears higher continuum flux red-ward of 8900\,\AA\ compared to bluer wavelengths. From the data in hand though, it is not possible to conclusively confirm the redshift of the SCP15C03 host galaxy, hence we do not report a redshift in Table~\ref{tab:nonia}.

\subsection{See Change SN~Ia hosts}\label{sec:clusteria}

The host redshifts of each SN~Ia found during the See Change survey, and emission line measurements, are summarised in Table~\ref{tab:z}. The individual galaxy spectra are presented and discussed below. Note that emission line fluxes quoted in this section are as measured, and not corrected for extinction. However, any line ratios or SFR constraints have been corrected for extinction, as outlined in Section~\ref{sec:galpar}.

\begin{figure*}
    \centering
    \begin{subfigure}[t]{\textwidth}
        \includegraphics[width=\linewidth]{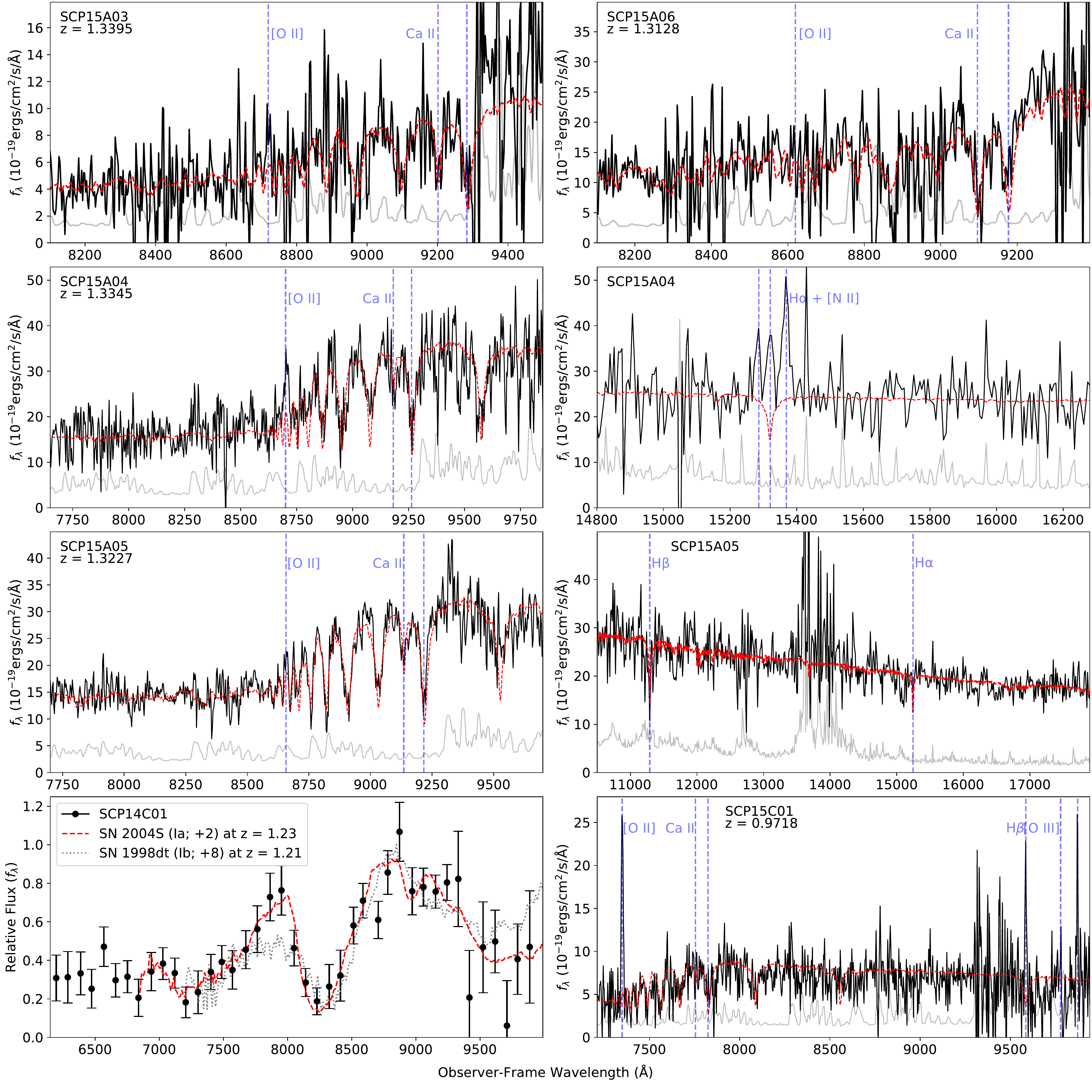}
    \end{subfigure}
\caption{The VLT spectra (FORS2 and/or X-Shooter) of the host galaxies of SN~Ia discovered during the See Change survey. The (solid) black line shows the data, with the errors on the spectrum shown as the (solid) grey line. The best fitting \texttt{FAST} SED is shown as a (dashed) red line. The positions of selected lines are also indicated. SCP14C01 had no detected host galaxy, this spectrum being a live spectrum of the transient itself. This is compared to spectra of SN~Ib~1998dt \citep{2001AJ....121.1648M,2012MNRAS.425.1789S} and SN~Ia~2004S \citep{2007AJ....133...58K}, with the days after peak and redshift of the comparison spectra indicated in the parentheses. The comparison spectra are simply shifted in redshift, and are not warped/adjusted in shape to match SCP14C01.}
\end{figure*}
\begin{figure*}\ContinuedFloat
    \begin{subfigure}[t]{\textwidth}
         \includegraphics[width=\linewidth]{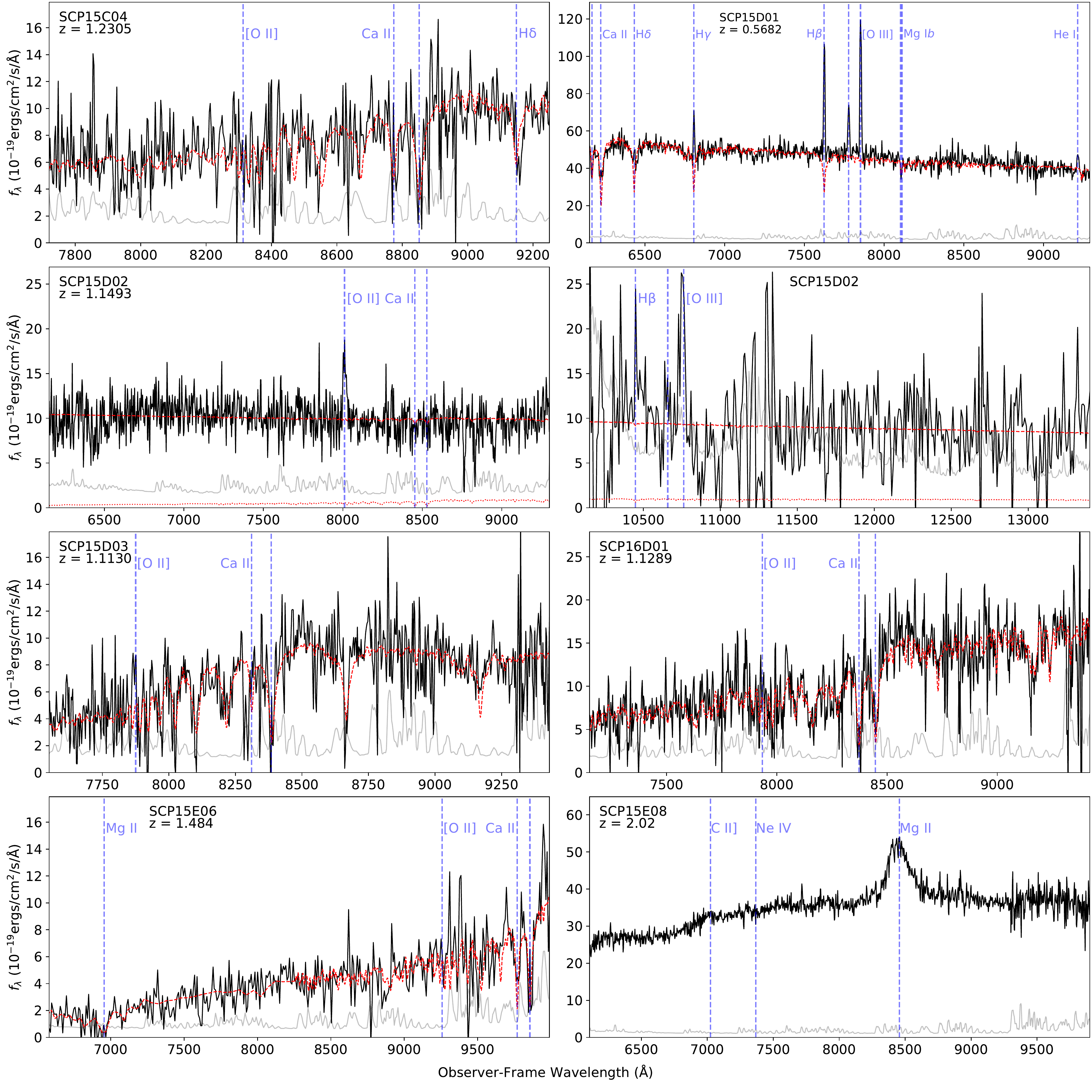}
    \end{subfigure}
\caption{\textbf{Continued.} The VLT spectra of the host galaxies of SNe~Ia discovered during the See Change survey. For the case of SCP15D02, the red dashed line shows the combined stellar+AGN best fit, with the stellar component of this fit shown by the red dotted line.}
\end{figure*}
\begin{figure*}\ContinuedFloat
    \begin{subfigure}[t]{\textwidth}
         \includegraphics[width=\linewidth]{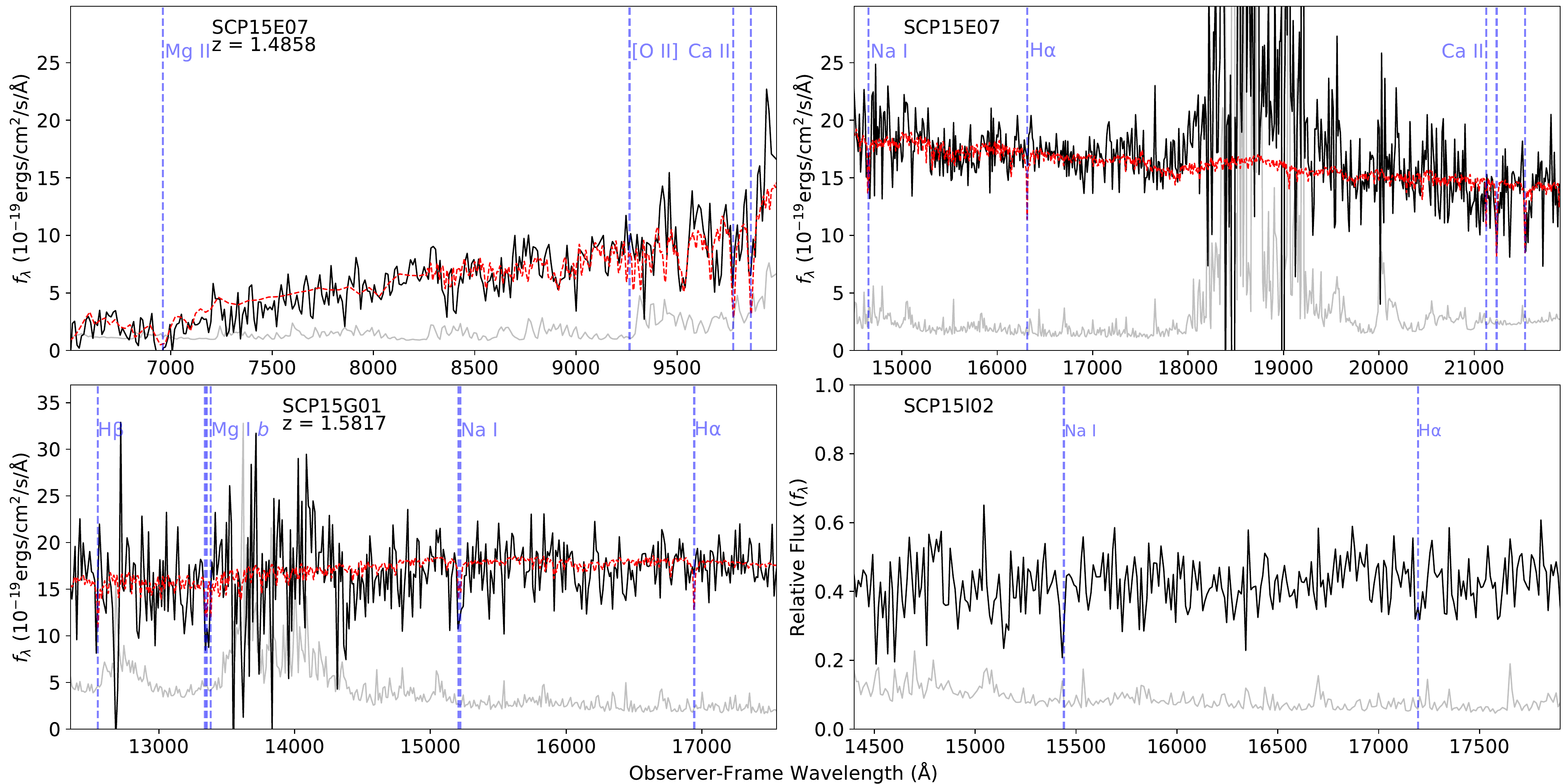}
    \end{subfigure}
\caption{\textbf{Continued.} The VLT spectra of the host galaxies of SNe~Ia discovered during the See Change survey. The spectrum of the host galaxy of possible SN~Ia SCP15I02 shows tentative evidence of Na~{\sc i}\,D and H$\alpha$ absorption at $z=1.62$, near the cluster redshift of $z=1.63$, but this is not a secure redshift.}
\label{fig:snia}
\end{figure*}

\subsubsection{SCP15A03}
The host galaxy of SN~Ia SCP15A03 was identified to be at $z=1.3395$, with the spectrum showing Ca~{\sc ii} H\,\&\,K and H9/CN absorption lines. The host spectrum, which is shown in Fig.~\ref{fig:snia}, shows weak evidence of [O~{\sc ii}] emission. Assuming this emission is associated with star formation, and not contaminated by other sources, implies a SFR of $12^{+7}_{-10}$,M$_{\odot}$\,yr$^{-1}$, whereas the \texttt{FAST} SED fitting finds a SFR of $0.0^{+0.3}_{0.0}$,M$_{\odot}$\,yr$^{-1}$. Given the large uncertainties in the emission line SFR measurement, this represents only a $\sim$1$\sigma$ disagreement. Using \texttt{FAST}, we derive a stellar age of $1.0^{+0.1}_{-0.3}$\,Gyr for this host galaxy.

\begin{table}
	\centering
	\caption{The \texttt{FAST} best-fit stellar parameters for the SCP15A04/SCP16A04 host galaxy, with and without including an AGN component in the fit.}
	\label{tab:15a04}
	\begin{tabular}{lcc}
\hline
	Parameter & Without AGN & With AGN\\
\hline
    log(M/M$_{\odot}$) &$10.67^{+0.14}_{-0.03}$ &$10.67^{+0.22}_{-0.10}$\\[0.1cm]
    $A_V$ &$1.2^{+0.2}_{-0.6}$ &$0.7^{+0.5}_{-0.1}$\\[0.1cm]
    SFR (M$_{\odot}$\,yr$^{-1}$) &$0.0^{+12.3}_{-0.0}$ &$0.0^{+21.4}_{-0.0}$\\[0.1cm]
    Stellar age (Gyr) &$0.2^{+0.5}_{-0.1}$ &$0.3^{+0.1}_{-0.2}$\\[0.1cm]
\hline
\end{tabular}
\end{table}

\subsubsection{SCP15A04 and SCP16A04}
During the See Change survey, one galaxy in SPT0205 produced two individual SNe~Ia, SCP15A04 and SCP16A04. The galaxy shows [O~{\sc ii}] emission in the optical (seen in both the FORS2 and X-Shooter spectra), along with Ca~{\sc ii} H\,\&\,K at $z=1.3345$. The X-Shooter NIR spectrum reveals emission lines from H$\alpha$, and [N~{\sc ii}] 6548 and 6583\,\AA. The measurement of these emission line fluxes is illustrated in Fig.~\ref{fig:em}, and we find log([N~{\sc ii}] $\lambda6583$/H$\alpha)=0.03^{+0.12}_{-0.17}$, which indicates there is likely at least some AGN contribution to the emission line flux (see e.g.\ \citealp{2001ApJ...556..121K,2003MNRAS.346.1055K}). H$\beta$ and [O~{\sc iii}] are in regions of very poor S/N, so no meaningful measurement can be made for those. If the H$\alpha$ flux is associated with star formation, a SFR of $23^{+11}_{-16}$\,M$_{\odot}$\,yr$^{-1}$ would be implied, although given the [N~{\sc ii}] $\lambda6583$/H$\alpha$ ratio, this should be considered an upper limit. The \texttt{FAST} SED fitting indicates a SFR of $0.0^{+12.3}_{-0.0}$\,M$_{\odot}$\,yr$^{-1}$, although this assumes that any AGN activity does not contribute significantly to the continuum. \texttt{FAST} implies a stellar age of $0.2^{+0.5}_{-0.1}$\,Gyr for this galaxy. Due to this evidence of AGN activity, we also run a modified version of \texttt{FAST} that includes an AGN component in the fitting \citep{2017MNRAS.465.3390A,2018MNRAS.474.1225A}. The results for with and without an AGN component in the fitting are summarised in Table~\ref{tab:15a04} and show that in this specific case, the parameters we derive for the two cases are consistent with each other.

\subsubsection{SCP15A05}
The host galaxy of SCP15A05 shows Ca~{\sc ii} H\,\&\,K at $z=1.3227$, along with several other absorption features at that redshift, including strong Balmer lines. We also detect Mg~{\sc ii} 2800\,\AA\ absorption. The FORS2 and X-Shooter spectra of this galaxy are shown in Fig.~\ref{fig:snia}. After removing the effects of a sky line in part of the expected emission profile as outlined in Section~\ref{sec:galpar}, we measure a H$\alpha$ flux of $(5.2\pm3.8)\times10^{-18}$\,erg\,s$^{-1}$. The H$\alpha$ measurement implies a SFR of $0.7^{+0.9}_{-0.6}$M$_{\odot}$\,yr$^{-1}$, consistent with the \texttt{FAST} estimate of $5.1^{+0.0}_{-5.1}$M$_{\odot}$\,yr$^{-1}$ from the stellar light. For this host galaxy we derive a stellar age of $1.0^{+0.3}_{-0.8}$\,Gyr. The detection of strong Balmer absorption lines indicates that A stars contribute a high fraction of the total optical light emitted by the galaxy, and thus helps to constrain the stellar age significantly.

\subsubsection{SCP15A06}
The host galaxy of SCP15A06 shows strong Ca~{\sc ii} H\,\&\,K absorption lines at $z=1.3128\pm0.0005$, similar to that found by \citet[][$z=1.3119\pm0.0005$]{2013ApJ...763...93S}, who also derived the redshift from the Ca~{\sc ii} lines. Our FORS2 spectrum is shown in Fig.~\ref{fig:snia}. The spectrum taken by \citet{2013ApJ...763...93S} can be seen in their fig.~3 (galaxy J020543.00--582936.4), which they used to derive a 3$\sigma$ upper limit on the star formation rate of $<0.63$\,M$_{\odot}$\,yr$^{-1}$ from the [O~{\sc ii}] flux, although note that this estimate does not include any correction for potential internal extinction (which we constrain to be $A_V=0.3^{+0.6}_{-0.1}$ for the stellar light). From our FORS2 spectrum we derive $\mathrm{SFR}<8.4$\,M$_{\odot}$\,yr$^{-1}$ from [O~{\sc ii}] constraints and $\mathrm{SFR} = 0.0^{+0.1}_{-0.0}$\,M$_{\odot}$\,yr$^{-1}$ from the \texttt{FAST} SED fitting. Using \texttt{FAST} we derive a stellar age of $1.3^{+1.9}_{-0.3}$\,Gyr for the galaxy.
 
\subsubsection{SCP14C01}
SCP14C01 was a SN discovered in our {\it HST} images taken 2014 Dec 26, with no detected host galaxy. As host galaxy spectroscopy to determine a precise redshift was not possible, we obtained live spectroscopy using director's discretionary time with FORS2 on 2015 Jan 20, which can be seen in Fig.~\ref{fig:snia}. We find the spectrum to be consistent with a SN~Ia at a redshift $z=1.23$. This makes it one of the most distant SNe~Ia with a live spectrum. Examples of higher redshift SNe~Ia with live spectra include SN~2002fw at $z=1.3$ \citep{2004ApJ...600L.163R}, SCP06G4 at $z = 1.35$ \citep{2010PASJ...62...19M}, and the lensed SNe HFF14Tom at $z=1.346$ \citep{2015ApJ...811...70R} and PS1-10afx at $z=1.388$ \citep{2013ApJ...767..162C,2013ApJ...768L..20Q}. SCP14C01 is the only See Change SN~Ia for which we were able to obtain a live spectrum.

Our FORS2 spectrum of SCP14C01 was analysed using the SN classification code \texttt{SNID}, which uses cross-correlation techniques to determine the type, age and redshift of SNe \citep{2007ApJ...666.1024B}. The results of this analysis show no confident matches. The FORS2 data were obtained between 57042.16 and 57042.27\,MJD, with SCP14C01 peaking $57041.9\pm0.8$\,MJD \cosmolp, meaning our spectra were taken when the SN was around peak brightness. When conservative peak constraints and loose redshift constraints ($0.1\leq z\leq2$) are placed on the \texttt{SNID} cross-correlation analysis, the best matching templates are then SNe~Ia in the region of $z=1.23$. We stress however that these matches all have values of the parameter $r{\mathrm{lap}}<5$ when \textit{lap} is constrained to $\geq0.4$, which \citet{2007ApJ...666.1024B} suggest are inconclusive. In Fig.~\ref{fig:snia}, we have compared our SCP14C01 spectrum to SN~Ia~2004S \citep{2007AJ....133...58K} and SN~Ib~1998dt \citep{2001AJ....121.1648M,2012MNRAS.425.1789S}\footnote{These were retrieved via the Open Supernova Catalog; \citet{2017ApJ...835...64G}; \url{https://sne.space}.}. From this analysis it seems clear that the transient is a SN~I at $z\sim1.23$, but from the spectra alone we cannot recover a conclusive classification for the SN type. However, when including light curve constraints we are able to confidently classify this as a SN~Ia \cosmolp. SCP14C01 could either be an intracluster SN~Ia, or one that resides in a low-luminosity host.

\subsubsection{SCP15C01} \label{sec:15c01}
SCP15C01 was a SN~Ia in the foreground of the MOO1014 cluster, with the redshift of $z=0.9718$. This redshift was derived mainly from [O~{\sc ii}], H$\beta$ and [O~{\sc iii}] emission lines, although higher order Balmer line absorption can also be seen in the continuum (see Fig.~\ref{fig:snia}). A SFR derived from the [O~{\sc ii}] line is very poorly constrained, due in large part to the very uncertain extinction internal to the host ($A_V=0.8^{+0.4}_{-0.8}$). For this host galaxy, we can also estimate the SFR from H$\beta$ which has a measured flux of $(2.16\pm0.30)\times10^{-17}$\,erg\,s$^{-1}$, after correcting for stellar absorption. This implies a SFR of $12^{+17}_{-10}$\,M$_{\odot}$\,yr$^{-1}$, which is still poorly constrained, even if better than is possible using [O~{\sc ii}]. The \texttt{FAST} SED fitting indicates a low star formation rate of $0.2^{+2.4}_{-0.2}$\,M$_{\odot}$\,yr$^{-1}$, which is still consistent with the poorly constrained emission line SFR. \texttt{FAST} gives a stellar age of $0.6^{+0.7}_{-0.4}$\,Gyr for the host galaxy of SCP15C01.

\subsubsection{SCP15C04} \label{sec:15c04}
The FORS2 spectrum of the SCP15C04 host galaxy, which is shown in  Fig.~\ref{fig:snia}, shows Ca~{\sc ii}  H\,\&\,K at $z=1.2305$. The \texttt{FAST} analysis indicates a small amount of star formation ($2.2^{+0.9}_{-1.4}$\,M$_{\odot}$\,yr$^{-1}$) and a stellar age of $4.0^{+0.0}_{-1.6}$\,Gyr. We do not detect [O~{\sc ii}] emission, although the resulting $2\sigma$ limit on the SFR of $<0.9$\,M$_{\odot}$\,yr$^{-1}$ is still broadly consistent with that determined from the SED. The SED indicates that this galaxy has very little internal extinction, with $A_V=0.2^{+0.1}_{-0.2}$.

\subsubsection{SCP16C03}
SCP16C03 was a background SN~Ia at $z=2.2216$, lensed by the MOO1014 cluster at $z=1.23$. This redshift was derived from our X-Shooter data, and makes SCP16C03 the highest redshift SN~Ia with a spectroscopic host-galaxy redshift. The photometric and spectroscopic follow-up, and a detailed analysis of SCP16C03 and its host galaxy is presented in \citet{2017arXiv170704606R}.

\subsubsection{SCP15D01}
SCP15D01 was a SN~Ia in the foreground of SPT2106. Multiple emission and absorption lines are visible in the FORS2 spectrum, from which we derive a redshift of $z=0.5682$ for the host galaxy. Both the SED fitting and the emission lines indicate some ongoing star formation. A SFR of $3.5^{+1.6}_{-1.3}$\,M$_{\odot}$\,yr$^{-1}$ was derived from the SED fitting, with a SFR of $6^{+4}_{-3}$\,M$_{\odot}$\,yr$^{-1}$ calculated from the strength of the H$\beta$ emission line. While SCP15D01 is a SN~Ia, it has a substantially lower redshift than our target SN~Ia sample of $z\gtrsim1$.

\subsubsection{SCP15D02}
SCP15D02 was a possible SN~Ia in the SPT2106 cluster. The host galaxy shows significant [O~{\sc ii}] emission at $z=1.1493$. Furthermore, the X-Shooter NIR spectrum shows an emission line consistent with  [O~{\sc iii}] 5007\,\AA\ at $z=1.1493$. At this redshift, H$\alpha$ is in a region of very high atmospheric absorption, so a measurement of that line is not possible. The FORS2 and X-Shooter spectra of the object can be seen in Fig.~\ref{fig:snia}. The [O~{\sc ii}] and [O~{\sc iii}] 5007\,\AA\ emission lines are both broad for a galaxy spectrum, with FWHM\,$\sim$\,700\,km\,s$^{-1}$. We measure log([O~{\sc iii}] $\lambda$5007/H$\beta)=0.3^{+0.2}_{-0.6}$, which does not tell us a great deal about potential AGN contribution (e.g.\ see Fig.~1 of \citealp{2003MNRAS.346.1055K}). Due to this, we use the version of the \texttt{FAST} code that includes an AGN component in the fitting. This indicates that potentially the majority of the continuum flux in our spectra could be from an AGN, which makes the derived stellar parameters very uncertain, with a stellar age of $2.5^{+2.4}_{-2.4}$\,Gyr, log(M/M$_{\odot}$) of $9.29^{+2.17}_{-0.60}$ and SFR of $0^{+513}_{-0}$\,M$_{\odot}$\,yr$^{-1}$. Due the likelihood of heavy AGN contamination of the emission lines, we do not calculate SFRs from the measured emission line fluxes.

\subsubsection{SCP15D03}
The host galaxy of SN~Ia SCP15D03 shows several absorption lines at $z=1.1130$ , including strong Balmer absorption. The spectrum has strong H$\delta$ absorption and is consistent with a relatively strong Balmer break. From visual inspection of Fig.~\ref{fig:snia}, it appears that Ca~{\sc ii}~K does not align with the redshift. However, the redshift does agree with the other Balmer lines and potential [O~{\sc ii}] emission. In galaxies with strong Balmer lines, Ca~{\sc ii}~K can be relatively weak (with Ca~{\sc ii}~H blended with H$\epsilon$), due to the domination of A stars over later type stars. The SED of the SCP15D03 host is consistent with no ongoing star formation ($0.0^{+0.1}_{-0.0}$\,M$_{\odot}$\,yr$^{-1}$). However there is a weak detection of [O~{\sc ii}] emission from the spectrum, which if all associated with star formation would imply a SFR of $4^{+4}_{-2}$\,M$_{\odot}$\,yr$^{-1}$. This is higher than the SED SFR, but still consistent within 2$\sigma$. The \texttt{FAST} analysis indicates a stellar age of $0.8^{+0.2}_{-0.1}$\,Gyr.

\subsubsection{SCP15D04}
\begin{figure}
\includegraphics[width=\columnwidth]{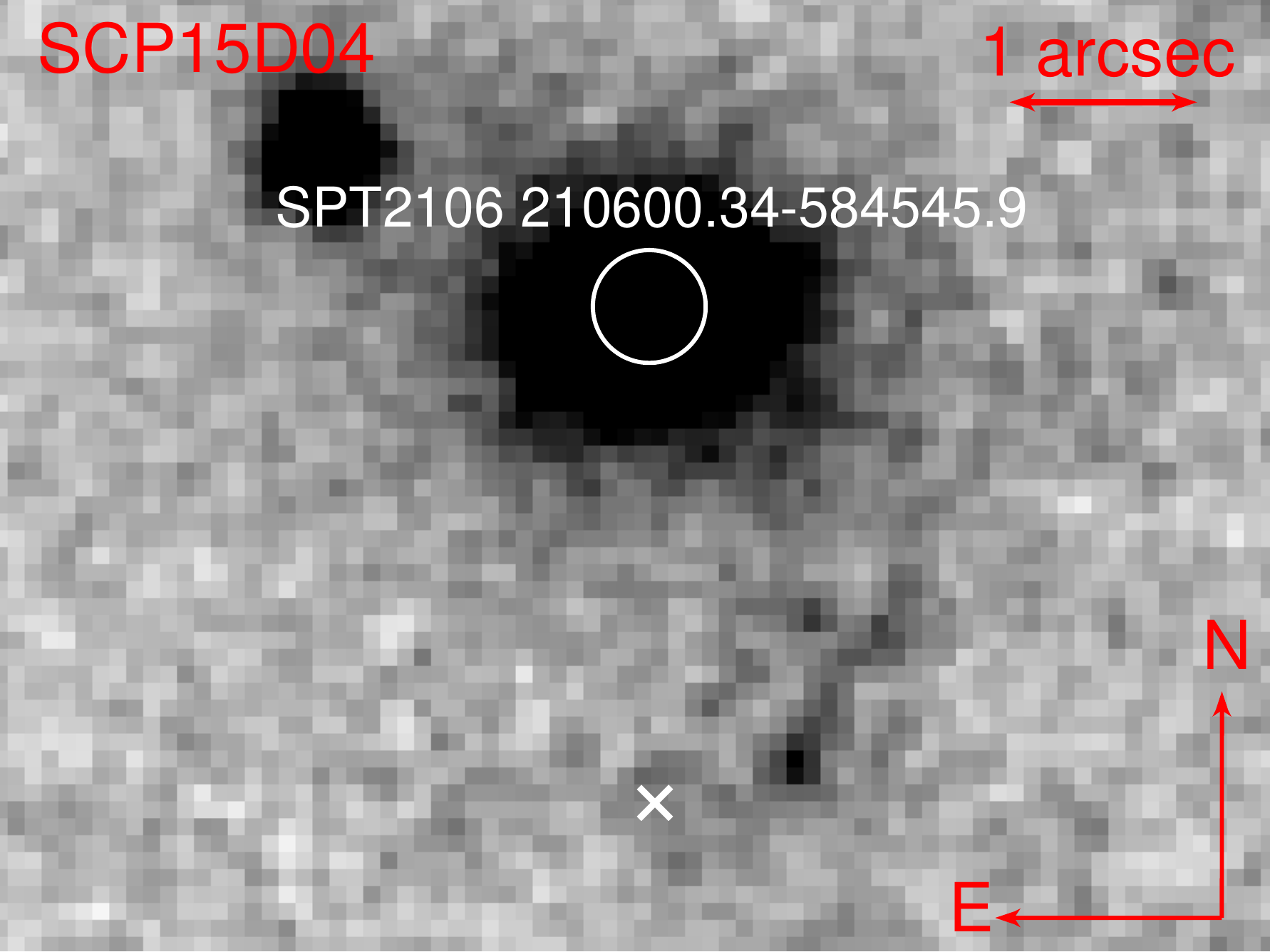}
\caption{Stacked {\it HST} WFC3/IR F105W image of the region around SCP15D04. The position of the SN is indicated by a white `$\times$'. The centre of the nearby galaxy SPT2106~210600.34--584545.9 at $z=1.135$ is indicated by the white circle.}\label{15d04hst}
\end{figure}

It is not immediately obvious which galaxy, if any, the transient SCP15D04 is associated with. Fig.~\ref{15d04hst} shows the position of the transient, with a faint extended source within $\sim1^{\prime\prime}$ and a much brighter source to the north. We obtained FORS2 spectroscopy of the very faint extended source slightly to the west of SCP15D04 to search for possible emission lines. However, none were found, so the redshift could not be determined. In some exposures, the slitlet went through the galaxy to the north of this faint source, with an exposure time of 3.33\,h. The spectrum shows that this galaxy, labelled here SPT2106~210600.34--584545.9, is in the SPT2106 cluster, displaying strong Ca~{\sc ii} H\,\&\,K absorption at $z=1.135$. Taking a redshift of $z=1.135$ would imply an apparent offset of 22\,kpc from SPT2106~210600.34--584545.9. SCP15D04 could therefore potentially be associated with that galaxy, even if the (apparently closer) faint extended source is not.

\subsubsection{SCP16D01}
The host galaxy of SCP16D01 shows strong Ca~{\sc ii} H\,\&\,K and G-band absorption, along with other absorption features. We find a redshift of $z=1.1289$ and it has a very strong cross-correlation peak, as would be expected given the strength and number of features. The spectrum of the host galaxy is shown in Fig.~\ref{fig:snia}, which shows a clear 4000\,\AA\ break. The stellar age of this galaxy derived from the SED fitting is $2.5^{+0.8}_{-0.3}$\,Gyr. This is significantly older than the stellar age of the SCP15D03 host in the same cluster. This can be seen in the spectra, where the SCP15D03 host shows strong Balmer absorption, rather than the strong 4000\,\AA\ break seen in the SCP16D01 host, due to the light of the former having a much stronger contribution from A stars. The SED fitting of the SCP16D01 host galaxy indicates a SFR of $0.0^{+0.3}_{-0.0}$\,M$_{\odot}$\,yr$^{-1}$, with a SFR of $4^{+4}_{-3}$\,M$_{\odot}$\,yr$^{-1}$ derived from the weak [O~{\sc ii}] detection.

\subsubsection{SCP15E06}
The FORS2 data of the SCP15E06 host galaxy shows a red continuum with Ca~{\sc ii} H~\&~K absorption lines, along with possible absorption from Mg~{\sc ii} 2800\,\AA\ also seen in the spectrum, which is shown in Fig.~\ref{fig:snia}. We adopt a redshift of 1.484 from the Ca~{\sc ii} H\,\&\,K lines. The cross-correlation prefers a slightly lower redshift of 1.481, however as this conflicts with Ca~{\sc ii} H~\&~K (the main features that can be seen by visual inspection), we use the redshift derived from Ca~{\sc ii} H\,\&\,K alone. Compared to most of the other objects, this has significant uncertainty in the host mass, with log$_{10}$(M/M$_{\odot})=10.94^{+0.29}_{-0.14}$, which in part is due to the uncertainty in the F814W \textit{HST} photometry. The stellar age of this host is also relatively uncertain at $1.6^{+2.4}_{-0.5}$\,Gyr, noting the age of the Universe at $z=1.48$ is only 4.3\,Gyr.

\subsubsection{SCP15E07}
We obtained both FORS2 and X-Shooter observations for the SCP15E07 host galaxy. A redshift is not immediately obvious by visual inspection. However, the cross-correlation yields a peak at $z=1.4858$. The Ca~{\sc ii} NIR triplet appears to be detected in the NIR data, as well as some optical features such as Ca~{\sc ii}~H\,\&\,K. The combined spectrum of the SCP15E07 host are shown in Fig.~\ref{xcorr}, with zoomed-in sections of the spectrum shown in Fig.~\ref{fig:snia}. We used \texttt{FAST} to derive a stellar age of $1.6^{+0.5}_{-0.7}$\,Gyr. Both the H$\alpha$ emission measurement and the \texttt{FAST} SED fitting indicate a low SFR for the SCP15E07 host galaxy, deriving $0.0^{+0.1}_{-0.0}$ and $0.6^{+0.4}_{-0.4}$\,M$_{\odot}$\,yr$^{-1}$ respectively for the two methods.

\subsubsection{SCP15E08}
The source near SCP15E08 has a bright continuum and a single emission line at $\sim8460$\,\AA. The emission line is very broad, with FWHM $\sim6500$\,km\,s$^{-1}$. The source was detected as an X-ray source by XMM in 2013 \citep{2016A&A...590A...1R}. Taken together, this implies that this galaxy is an AGN. As only one line is identified, the redshift is not completely secure, but the line is most likely to be Mg~{\sc ii} 2800\,\AA\ at $z=2.02$, which is consistent with some weaker features expected in a quasar spectrum. The continuum appears red and is likely subject to significant internal reddening. The spectrum, which is shown in Fig.~\ref{fig:snia}, shows the possible prescence of C~{\sc ii}] 2326\,\AA, and the position of [Ne~{\sc iv}] is added for reference. Note that the 2175\,\AA\ extinction bump would be blueward of C~{\sc ii}] 2326\,\AA. This is often not seen in quasar spectra \citep{2000PASP..112..537P}, but has been detected in some case (e.g.\ \citealp{2011ApJ...732..110J}), so caution should be excersised in interpreting broad changes in the continuum here. The photometry of SCP15E08 is consistent with a SN~Ia, likely at a higher redshift than the cluster ($z>1.6$), and is indeed consistent with a SN~Ia at $z=2.02$. The transient is not coincident with the centre of the AGN, so the new object is not due to variability in the central source itself.

\subsubsection{SCP15G01}
A redshift of $z=1.5817$ was obtained for the SCP15G01 host galaxy, mainly from Keck data (Hayden et al. in prep). Using this redshift, we can put limits on the H$\alpha$ emission with our VLT data. At this redshift, the region of the Ca~{\sc ii} H\,\&\,K lines has very poor sensitivity in X-Shooter. However, using this redshift, there does appear to be possible absorption from Na~{\sc i}\,D and Mg\,{\it b}. The \texttt{FAST} SED analysis indicates that this is a massive galaxy, with log$_{10}$(M/M$_{\odot}$)~=~11.44$^{+0.03}_{-0.15}$. The spectrum shows no obvious evidence of H$\alpha$ emission. A forced measurement of the excess H$\alpha$ flux over the fitted continuum yields $5.5\pm3.8\times10^{-18}$\,erg\,s$^{-1}$, implying SFR of $1.6^{+2.7}_{-1.3}$\,M$_{\odot}$\,yr$^{-1}$, consistent with the \texttt{FAST} rate of $0.0^{+17.4}_{-0.0}$\,M$_{\odot}$\,yr$^{-1}$. The VLT spectrum is shown in Fig.~\ref{fig:snia}. The SED indicates a stellar age of $4.0^{+0.0}_{-2.0}$\,Gyr, where the age of the Universe at $z=1.5817$ sets the upper limit on the stellar age.

\begin{figure}
\includegraphics[width=\columnwidth]{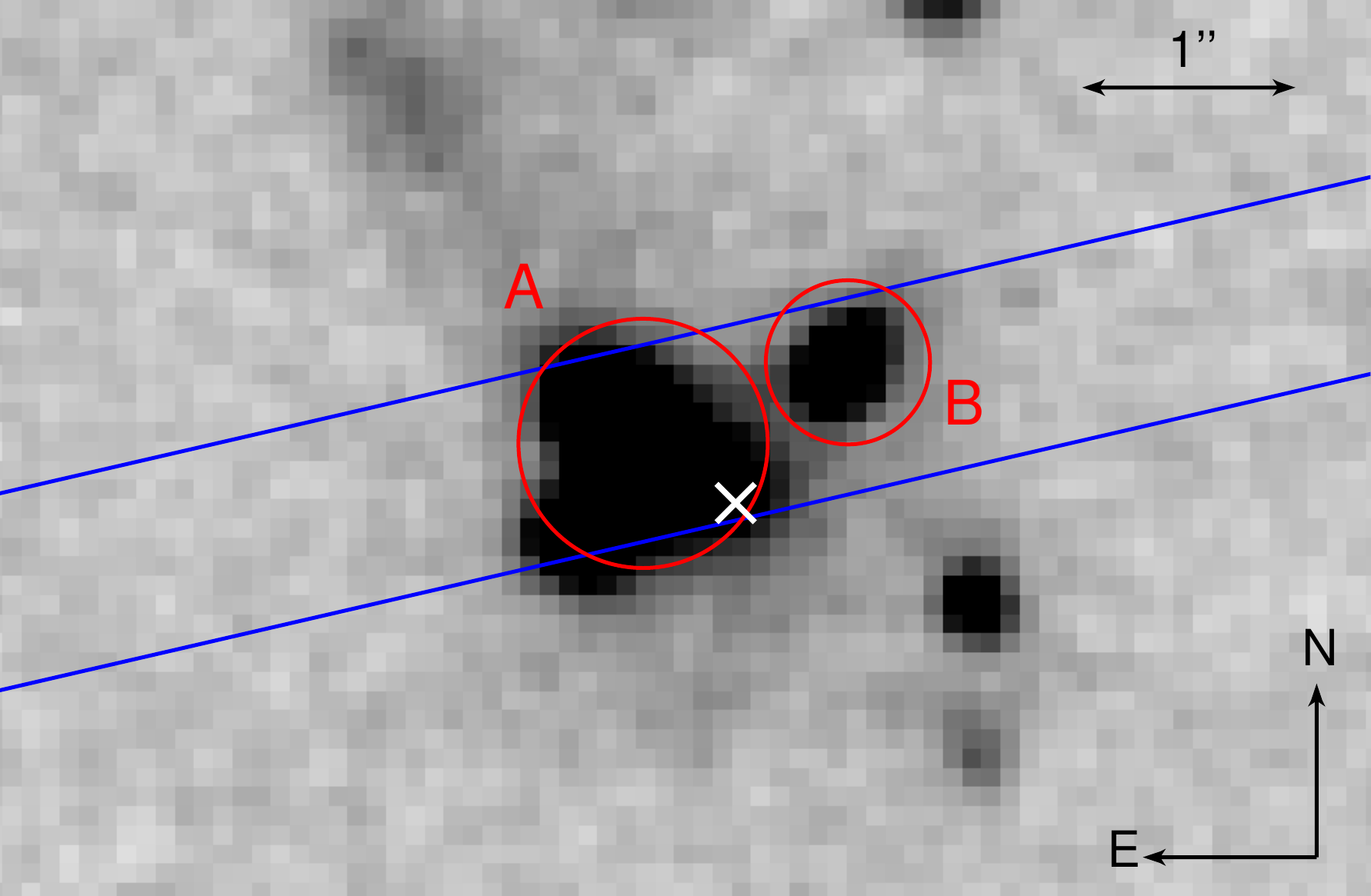}\vspace{0.2cm}
\includegraphics[width=\columnwidth]{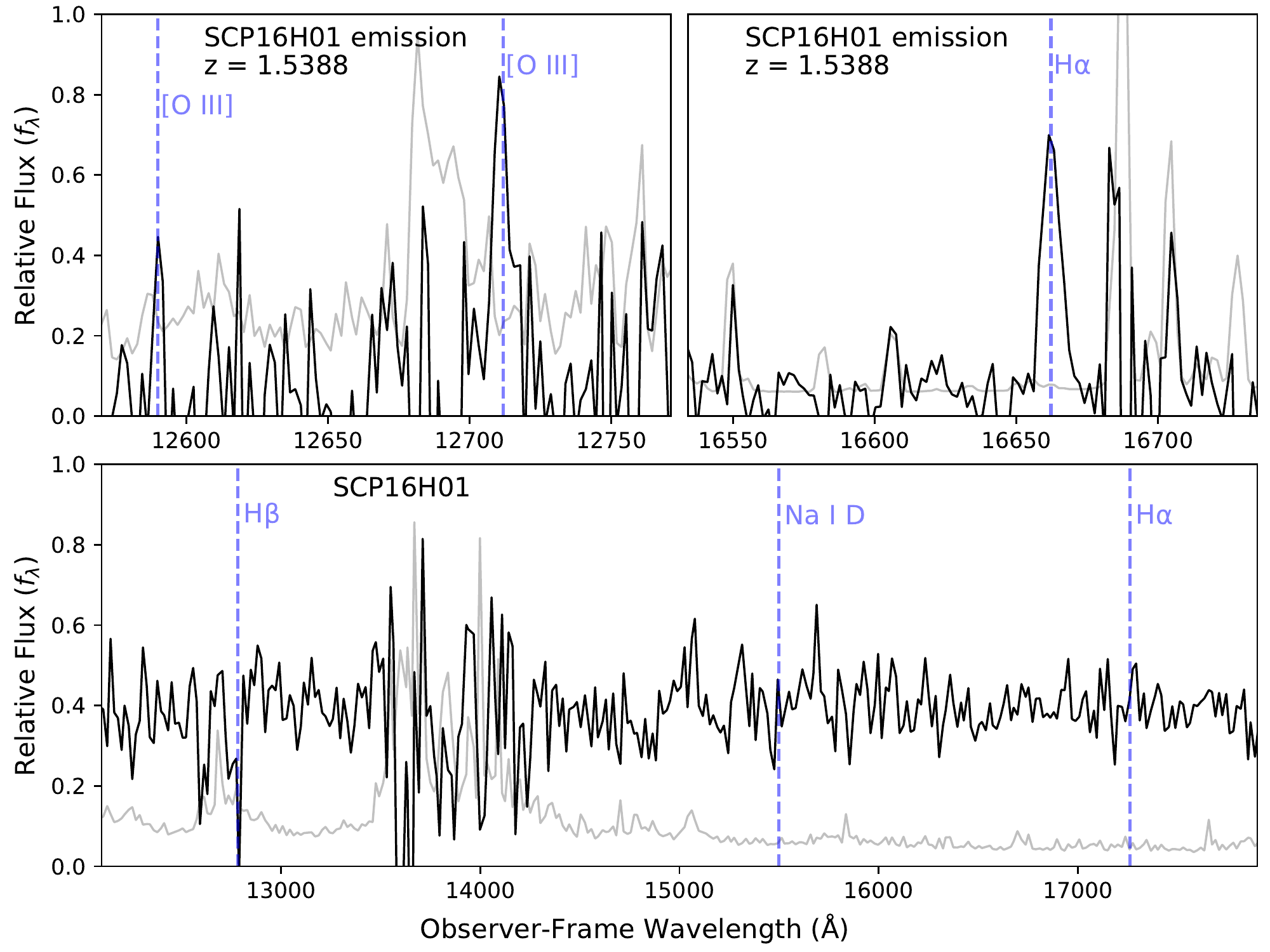}
\caption{\textbf{Upper:} Stacked {\it HST} WFC3/IR F105W image showing the host structure of SCP16H01. The two components with spectra are indicated in red (see text), the position of the X-Shooter slit is in blue and the position of SN~Ia SCP16H01 is indicated by a white `$\times$'. \textbf{Middle:} X-Shooter spectrum of Component~B of the SCP16H01 host structure, showing H$\alpha$ and [O~{\sc iii}] 5007\,\AA\ emission at $z=1.5388$. \textbf{Lower:} X-Shooter spectrum of Component~A of the SCP16H01 host structure, showing no secure redshift, with the expected positions of lines placed at the cluster redshift ($z=1.63$) indicated for reference. This spectrum is corrected for contamination by Component~B. Key as given in Fig.~\ref{moocluster}.}
\label{16h01}
\end{figure}

\subsubsection{SCP16H01}
The X-Shooter spectrum of the SCP16H01 host galaxy shows two components; the main continuum and emission lines slightly offset (in the spatial direction) from the continuum. The structure can be seen in Fig.~\ref{16h01}, with the position of the SN also indicated. The continuum is from component `A' and the H$\alpha$ and [O~{\sc iii}] 5007\,\AA\ emission lines are from component `B'. The accompanying [O~{\sc iii}] 4959\,\AA\ line is possibly present as well, but the larger errors and weakness of the line make this only tentative. This gives a redshift for component B of $z=1.5388$, which places the galaxy in the foregound of the cluster. These emission lines from component B are very narrow, with the strongest line, H$\alpha$, measured at FWHM $\sim100$\,km\,s$^{-1}$ after correcting for spectral resolution. This spectrum can be seen in Fig.~\ref{16h01}. Any possible contamination from the emission line object was subtracted from the main trace using the half of the emission line trace furthest spatially from the main trace. This may also subtract a small percentage of the main trace flux ($\lesssim$\,5\%). The errors are fully propagated through this subtraction, but do not increase significantly due to the optimal extractions employed; i.e.\ where the flux from the emission-line object is strongest spatially has a reasonable offset from the centre of the main trace and thus the flux/error at this pixel has a low weighting in the optimal combining. The spectrum of component A, after removal of the contamination, shows no evidence of features at the same redshift as component B. In conclusion, from the imaging we can see the SN is likely associated with component A, but the spectroscopic redshift of that component remains inconclusive.

\subsubsection{SCP15I02}
The host galaxy of possible SN~Ia SCP15I02 has a detected continuum. However, we have not been able to obtain a secure redshift from the data. The spectrum is shown in Fig.~\ref{fig:snia} and shows possible evidence of Na~{\sc i}\,D absorption near the cluster redshift of 1.63. At this redshift, X-Shooter has low sensitivity in the region of Ca~{\sc ii}~H\,\&\,K.

\section{Discussion} \label{sec:dis}

Our observing strategy for obtaining spectroscopic redshifts was partially focused on emission line fluxes; H$\alpha$ with X-Shooter and [O~{\sc ii}] with FORS2. The lack of strong emission lines found in these $z>1$ galaxies meant that obtaining the redshifts was more challenging, as we had to rely on absorption lines and therefore required higher S/N on the galaxy continuum. In the cases of the lowest luminosity hosts, obtaining a spectroscopic redshift was not possible. We have however been been able to obtain redshifts for 15 SNe~Ia and one possible SN~Ia at $z\gtrsim1$. Thirteen of these sixteen transients were in the target clusters, in addition to one in the foreground and two in the background. Looking towards future surveys, this shows that it is possible to obtain spectroscopic redshifts for the majority of SNe~Ia in cluster environments through host spectroscopy in the $1<z<1.5$ range. This often requires several hours of exposure, even on $8-10$\,m class telescopes.

As well as their use in the cosmological analysis, these redshifts were often needed to classify the transients. At these high redshifts, a live transient spectrum was typically not feasible, due to a combination of the heavy telescope time investment that would have been required, the difficulty scheduling such long observations when the SN was at peak and coinciding with optimal observing conditions. The full transient classifications will be presented in \cosmolt. We also obtained redshifts for most of the other transients discovered through See Change. Indeed the redshifts for these objects were sometimes needed to rule them out as cluster SNe~Ia, for example, to break a degeneracy from a light curve that could be consistent with a cluster SN~Ia or a foreground CC SN (e.g.\ SCP15C02; Hayden et al.\ in prep).

\subsection{Cluster Envirnoments of the SNe~Ia}
A clear result of our $z\gtrsim1$ cluster SN~Ia host spectroscopy is the apparent lack of star formation from the \texttt{FAST} SED fitting. There are only two cluster SN~Ia host galaxies with emission lines detected with $>$3$\sigma$ significance, the hosts of SCP15A04/16A04 and SCP15D02, both of which show evidence of AGN activity in the spectrum. The foreground $z=0.9718$ SCP15C01 host galaxy also shows significant [O~{\sc ii}] and H$\beta$ emission.

Another interesting result of the See Change programme is that some clusters show higher rates of detected SNe~Ia than others. The three SPT clusters have produced more than half of the cluster SNe~Ia, likely SNe~Ia or possible SNe~Ia found by See Change. The number of SNe~Ia in each See Change cluster that was also a potential VLT target are summarised in Table~\ref{tab:clust}. Of the 39 See Change transients that we attempted to obtain a  VLT spectroscopic redshfit, we were successful for 26 of those. For each cluster, we summarise the galaxy parameters that we derive for the cluster SN~Ia host galaxies below, and how they compare to other cluster members. The majority of the spectroscopic follow-up of the objects not feasible to target with the VLT (i.e.\ Dec. $\gg 0^{\circ}$) was done at the Keck observatory. The redshifts and stellar masses of these objects will be presented in a future publication.

\begin{table}
	\centering
	\caption{Summary of the number of SNe~Ia, likely SN~Ia or possible SNe~Ia in each cluster (not including those in the foreground or background), and the number of those for which we were able to derive a spectroscopic redshift.}
	\label{tab:clust}
	\begin{tabular}{lccc} 
\hline
	Cluster &Redshift &SNe~Ia & SNe~Ia with spec-\textit{z}\\
\hline
	SPT2106 (D) &1.13 &4 &3\\
	MOO1014 (C) &1.23 &2 &2\\
	SPT0205 (A) &1.32 &6 &5\\
	SpARCS0035 (F) &1.34 &0 &0\\
    	SPT2040 (E) &1.48 &4 &2\\
    	XDCP0044 (G) &1.58 & 1 &1\\
    	SpARCS0224 (I) &1.63 & 2 &1\\
    	SpARCS0330 (H) &1.63 & 1 &0\\
\hline
    Total & &20 &14\\
\hline
\end{tabular}
\subcaption*{\textbf{Notes.} The numbers for the cluster SNe~Ia \textit{without} a spectroscopic redshift are from the number of SNe~Ia that have a photometric redshift consistent with the SN~Ia residing in the observed galaxy cluster. We note that the one host galaxy with a spectroscopic redshift in cluster SpARCS0224 was not observed with VLT.}
\end{table}

{\it SPT0205} had six SNe~Ia detected. None of the SN~Ia host galaxies in SPT0205 show strong emission lines. The strongest lines are those seen in the host of SCP15A04/SCP16A04. The high [N~{\sc ii}] $\lambda$6583/H$\alpha$ ratio in this galaxy indicates they are likely influenced by AGN activity. In the four cluster SN~Ia host galaxies analysed in this work, we find stellar ages of $0.2^{+0.5}_{-0.1}$, $1.0^{+1.0}_{-0.3}$, $1.0^{+0.3}_{-0.8}$ and $1.3^{+1.9}_{-0.3}$\,Gyr. In their analysis of this cluster, \citet{2013ApJ...763...93S} concluded that the majority of the star formation took place at $z>2.5$. At the redshift of SPT0205, $z=2.5$ corresponds to an age of $\sim$2.1\,Gyr. Our results therefore imply that there has been more recent star formation in at least some members of this cluster. The hosts of SCP15A03, SCP15A04 (+ SCP16A04) and SCP15A05 all show strong high-order Balmer absorption. The SCP15A04 host is the only galaxy where [O~{\sc ii}] is detected at $>$3$\sigma$ significance, but we know from the [N~{\sc ii}]/H$\alpha$ ratio that at least some of this will likely come from AGN activity. The spectra of these three host galaxies are consistent with post-starburst spectra. The only spectrum where the Ca~{\sc ii}~H\,\&\,K/4000\AA\ break is more prominent in the spectra than the Balmer absorption lines is the host of SCP15A06. This is different to the spectra published by \citet[][see their fig.~3]{2013ApJ...763...93S}, where in general the Ca~{\sc ii}~H\,\&\,K/4000\AA\ break were the strongest features seen in the galaxies they observed. This could be a selection effect, given that the 9 galaxy spectra published by \citet{2013ApJ...763...93S} were the only ones confirmed from 47 slits of their spectroscopic observations and thus likely to be brighter/more massive. In contrast, our target was to observe the specific galaxies that happened to host a SN~Ia during the See Change survey. Another possible contributing factor is the delay time distribution of SNe~Ia, where the SN~Ia rate is dependent on the time since the hosting stellar population was formed (see e.g.\ \citealp{2008PASJ...60.1327T}). This in turn would cause a correlation between the global stellar age of a galaxy and the relative SN~Ia rate.

{\it SPT2106} produced two SNe~Ia, plus a possible SN~Ia. As with SPT0205, the majority of confirmed cluster members in SPT2106 appear to be passive, with no obvious [O~{\sc ii}] detection \citep{2011ApJ...731...86F}. Of our three SN host galaxies, only SCP15D02 shows strong [O~{\sc ii}] and [O~{\sc iii}] emission, with the emission lines being relatively broad and likely affected by AGN activity. The hosts of SCP15D03 and SCP16D01 show significantly different stellar populations. The spectrum of the SCP15D03 host galaxy shows a much younger population ($0.8^{+0.2}_{-0.1}$\,Gyr) than that of SCP16D01 ($2.5^{+0.8}_{-0.3}$\,Gyr), with the latter being $\sim$8 times more massive than the former.

{\it SPT2040} had two cluster SNe~Ia and one background SN~Ia that we were able to obtain redshifts for. In addition, there was a likely SN~Ia and possible SN~Ia for which we were not able to obtain a redshift. The FORS2 follow-up of galaxies in this cluster was challenging, as with the cluster redshift at $z=1.478$, the positions of Ca~{\sc ii} H\,\&\,K are red enough for the efficiency of the detector to be significantly reduced. However, we have been able to obtain redshifts for cluster SNe~Ia SCP15E06 and SCP15E07, the latter with the aid of X-Shooter data. The hosts of both SCP15E06 and SCP15E07 appear passive from the SED fitting, and have no detected [O~{\sc ii}] emission. A weak H$\alpha$ detection indicates a low SFR of $0.6\pm0.4$\,M$_{\odot}$\,yr$^{-1}$ for the host galaxy of SCP15E07. This is in contrast to the other confirmed cluster members, with \citet{2014ApJ...794...12B} finding significant ongoing star formation in SPT2040. They identified 15 cluster members, all of which showed [O~{\sc ii}] emission. This is from observations from 59 slits, implying a successful detection rate of [O~{\sc ii}] of $25\pm7\%$, significantly higher than that found in SPT0205 and SPT2106, which both had [O~{\sc ii}] detection rates of $2\pm2\%$ \citep{2014ApJ...794...12B}. However, the authors note that their SPT2040 spectra were not sufficiently sensitive to pick up Ca~{\sc ii} H\,\&\,K, implying a large bias towards finding emission line galaxies -- i.e.\ the 15 galaxies confirmed are not necessarily a representative sample of the average cluster galaxy. Note that this caveat does not affect the above [O~{\sc ii}] detection rate comparison made by \citet{2014ApJ...794...12B}. However, it does show that no conclusion can be drawn from the fact that, unlike the previously confirmed members, our two SN~Ia hosts in SPT2040 appear relatively passive, and indeed a meaningful comparison to our spectra is not possible. The implied stellar ages of the two galaxies are $1.6^{+0.5}_{-0.7}$ and $1.6^{+2.4}_{-0.5}$\,Gyr. In addition to the cluster SNe~Ia, we have also identified a background SN~Ia in the field of SPT2040, occurring in a broad-lined AGN host galaxy at $z=2.02$, SCP15E08.

\textit{MOO1014} only has one cluster SN for which we have been able obtain a host spectrum. The spectrum shows Ca~{\sc ii}~H\,\&\,K with evidence of a 4000\,\AA\ break. The SED fitting indicates an old stellar population, with the best-fit stellar age of $4.0^{+0.0}_{-1.6}$\,Gyr. The spectra of the two additional cluster galaxies that we confirm (but did not host a SN~Ia; shown if Fig.~\ref{moocluster}) also show strong Ca~{\sc ii}~H\,\&\,K absorption and evidence of a 4000\,\AA\ break.

The host galaxy of SCP15G01 does not show strong emission lines and the SED is consistent with an old stellar population, with a stellar age of $4.0^{+0.0}_{-2.0}$\,Gyr. Of the three spectra of cluster XDCP0044 presented by \citet{2011A&A...531L..15S}, two show [O~{\sc ii}] emission.

\subsection{Stellar Mass and SFR Measurements}
\begin{figure}
\includegraphics[width=\columnwidth]{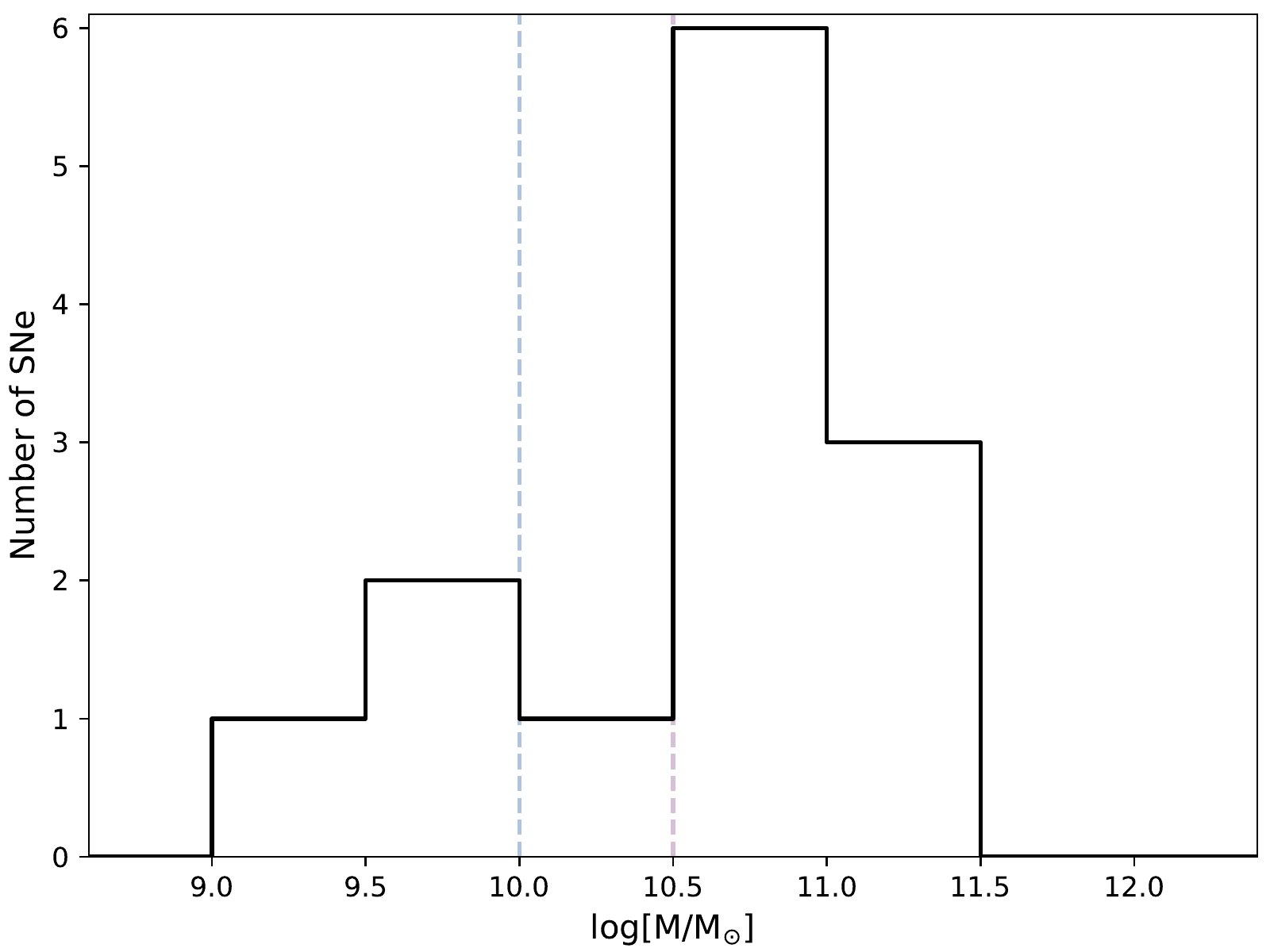}
\includegraphics[width=\columnwidth]{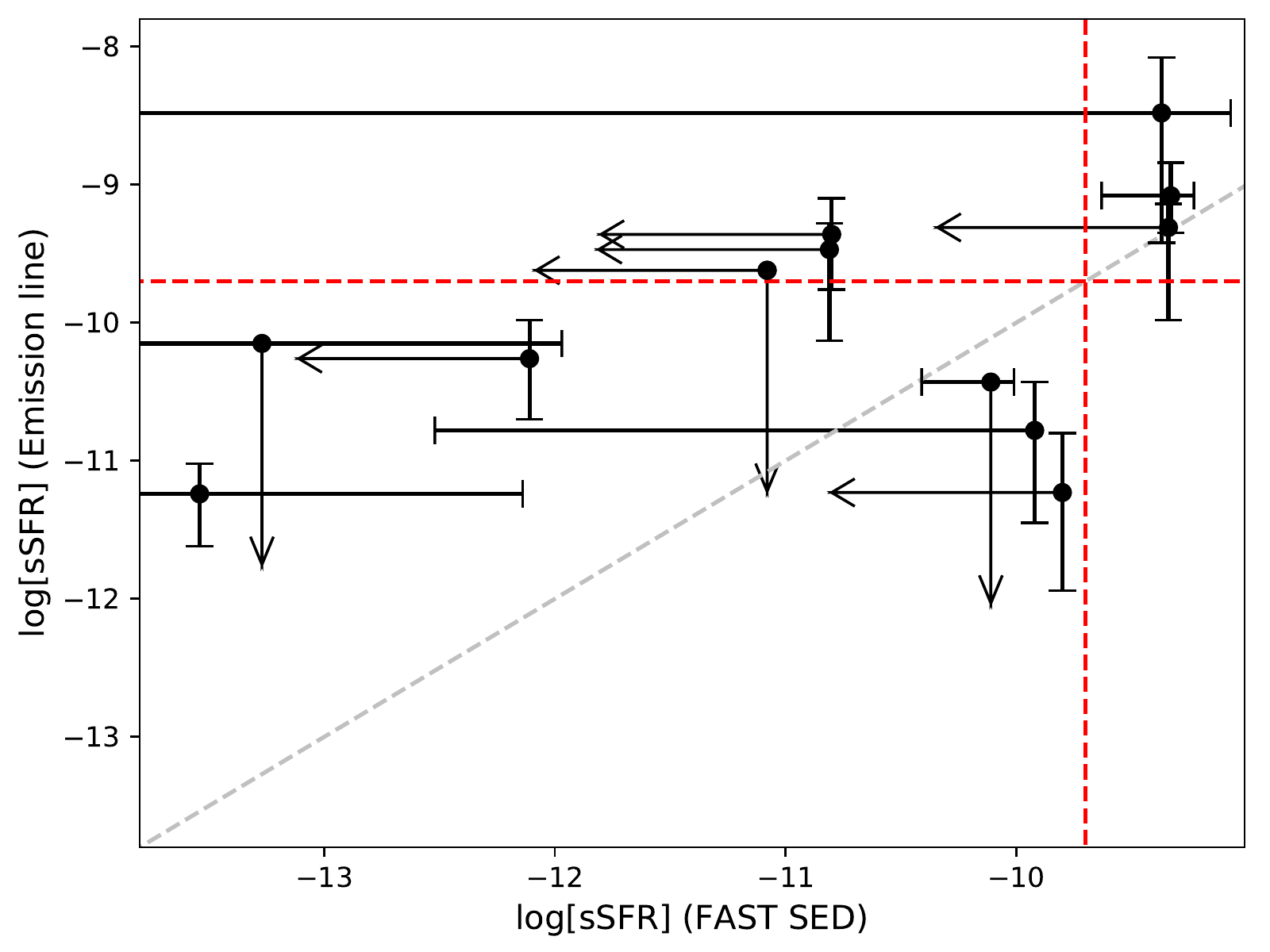}
\caption{\textit{Top:} Histogram (solid black line) of the masses derived by \texttt{FAST}, with the cut-offs in the `mass step' of log[M/M$_{\odot}$] of 10 \citep[light blue]{2010MNRAS.406..782S} and 10.5 \citep[light purple]{2018A&A...615A..68R} also indicated. As it hosted two SNe~Ia, SCP15A04 is counted twice here. \textit{Bottom:} Comparison between the sSFR derived from the emission lines, and those derived from the \texttt{FAST} SED fitting. The grey dashed line indicates where the two sSFR measurements are equal and the red dashed lines indicate the cutoff between the two log[sSFR] bins of $-9.7$ \citep{2010MNRAS.406..782S}. We include the SCP15D01 host galaxy in the sSFR comparison, but as it is at a comparatively low-$z$, we do not include it in the stellar-mass histogram. \label{fig:mass}}
\end{figure}

We used \texttt{FAST} SED fitting on a combination of the \textit{HST} photometry and VLT spectroscopy for each of the SN~Ia host galaxies for which a secure spectroscopic redshift has been obtained. The derived stellar mass, internal extinction ($A_V$), SFR, sSFR and stellar age for each host is listed in Table~\ref{tab:param}, along with SFR parameters as derived from the emission line constraints on each host galaxy. We do not analyse the host of SCP15E08, as much of the light detected in the spectrum will originate from the central source, rather than the stellar population. The upper panel of Fig.~\ref{fig:mass} shows a histogram of the masses derived from \texttt{FAST} for the $z\gtrsim1$ SNe~Ia. If we take the cut-off in the `mass step' to be log[M/M$_{\odot}$]~=~10 from \citet{2010MNRAS.406..782S} we find $10.5\pm0.5$ (SCP5D03 could be in either bin) of the 13 SNe~Ia with a host spectroscopic redshift are in the upper mass bin. If we take the cut-off as 10.5 from \citet{2018A&A...615A..68R}, we find $9\pm1$ of the 13 are in the upper mass bin. These numbers do not include possible SN~Ia SCP15D02 due to its highly uncertain stellar mass. If we include the apparently hostless SCP14C01 in here, this of course adds one to the low stellar mass bin. The uncertainties in the derived stellar masses are generally around 0.1--0.2 dex, sufficiently accurate to unambiguously determine which side of the mass-step an object lies in most cases.

\begin{landscape}
\begin{table}
	\centering
	\caption{Summary of the properties of the galaxies that hosted SNe~Ia (note SCP15D02 is only classified as a possible SN~Ia), where a redshift could be determined, excluding the host of SCP15E08.}
	\label{tab:param}
	\begin{tabular}{l c c c c c c c c} 
\hline\\[-0.3cm]
	Host& Redshift &${\mathrm{log_{10}(M/M_{\odot}}})$  &$A_V$ &SED SFR (${\mathrm{M_{\odot}\,yr^{-1}}}$)&Emission SFR (${\mathrm{M_{\odot}\,yr^{-1}}}$) &${\mathrm{log_{10}(sSFR_{\mathrm{SED}})}}$ & ${\mathrm{log_{10}(sSFR_{\mathrm{EL}})}}$ &Stellar Age (Gyr)\\[0.1cm]
\hline\\[-0.25cm]
	SCP15A03 &$1.3395\pm0.0003$ &$10.56^{+0.05}_{-0.11}$ &$0.8^{+0.1}_{-0.4}$ &$0.0^{+0.3}_{-0.0}$ &$12^{+7}_{-10}$ ([O~{\sc ii}]) &$<-10.8$ &$-9.5^{+0.2}_{-0.7}$ &$1.0^{+1.0}_{-0.3}$\\[0.15cm]
    SCP15A04 & $1.3345\pm0.0001$ &$10.67^{+0.14}_{-0.03}$ &$1.2^{+0.2}_{-0.6}$ &$0.0^{+12.3}_{-0.0}$ &$23^{+11}_{-16}$ (H$\alpha$) &$<-9.3$ &$-9.3^{+0.2}_{-0.7}$ &$0.2^{+0.5}_{-0.1}$\\[0.15cm]
	SCP15A05 & $1.3227\pm0.0002$ &$10.63^{+0.04}_{-0.12}$ &$0.5^{+0.4}_{-0.2}$ & $5.1^{+0.0}_{-5.1}$ &$0.7^{+0.9}_{-0.6}$ (H$\alpha$) &$-9.9^{+0.0}_{-2.6}$ &$-10.8^{+0.4}_{-0.7}$ &$1.0^{+0.3}_{-0.8}$\\[0.15cm]
    SCP15A06 &$1.3128\pm0.0005$ &$11.07^{+0.30}_{-0.01}$ &$0.3^{+0.6}_{-0.1}$ &$0.0^{+0.1}_{-0.0}$ &$<8.4$ ([O~{\sc ii}])&$-13.3^{+1.3}_{-{\mathrm{unc.}}}$ &$<-10.1$ &$1.3^{+1.9}_{-0.3}$\\[0.15cm]
    SCP15C01 &$0.9718\pm0.0002$ &$9.56^{+0.07}_{-0.19}$ &$0.8^{+0.4}_{-0.8}$ &$0.2^{+2.4}_{-0.2}$ & $12^{+17}_{-10}$ (H$\beta$) &$-9.4^{+0.3}_{-\mathrm{unc.}}$ & $-8.5^{+0.4}_{-0.9}$ &$0.6^{+0.7}_{-0.4}$\\[0.15cm]
    SCP15C04 &$1.2305\pm0.0003$ &$10.45^{+0.09}_{-0.16}$ &$0.2^{+0.1}_{-0.2}$ &$2.2^{+0.9}_{-1.4}$ &$<0.9$ ([O~{\sc ii}])&$-10.1^{+0.1}_{-0.3}$ &$<-10.4$ &$4.0^{+0.0}_{-1.6}$\\[0.15cm]
    SCP15D01 &$0.5682\pm0.0001$ &$9.87^{+0.08}_{-0.06}$ &$0.5^{+0.2}_{-0.2}$ &$3.5^{+1.6}_{-1.3}$ &$6^{+4}_{-3}$ (H$\beta$) &$-9.3^{+0.2}_{-0.2}$ &$-9.1^{+0.2}_{-0.3}$ &$4.0^{+3.9}_{-2.3}$\\[0.15cm]
    SCP15D02 &$1.1493\pm0.0002$ &$9.29^{+2.17}_{-0.60}$ &$1.6^{+0.3}_{-0.3}$ & $0^{+513}_{-0}$ &... &$<-7.1$ &... &$2.5^{+2.4}_{-2.4}$\\[0.15cm]
    SCP15D03 &$1.1130\pm0.0003$ &$9.98^{+0.06}_{-0.03}$ &$0.5^{+0.2}_{-0.1}$ &0.0$^{+0.1}_{-0.0}$ &$4^{+4}_{-2}$ ([O~{\sc ii}]) &$<-10.8$ &$-9.4^{+0.3}_{-0.4}$ &$0.8^{+0.2}_{-0.1}$\\[0.15cm]
    SCP16D01 &$1.1289\pm0.0002$ &$10.89^{+0.04}_{-0.05}$ &$0.5^{+0.2}_{-0.1}$ &$0.0^{+0.3}_{-0.0}$ &$4^{+4}_{-3}$ ([O~{\sc ii}]) &$<-12.1$ &$-10.3^{+0.3}_{-0.4}$ &$2.5^{+0.8}_{-0.3}$\\[0.15cm]
    SCP15E06 & $1.484\pm0.003$ &$10.94^{+0.29}_{-0.14}$ &$0.5^{+0.7}_{-0.3}$ &$0.0^{+0.4}_{-0.0}$ & $<21$ ([O~{\sc ii}]) &$<-11.1$ &$<-9.6$ &$1.6^{+2.4}_{-0.5}$\\[0.15cm]
    SCP15E07 &$1.4858\pm0.0003$ &$11.03^{+0.02}_{-0.10}$ &$0.2^{+0.1}_{-0.1}$ &$0.0^{+0.1}_{-0.0}$ &$0.6^{+0.4}_{-0.4}$ (H$\alpha$) &$-13.5^{+1.4}_{-{\mathrm{unc.}}}$ &$-11.2^{+0.2}_{-0.4}$ &$1.6^{+0.5}_{-0.7}$\\[0.15cm]
    SCP15G01 &$1.5817\pm0.0004$ & $11.44^{+0.03}_{-0.15}$ &$0.7^{+0.5}_{-0.3}$ &$0.0^{+17.4}_{-0.0}$ & $1.6^{+2.7}_{-1.3}$ (H$\alpha$) &$<-9.8$ &$-11.2^{+0.4}_{-0.7}$ &$4.0^{+0.0}_{-2.0}$\\[0.15cm]
\hline
\end{tabular}
\subcaption*{\textbf{Notes.} The uncertainties on the redshifts are derived from the cross-correlation, with the exception of SCP15E06 (difference between visual inspection redshift and that from cross-correlation). We take the minimum uncertainty to be $\Delta{z}=\pm0.0001$, even though the cross-correlation gives a more precise result in a few cases. The parameters are derived from emission line measurements from the VLT spectra and the SED fitting performed by \texttt{FAST}. sSFR$_{\mathrm{SED}}$ is derived from the SFR determined by the \texttt{FAST} SED fitting of the stellar spectra and photometry, whereas sSFR$_{\mathrm{EL}}$ is derived from the emission-line SFR measurement. Both measurements use the stellar mass from \texttt{FAST}. Upper limits indicate 95\% confidence and `unc.' = unconstrained. All SFR measurements are corrected for extinction, but do not consider contamination by AGN activity or post-AGB stars.}
\end{table}
\end{landscape}

In the lower panel of Fig.~\ref{fig:mass} we show a comparison of the sSFRs derived via the emission lines and those derived via \texttt{FAST} SED fitting. In four cases the emission-line sSFRs are higher (by $>$1$\sigma$) than those derived from the SED fitting. In each of these cases (SCP15A03, SCP15D03, SCP16D01 and SCP15E07), the emission line measurements themselves represent a $<$3$\sigma$ detection, however, when taken together, they could indicate emission-line contamination from sources unrelated to recent star formation. If one is adding a step-type function into SN~Ia standardisation, it only matters which side of this step they lie on. For example, significant differences in the sSFR derived from the SED and emission lines would not matter if both indicated log[sSFR]~$<-9.7$. The sSFRs are generally more poorly constrained than the stellar masses, particularly the sSFR measurements derived from the emission lines.

We have shown that the combination of \textit{HST} photometry and VLT spectroscopy gives sufficiently good constraints on the stellar mass of each host galaxy to reveal which `mass bin' the host lies in, and thus allowing a correction for the Hubble residuals to be made in the cosmological analysis. This could be particularly important for the See Change survey due the passive environments of the majority of galaxies, meaning the distribution of host galaxy environments will likely differ significantly to that explored in many other surveys. 

We find the sSFRs to generally be well constrained by the SED fitting of the VLT spectra and \textit{HST} photometry, and which side of the sSFR step they reside can usually be determined. The SFRs derived from the emission line measurements are generally more poorly constrained than from the SED fitting. This is in part due to the fact that we often had to rely on an [O~{\sc ii}] measurement for the SFR estimate. At 3727\,\AA, [O~{\sc ii}] is more sensitive to extinction, and in many galaxies the extinction itself is not tightly constrained, making the uncertainty on the SFR much higher than the direct measurement on the observed emission line strength. There is evidence that emission lines in some of the host galaxies may be contaminated by sources not associated with star formation, although for individual galaxies the evidence is not significant. For these reasons we conclude that the SFRs derived from the SED fitting are likely to be generally more accurate and precise.

The average uncertainty in the distance modulus for the See Change SNe~Ia is $\sim$0.2\,mag \citep[][submitted]{see-change-intro}. The mass step was measured as 0.07\,mag by \citet{2018A&A...615A..68R}. With a sample of 27 cosmologically-useful SNe~Ia or likely SNe~Ia, and with the host environments being different to many other SN~Ia surveys, it will therefore be important to consider the host masses when conducting the final cosmological analysis. As an example, we take the uncertainty on each SN to be $\sim$0.2\,mag, plus an intrinsic uncertainty on the distance modulus (i.e.\ that which arises from SNe~Ia not being perfectly standardisable) to be $\sim$0.1\,mag. If we combine 27 such SNe~Ia (at the same redshift) we would get an combined uncertainty on the distance modulus of $\sim0.04$\,mag. Thus if all SNe were on one side of the mass step (and assuming the mass step does not evolve with redshift), the offset arising from the mass step would be of similar order to the uncertainly. This is obviously a very simplistic picture as: a) the See Change SNe~Ia range from $z=0.86$ to $z=2.29$ \citep[][submitted]{see-change-intro}, and b) the hosts are not \textit{all} on one side of the mass step, as we have shown in this work. However it does illustrate that such a correction needs to be considered for a sample the size of See Change.

\begin{figure}
\includegraphics[width=\columnwidth]{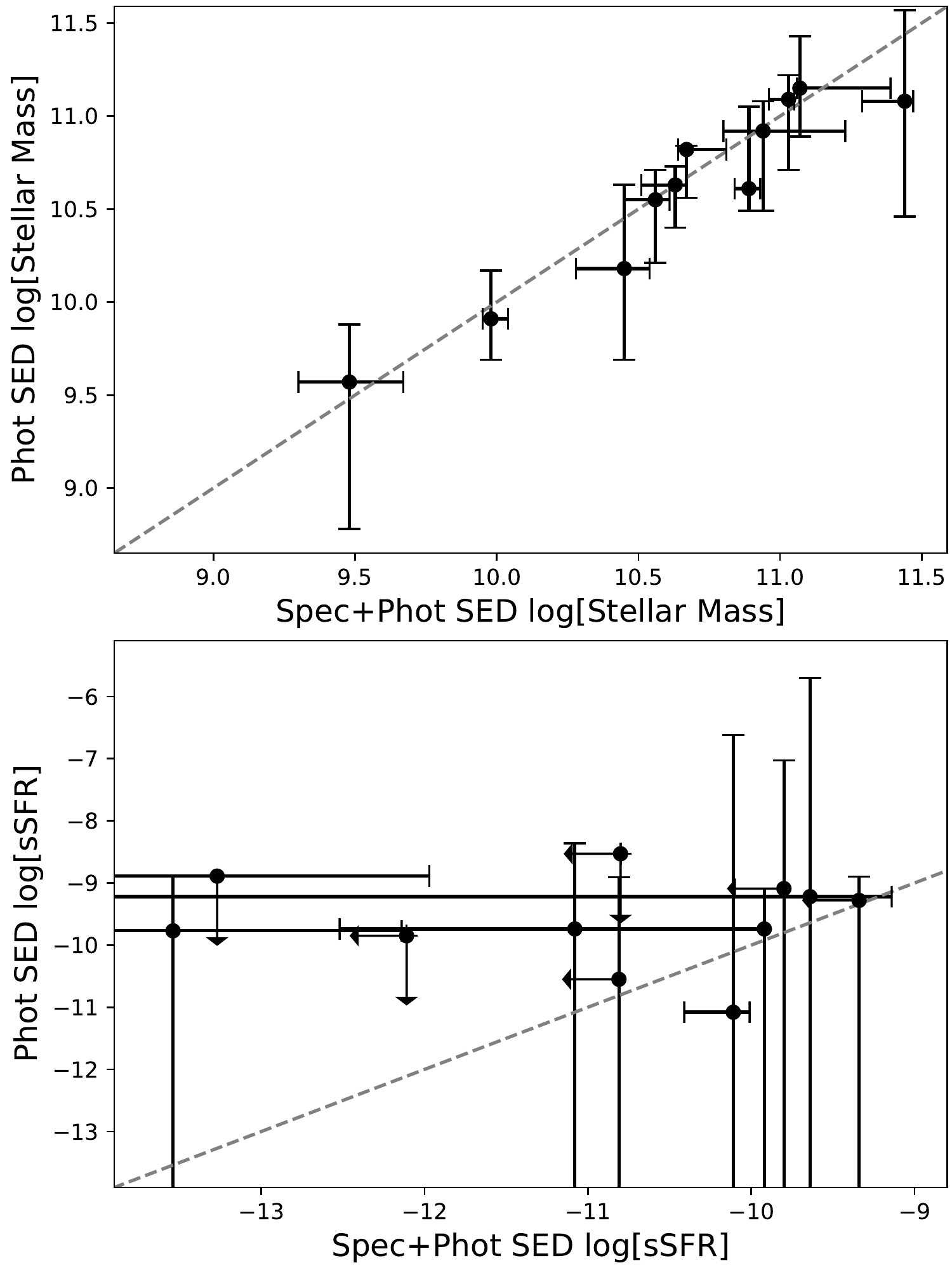}
\caption{Comparison between galaxy parameters derived from SED fitting using our spectra and the \textit{HST} photometry to those using just the \textit{HST} photometry and spectroscopic redshift (but not the spectra themselves). We show comparisons for stellar mass (upper panel) and sSFR (lower panel). Due to the highly uncertain stellar parameters, the SCP15D02 host galaxy is not included in these plots. We also do not include SCP15D01 as  it has a substantially lower redshift than our target range.\label{fig:comp}}
\end{figure}

In this work we derived galaxy parameters from SED fitting of a combination of our spectra and the \textit{HST} photometry. In Fig.~\ref{fig:comp} we compare these values to those derived using a combination of the \textit{HST} photometry and the spectroscopic redshift, but not using the spectra themselves in the fitting. This shows that the uncertainty in the parameters is significantly reduced using the spectra. The stellar ages and reddening are very poorly constrained from the photometry (not included in Fig.~\ref{fig:comp}), as one would expect given the degeneracy between the two parameters from photometry alone. The estimates on the sSFRs with \textit{HST} photometry and spectroscopic redshifts are poorly constrained, with it generally not being possible to determine which side of any sSFR `step' that an individual galaxy would lie. The \textit{HST} photometry and spectroscopic redshifts do however provide reasonable estimates on the stellar masses. In terms of reducing the telescope time used, this would be useful in the case of emission-line galaxies, where one could potentially confirm the redshift without significant S/N on the continuum. However, in the case of low-SFR environments like most of our sample, reasonable S/N is needed on the continuum to detect absorption lines and confirm the redshift. Therefore, in such cases, no extra telescope time is actually required to significantly improve the stellar mass estimates and other parameters.

\section{Summary and Conclusions} \label{sec:sum}
In this paper we have presented VLT spectra of the host galaxies of transients discovered in the See Change survey, including the hosts of 15 SNe~Ia and one possible SN~Ia at redshifts $z>0.97$. We now summarise our main findings:
\begin{itemize}
\item The See Change survey targeted massive galaxy clusters at $1.13 \leq z \leq 1.75$ to observe SNe~Ia.
\item The primary aim of the work presented here was to obtain secure redshifts of the host galaxies of the $z\gtrsim1$ SNe~Ia found in the survey for use in a cosmological analysis and SN typing of the objects themselves.
\item We targeted 39 transients with the VLT and successfully obtained redshifts for 26 of those. Of these, 15 are classified as $z\gtrsim1$ SNe~Ia, plus one possible SN~Ia, which is approximately two thirds of the See Change SNe~Ia with spectroscopic redshifts.
\item The majority of host galaxies did not show strong [O~{\sc ii}] 3727\,\AA\ emission, consistent with many of the other confirmed galaxies in these massive clusters also being mainly passive.
\item Combining these spectra with \textit{HST} photometry, we have used \texttt{FAST} to derive parameters for each SN~Ia host galaxy, including stellar mass, SFR and stellar age.
\item The uncertainties in the \texttt{FAST} stellar masses are generally relatively small, allowing us to determine which side of the mass step each object lies. This will allow a correction to be made when performing a cosmological analysis.
\item In the case of SNe~Ia (plus the possible SN~Ia SCP15D02) at $0.97\leq{z}\leq1.5$ (i.e. where [O~{\sc ii}] and Ca~{\sc ii} H\,\&\,K are still at $<1\mu$m), we successfully obtained a redshift for 13 out of the 17 attempted with the VLT. We have therefore shown that even for passive galaxies, it is possible to obtain secure redshifts for the majority of SNe~Ia found in galaxy clusters out to $z=1.5$.
\item Confirming the redshifts of these cluster SN~Ia host galaxies does however require significant time, even on $8-10$\,m class telescopes, with typical required total exposure times around $3-4$\,hrs.
\item 15 objects for which we obtained a redshift are able to be classified as $z\gtrsim1$ SNe~Ia. This includes one in a broad-lined AGN host galaxy at $z=2.02$ and a lensed SN~Ia in a passive galaxy at $z=2.22$ (reported by \citealp{2017arXiv170704606R}).
\end{itemize}

From the See Change survey, we have already published a paper on a $z=2.22$ SN~Ia, lensed by the MOO1014 cluster \citep{2017arXiv170704606R}. Several papers that utilise See Change data to study the clusters themselves have also been published. An overview of the See Change survey itself will be presented by \citet[][submitted]{see-change-intro}. The cosmological analysis in currently ongoing and will be presented by \cosmolt.

\section*{Acknowledgements}

SCW and IMH acknowledge support from STFC consolidated grants ST/P00038X/1 and ST/R000514/1. We acknowledge an anonymous referee for comments which improved the quality of this paper. This work is based on observations collected at the European Organisation for Astronomical Research in the Southern Hemisphere under ESO programmes 294.A-5025(A), 095.A-0830(A, B, C), 096.A-0926(B, C), 097.A-0442(A, B, C) and 0100.A-0851(A). We thank the ESO staff for assisting with observation planning. This work is also based on observations made with the NASA/ESA \textit{Hubble Space Telescope}, obtained at the Space Telescope Science Institute, which is operated by the Association of Universities for Research in Astronomy, Inc., under NASA contract NAS 5-26555. These observations are associated with programs GO-13677 and GO-14327, along with GO-13747. This paper makes use of data obtained at the W.~M.\ Keck Observatory, which is operated as a scientific partnership among the California Institute of Technology, the University of California and the National Aeronautics and Space Administration. The Observatory was made possible by the generous financial support of the W. M. Keck Foundation.





\begin{thebibliography}{}
\makeatletter
\relax
\def\mn@urlcharsother{\let\do\@makeother \do\$\do\&\do\#\do\^\do\_\do\%\do\~}
\def\mn@doi{\begingroup\mn@urlcharsother \@ifnextchar [ {\mn@doi@}
  {\mn@doi@[]}}
\def\mn@doi@[#1]#2{\def\@tempa{#1}\ifx\@tempa\@empty \href
  {http://dx.doi.org/#2} {doi:#2}\else \href {http://dx.doi.org/#2} {#1}\fi
  \endgroup}
\def\mn@eprint#1#2{\mn@eprint@#1:#2::\@nil}
\def\mn@eprint@arXiv#1{\href {http://arxiv.org/abs/#1} {{\tt arXiv:#1}}}
\def\mn@eprint@dblp#1{\href {http://dblp.uni-trier.de/rec/bibtex/#1.xml}
  {dblp:#1}}
\def\mn@eprint@#1:#2:#3:#4\@nil{\def\@tempa {#1}\def\@tempb {#2}\def\@tempc
  {#3}\ifx \@tempc \@empty \let \@tempc \@tempb \let \@tempb \@tempa \fi \ifx
  \@tempb \@empty \def\@tempb {arXiv}\fi \@ifundefined
  {mn@eprint@\@tempb}{\@tempb:\@tempc}{\expandafter \expandafter \csname
  mn@eprint@\@tempb\endcsname \expandafter{\@tempc}}}

\bibitem[\protect\citeauthoryear{{Abbott} et~al.,}{{Abbott}
  et~al.}{2019}]{2019ApJ...872L..30A}
{Abbott} T.~M.~C.,  et~al., 2019, \mn@doi [\apj] {10.3847/2041-8213/ab04fa},
  \href {https://ui.adsabs.harvard.edu/abs/2019ApJ...872L..30A} {872, L30}

\bibitem[\protect\citeauthoryear{{Adelman-McCarthy} et~al.,}{{Adelman-McCarthy}
  et~al.}{2007}]{2007ApJS..172..634A}
{Adelman-McCarthy} J.~K.,  et~al., 2007, \mn@doi [\apjs] {10.1086/518864},
  \href {https://ui.adsabs.harvard.edu/abs/2007ApJS..172..634A} {172, 634}

\bibitem[\protect\citeauthoryear{{Aird}, {Coil}  \& {Georgakakis}}{{Aird}
  et~al.}{2017}]{2017MNRAS.465.3390A}
{Aird} J.,  {Coil} A.~L.,   {Georgakakis} A.,  2017, \mn@doi [\mnras]
  {10.1093/mnras/stw2932}, \href
  {http://adsabs.harvard.edu/abs/2017MNRAS.465.3390A} {465, 3390}

\bibitem[\protect\citeauthoryear{{Aird}, {Coil}  \& {Georgakakis}}{{Aird}
  et~al.}{2018}]{2018MNRAS.474.1225A}
{Aird} J.,  {Coil} A.~L.,   {Georgakakis} A.,  2018, \mn@doi [\mnras]
  {10.1093/mnras/stx2700}, \href
  {http://adsabs.harvard.edu/abs/2018MNRAS.474.1225A} {474, 1225}

\bibitem[\protect\citeauthoryear{{Appenzeller} et~al.,}{{Appenzeller}
  et~al.}{1998}]{1998Msngr..94....1A}
{Appenzeller} I.,  et~al., 1998, The Messenger, \href
  {http://adsabs.harvard.edu/abs/1998Msngr..94....1A} {94, 1}

\bibitem[\protect\citeauthoryear{{Barbary}}{{Barbary}}{2016}]{2016JOSS....1...58B}
{Barbary} K.,  2016, \mn@doi [The Journal of Open Source Software]
  {10.21105/joss.00058}, \href
  {https://ui.adsabs.harvard.edu/#abs/2016JOSS....1...58B} {1, 58}

\bibitem[\protect\citeauthoryear{{Barbary} et~al.,}{{Barbary}
  et~al.}{2012}]{2012ApJ...745...32B}
{Barbary} K.,  et~al., 2012, \mn@doi [\apj] {10.1088/0004-637X/745/1/32}, \href
  {http://adsabs.harvard.edu/abs/2012ApJ...745...32B} {745, 32}

\bibitem[\protect\citeauthoryear{{Bayliss} et~al.,}{{Bayliss}
  et~al.}{2014}]{2014ApJ...794...12B}
{Bayliss} M.~B.,  et~al., 2014, \mn@doi [\apj] {10.1088/0004-637X/794/1/12},
  \href {http://adsabs.harvard.edu/abs/2014ApJ...794...12B} {794, 12}

\bibitem[\protect\citeauthoryear{{Belfiore} et~al.,}{{Belfiore}
  et~al.}{2016}]{2016MNRAS.461.3111B}
{Belfiore} F.,  et~al., 2016, \mn@doi [\mnras] {10.1093/mnras/stw1234}, \href
  {http://adsabs.harvard.edu/abs/2016MNRAS.461.3111B} {461, 3111}

\bibitem[\protect\citeauthoryear{{Bertin} \& {Arnouts}}{{Bertin} \&
  {Arnouts}}{1996}]{1996A&AS..117..393B}
{Bertin} E.,  {Arnouts} S.,  1996, \aaps, \href
  {http://adsabs.harvard.edu/abs/1996A\%26AS..117..393B} {117, 393}

\bibitem[\protect\citeauthoryear{{Bleem} et~al.,}{{Bleem}
  et~al.}{2015}]{2015ApJS..216...27B}
{Bleem} L.~E.,  et~al., 2015, \mn@doi [\apjs] {10.1088/0067-0049/216/2/27},
  \href {http://adsabs.harvard.edu/abs/2015ApJS..216...27B} {216, 27}

\bibitem[\protect\citeauthoryear{{Blondin} \& {Tonry}}{{Blondin} \&
  {Tonry}}{2007}]{2007ApJ...666.1024B}
{Blondin} S.,  {Tonry} J.~L.,  2007, \mn@doi [\apj] {10.1086/520494}, \href
  {http://adsabs.harvard.edu/abs/2007ApJ...666.1024B} {666, 1024}

\bibitem[\protect\citeauthoryear{{Brammer}, {van Dokkum}  \& {Coppi}}{{Brammer}
  et~al.}{2008}]{2008ApJ...686.1503B}
{Brammer} G.~B.,  {van Dokkum} P.~G.,   {Coppi} P.,  2008, \mn@doi [\apj]
  {10.1086/591786}, \href {http://adsabs.harvard.edu/abs/2008ApJ...686.1503B}
  {686, 1503}

\bibitem[\protect\citeauthoryear{{Brodwin} et~al.,}{{Brodwin}
  et~al.}{2015}]{2015ApJ...806...26B}
{Brodwin} M.,  et~al., 2015, \mn@doi [\apj] {10.1088/0004-637X/806/1/26}, \href
  {http://adsabs.harvard.edu/abs/2015ApJ...806...26B} {806, 26}

\bibitem[\protect\citeauthoryear{{Calzetti}, {Armus}, {Bohlin}, {Kinney},
  {Koornneef}  \& {Storchi-Bergmann}}{{Calzetti}
  et~al.}{2000}]{2000ApJ...533..682C}
{Calzetti} D.,  {Armus} L.,  {Bohlin} R.~C.,  {Kinney} A.~L.,  {Koornneef} J.,
   {Storchi-Bergmann} T.,  2000, \mn@doi [\apj] {10.1086/308692}, \href
  {https://ui.adsabs.harvard.edu/abs/2000ApJ...533..682C} {533, 682}

\bibitem[\protect\citeauthoryear{{Calzetti} et~al.,}{{Calzetti}
  et~al.}{2010}]{calzetti10}
{Calzetti} D.,  et~al., 2010, \mn@doi [\apj] {10.1088/0004-637X/714/2/1256},
  \href {http://adsabs.harvard.edu/abs/2010ApJ...714.1256C} {714, 1256}

\bibitem[\protect\citeauthoryear{{Cardelli}, {Clayton}  \& {Mathis}}{{Cardelli}
  et~al.}{1989}]{1989ApJ...345..245C}
{Cardelli} J.~A.,  {Clayton} G.~C.,   {Mathis} J.~S.,  1989, \mn@doi [\apj]
  {10.1086/167900}, \href {http://adsabs.harvard.edu/abs/1989ApJ...345..245C}
  {345, 245}

\bibitem[\protect\citeauthoryear{{Chornock} et~al.,}{{Chornock}
  et~al.}{2013}]{2013ApJ...767..162C}
{Chornock} R.,  et~al., 2013, \mn@doi [\apj] {10.1088/0004-637X/767/2/162},
  \href {https://ui.adsabs.harvard.edu/abs/2013ApJ...767..162C} {767, 162}

\bibitem[\protect\citeauthoryear{{Decker} et~al.,}{{Decker}
  et~al.}{2019}]{2019ApJ...878...72D}
{Decker} B.,  et~al., 2019, \mn@doi [\apj] {10.3847/1538-4357/ab12d7}, \href
  {https://ui.adsabs.harvard.edu/abs/2019ApJ...878...72D} {878, 72}

\bibitem[\protect\citeauthoryear{{Delahaye} et~al.,}{{Delahaye}
  et~al.}{2017}]{2017ApJ...843..126D}
{Delahaye} A.~G.,  et~al., 2017, \mn@doi [\apj] {10.3847/1538-4357/aa756a},
  \href {http://adsabs.harvard.edu/abs/2017ApJ...843..126D} {843, 126}

\bibitem[\protect\citeauthoryear{{Fassbender} et~al.,}{{Fassbender}
  et~al.}{2014}]{2014A&A...568A...5F}
{Fassbender} R.,  et~al., 2014, \mn@doi [\aap] {10.1051/0004-6361/201423941},
  \href {http://cdsads.u-strasbg.fr/abs/2014A\%26A...568A...5F} {568, A5}

\bibitem[\protect\citeauthoryear{{Foley} et~al.,}{{Foley}
  et~al.}{2011}]{2011ApJ...731...86F}
{Foley} R.~J.,  et~al., 2011, \mn@doi [\apj] {10.1088/0004-637X/731/2/86},
  \href {http://adsabs.harvard.edu/abs/2011ApJ...731...86F} {731, 86}

\bibitem[\protect\citeauthoryear{{Foltz} et~al.,}{{Foltz}
  et~al.}{2018}]{2018ApJ...866..136F}
{Foltz} R.,  et~al., 2018, \mn@doi [\apj] {10.3847/1538-4357/aad80d}, \href
  {http://adsabs.harvard.edu/abs/2018ApJ...866..136F} {866, 136}

\bibitem[\protect\citeauthoryear{{Freudling}, {Romaniello}, {Bramich},
  {Ballester}, {Forchi}, {Garc{\'{\i}}a-Dabl{\'o}}, {Moehler}  \&
  {Neeser}}{{Freudling} et~al.}{2013}]{2013A&A...559A..96F}
{Freudling} W.,  {Romaniello} M.,  {Bramich} D.~M.,  {Ballester} P.,  {Forchi}
  V.,  {Garc{\'{\i}}a-Dabl{\'o}} C.~E.,  {Moehler} S.,   {Neeser} M.~J.,  2013,
  \mn@doi [\aap] {10.1051/0004-6361/201322494}, \href
  {http://adsabs.harvard.edu/abs/2013A\%26A...559A..96F} {559, A96}

\bibitem[\protect\citeauthoryear{{Ganeshalingam}, {Li}  \&
  {Filippenko}}{{Ganeshalingam} et~al.}{2013}]{2013MNRAS.433.2240G}
{Ganeshalingam} M.,  {Li} W.,   {Filippenko} A.~V.,  2013, \mn@doi [\mnras]
  {10.1093/mnras/stt893}, \href
  {http://adsabs.harvard.edu/abs/2013MNRAS.433.2240G} {433, 2240}

\bibitem[\protect\citeauthoryear{{Gonzalez} et~al.,}{{Gonzalez}
  et~al.}{2015}]{2015ApJ...812L..40G}
{Gonzalez} A.~H.,  et~al., 2015, \mn@doi [\apjl] {10.1088/2041-8205/812/2/L40},
  \href {http://adsabs.harvard.edu/abs/2015ApJ...812L..40G} {812, L40}

\bibitem[\protect\citeauthoryear{{Gonzalez} et~al.,}{{Gonzalez}
  et~al.}{2019}]{2018arXiv180906820G}
{Gonzalez} A.~H.,  et~al., 2019, \mn@doi [\apjs] {10.3847/1538-4365/aafad2},
  \href {https://ui.adsabs.harvard.edu/abs/2019ApJS..240...33G} {240, 33}

\bibitem[\protect\citeauthoryear{{Guillochon}, {Parrent}, {Kelley}  \&
  {Margutti}}{{Guillochon} et~al.}{2017}]{2017ApJ...835...64G}
{Guillochon} J.,  {Parrent} J.,  {Kelley} L.~Z.,   {Margutti} R.,  2017,
  \mn@doi [\apj] {10.3847/1538-4357/835/1/64}, \href
  {http://adsabs.harvard.edu/abs/2017ApJ...835...64G} {835, 64}

\bibitem[\protect\citeauthoryear{{Guy} et~al.,}{{Guy}
  et~al.}{2007}]{2007A&A...466...11G}
{Guy} J.,  et~al., 2007, \mn@doi [\aap] {10.1051/0004-6361:20066930}, \href
  {https://ui.adsabs.harvard.edu/abs/2007A\%26A...466...11G} {466, 11}

\bibitem[\protect\citeauthoryear{{Guy} et~al.,}{{Guy}
  et~al.}{2010}]{2010A&A...523A...7G}
{Guy} J.,  et~al., 2010, \mn@doi [\aap] {10.1051/0004-6361/201014468}, \href
  {https://ui.adsabs.harvard.edu/abs/2010A&A...523A...7G} {523, A7}

\bibitem[\protect\citeauthoryear{{Hamuy}, {Phillips}, {Maza}, {Suntzeff},
  {Schommer}  \& {Aviles}}{{Hamuy} et~al.}{1995}]{1995AJ....109....1H}
{Hamuy} M.,  {Phillips} M.~M.,  {Maza} J.,  {Suntzeff} N.~B.,  {Schommer}
  R.~A.,   {Aviles} R.,  1995, \mn@doi [\aj] {10.1086/117251}, \href
  {http://adsabs.harvard.edu/abs/1995AJ....109....1H} {109, 1}

\bibitem[\protect\citeauthoryear{{Hayden}, {Rubin}, {Boone}, {Aldering},
  {Nordin}, {Brodwin}  et~al.}{{Hayden} et~al.}{2020}]{see-change-intro}
{Hayden} B.,  {Rubin} D.,  {Boone} K.,  {Aldering} G.,  {Nordin} J.,  {Brodwin}
  M.,   et~al., 2020, \apj, submitted

\bibitem[\protect\citeauthoryear{{Hicken}, {Wood-Vasey}, {Blondin}, {Challis},
  {Jha}, {Kelly}, {Rest}  \& {Kirshner}}{{Hicken}
  et~al.}{2009}]{2009ApJ...700.1097H}
{Hicken} M.,  {Wood-Vasey} W.~M.,  {Blondin} S.,  {Challis} P.,  {Jha} S.,
  {Kelly} P.~L.,  {Rest} A.,   {Kirshner} R.~P.,  2009, \mn@doi [\apj]
  {10.1088/0004-637X/700/2/1097}, \href
  {http://adsabs.harvard.edu/abs/2009ApJ...700.1097H} {700, 1097}

\bibitem[\protect\citeauthoryear{{Hildebrandt} et~al.,}{{Hildebrandt}
  et~al.}{2010}]{2010A&A...523A..31H}
{Hildebrandt} H.,  et~al., 2010, \mn@doi [\aap] {10.1051/0004-6361/201014885},
  \href {https://ui.adsabs.harvard.edu/abs/2010A&A...523A..31H} {523, A31}

\bibitem[\protect\citeauthoryear{{Horne}}{{Horne}}{1986}]{1986PASP...98..609H}
{Horne} K.,  1986, \mn@doi [\pasp] {10.1086/131801}, \href
  {http://adsabs.harvard.edu/abs/1986PASP...98..609H} {98, 609}

\bibitem[\protect\citeauthoryear{{Hubble}}{{Hubble}}{1929}]{1929PNAS...15..168H}
{Hubble} E.,  1929, \mn@doi [Proceedings of the National Academy of Science]
  {10.1073/pnas.15.3.168}, \href
  {http://adsabs.harvard.edu/abs/1929PNAS...15..168H} {15, 168}

\bibitem[\protect\citeauthoryear{{Jee}, {Ko}, {Perlmutter}, {Gonzalez},
  {Brodwin}, {Linder}  \& {Eisenhardt}}{{Jee}
  et~al.}{2017}]{2017ApJ...847..117J}
{Jee} M.~J.,  {Ko} J.,  {Perlmutter} S.,  {Gonzalez} A.,  {Brodwin} M.,
  {Linder} E.,   {Eisenhardt} P.,  2017, \mn@doi [\apj]
  {10.3847/1538-4357/aa88bc}, \href
  {http://adsabs.harvard.edu/abs/2017ApJ...847..117J} {847, 117}

\bibitem[\protect\citeauthoryear{{Jiang}, {Ge}, {Zhou}, {Wang}  \&
  {Wang}}{{Jiang} et~al.}{2011}]{2011ApJ...732..110J}
{Jiang} P.,  {Ge} J.,  {Zhou} H.,  {Wang} J.,   {Wang} T.,  2011, \mn@doi
  [\apj] {10.1088/0004-637X/732/2/110}, \href
  {http://adsabs.harvard.edu/abs/2011ApJ...732..110J} {732, 110}

\bibitem[\protect\citeauthoryear{{Jones} et~al.,}{{Jones}
  et~al.}{2018}]{2018ApJ...867..108J}
{Jones} D.~O.,  et~al., 2018, \mn@doi [\apj] {10.3847/1538-4357/aae2b9}, \href
  {https://ui.adsabs.harvard.edu/abs/2018ApJ...867..108J} {867, 108}

\bibitem[\protect\citeauthoryear{{Kauffmann} et~al.,}{{Kauffmann}
  et~al.}{2003}]{2003MNRAS.346.1055K}
{Kauffmann} G.,  et~al., 2003, \mn@doi [\mnras]
  {10.1111/j.1365-2966.2003.07154.x}, \href
  {http://adsabs.harvard.edu/abs/2003MNRAS.346.1055K} {346, 1055}

\bibitem[\protect\citeauthoryear{{Kausch} et~al.,}{{Kausch}
  et~al.}{2015}]{2015A&A...576A..78K}
{Kausch} W.,  et~al., 2015, \mn@doi [\aap] {10.1051/0004-6361/201423909}, \href
  {http://adsabs.harvard.edu/abs/2015A\%26A...576A..78K} {576, A78}

\bibitem[\protect\citeauthoryear{{Kelson}, {Martini}  \& {Mulchaey}}{{Kelson}
  et~al.}{2003}]{Kelson}
{Kelson} D.~D.,  {Martini} P.,   {Mulchaey} J.~S.,  2003, Optimal Measurements
  of Redshifts Using the Weighted Cross-Correlation,
  \url{https://code.obs.carnegiescience.edu/Algorithms/realcc/}

\bibitem[\protect\citeauthoryear{{Kessler} et~al.,}{{Kessler}
  et~al.}{2009}]{2009PASP..121.1028K}
{Kessler} R.,  et~al., 2009, \mn@doi [\pasp] {10.1086/605984}, \href
  {https://ui.adsabs.harvard.edu/abs/2009PASP..121.1028K} {121, 1028}

\bibitem[\protect\citeauthoryear{{Kewley}, {Dopita}, {Sutherland}, {Heisler}
  \& {Trevena}}{{Kewley} et~al.}{2001}]{2001ApJ...556..121K}
{Kewley} L.~J.,  {Dopita} M.~A.,  {Sutherland} R.~S.,  {Heisler} C.~A.,
  {Trevena} J.,  2001, \mn@doi [\apj] {10.1086/321545}, \href
  {http://adsabs.harvard.edu/abs/2001ApJ...556..121K} {556, 121}

\bibitem[\protect\citeauthoryear{{Kewley}, {Geller}  \& {Jansen}}{{Kewley}
  et~al.}{2004}]{2004AJ....127.2002K}
{Kewley} L.~J.,  {Geller} M.~J.,   {Jansen} R.~A.,  2004, \mn@doi [\aj]
  {10.1086/382723}, \href {http://adsabs.harvard.edu/abs/2004AJ....127.2002K}
  {127, 2002}

\bibitem[\protect\citeauthoryear{{Kriek}, {van Dokkum}, {Labb{\'e}}, {Franx},
  {Illingworth}, {Marchesini}  \& {Quadri}}{{Kriek}
  et~al.}{2009}]{2009ApJ...700..221K}
{Kriek} M.,  {van Dokkum} P.~G.,  {Labb{\'e}} I.,  {Franx} M.,  {Illingworth}
  G.~D.,  {Marchesini} D.,   {Quadri} R.~F.,  2009, \mn@doi [\apj]
  {10.1088/0004-637X/700/1/221}, \href
  {http://adsabs.harvard.edu/abs/2009ApJ...700..221K} {700, 221}

\bibitem[\protect\citeauthoryear{{Krisciunas} et~al.,}{{Krisciunas}
  et~al.}{2007}]{2007AJ....133...58K}
{Krisciunas} K.,  et~al., 2007, \mn@doi [\aj] {10.1086/509126}, \href
  {http://adsabs.harvard.edu/abs/2007AJ....133...58K} {133, 58}

\bibitem[\protect\citeauthoryear{{Lidman} et~al.,}{{Lidman}
  et~al.}{2012}]{2012MNRAS.427..550L}
{Lidman} C.,  et~al., 2012, \mn@doi [\mnras]
  {10.1111/j.1365-2966.2012.21984.x}, \href
  {http://adsabs.harvard.edu/abs/2012MNRAS.427..550L} {427, 550}

\bibitem[\protect\citeauthoryear{{Matheson}, {Filippenko}, {Li}, {Leonard}  \&
  {Shields}}{{Matheson} et~al.}{2001}]{2001AJ....121.1648M}
{Matheson} T.,  {Filippenko} A.~V.,  {Li} W.,  {Leonard} D.~C.,   {Shields}
  J.~C.,  2001, \mn@doi [\aj] {10.1086/319390}, \href
  {http://adsabs.harvard.edu/abs/2001AJ....121.1648M} {121, 1648}

\bibitem[\protect\citeauthoryear{{Meyers} et~al.,}{{Meyers}
  et~al.}{2012}]{meyers12}
{Meyers} J.,  et~al., 2012, \mn@doi [\apj] {10.1088/0004-637X/750/1/1}, \href
  {http://adsabs.harvard.edu/abs/2012ApJ...750....1M} {750, 1}

\bibitem[\protect\citeauthoryear{{Miknaitis} et~al.,}{{Miknaitis}
  et~al.}{2007}]{2007ApJ...666..674M}
{Miknaitis} G.,  et~al., 2007, \mn@doi [\apj] {10.1086/519986}, \href
  {https://ui.adsabs.harvard.edu/abs/2007ApJ...666..674M} {666, 674}

\bibitem[\protect\citeauthoryear{{Morokuma} et~al.,}{{Morokuma}
  et~al.}{2010}]{2010PASJ...62...19M}
{Morokuma} T.,  et~al., 2010, \mn@doi [\pasj] {10.1093/pasj/62.1.19}, \href
  {http://adsabs.harvard.edu/abs/2010PASJ...62...19M} {62, 19}

\bibitem[\protect\citeauthoryear{{Mosher} et~al.,}{{Mosher}
  et~al.}{2014}]{2014ApJ...793...16M}
{Mosher} J.,  et~al., 2014, \mn@doi [\apj] {10.1088/0004-637X/793/1/16}, \href
  {https://ui.adsabs.harvard.edu/abs/2014ApJ...793...16M} {793, 16}

\bibitem[\protect\citeauthoryear{{Muzzin} et~al.,}{{Muzzin}
  et~al.}{2009}]{2009ApJ...698.1934M}
{Muzzin} A.,  et~al., 2009, \mn@doi [\apj] {10.1088/0004-637X/698/2/1934},
  \href {http://adsabs.harvard.edu/abs/2009ApJ...698.1934M} {698, 1934}

\bibitem[\protect\citeauthoryear{{Muzzin}, {Wilson}, {Demarco}, {Lidman},
  {Nantais}, {Hoekstra}, {Yee}  \& {Rettura}}{{Muzzin}
  et~al.}{2013}]{2013ApJ...767...39M}
{Muzzin} A.,  {Wilson} G.,  {Demarco} R.,  {Lidman} C.,  {Nantais} J.,
  {Hoekstra} H.,  {Yee} H.~K.~C.,   {Rettura} A.,  2013, \mn@doi [\apj]
  {10.1088/0004-637X/767/1/39}, \href
  {http://adsabs.harvard.edu/abs/2013ApJ...767...39M} {767, 39}

\bibitem[\protect\citeauthoryear{{Neill} et~al.,}{{Neill}
  et~al.}{2009}]{2009ApJ...707.1449N}
{Neill} J.~D.,  et~al., 2009, \mn@doi [\apj] {10.1088/0004-637X/707/2/1449},
  \href {http://adsabs.harvard.edu/abs/2009ApJ...707.1449N} {707, 1449}

\bibitem[\protect\citeauthoryear{{Noble} et~al.,}{{Noble}
  et~al.}{2017}]{2017ApJ...842L..21N}
{Noble} A.~G.,  et~al., 2017, \mn@doi [\apjl] {10.3847/2041-8213/aa77f3}, \href
  {http://adsabs.harvard.edu/abs/2017ApJ...842L..21N} {842, L21}

\bibitem[\protect\citeauthoryear{{Perlmutter} et~al.,}{{Perlmutter}
  et~al.}{1999}]{1999ApJ...517..565P}
{Perlmutter} S.,  et~al., 1999, \mn@doi [\apj] {10.1086/307221}, \href
  {http://adsabs.harvard.edu/abs/1999ApJ...517..565P} {517, 565}

\bibitem[\protect\citeauthoryear{{Phillips}}{{Phillips}}{1993}]{1993ApJ...413L.105P}
{Phillips} M.~M.,  1993, \mn@doi [\apjl] {10.1086/186970}, \href
  {http://adsabs.harvard.edu/abs/1993ApJ...413L.105P} {413, L105}

\bibitem[\protect\citeauthoryear{{Pitman}, {Clayton}  \& {Gordon}}{{Pitman}
  et~al.}{2000}]{2000PASP..112..537P}
{Pitman} K.~M.,  {Clayton} G.~C.,   {Gordon} K.~D.,  2000, \mn@doi [\pasp]
  {10.1086/316551}, \href {http://adsabs.harvard.edu/abs/2000PASP..112..537P}
  {112, 537}

\bibitem[\protect\citeauthoryear{{Poznanski}, {Gal-Yam}, {Maoz}, {Filippenko},
  {Leonard}  \& {Matheson}}{{Poznanski} et~al.}{2002}]{2002PASP..114..833P}
{Poznanski} D.,  {Gal-Yam} A.,  {Maoz} D.,  {Filippenko} A.~V.,  {Leonard}
  D.~C.,   {Matheson} T.,  2002, \mn@doi [\pasp] {10.1086/341741}, \href
  {http://adsabs.harvard.edu/abs/2002PASP..114..833P} {114, 833}

\bibitem[\protect\citeauthoryear{{Quimby} et~al.,}{{Quimby}
  et~al.}{2013}]{2013ApJ...768L..20Q}
{Quimby} R.~M.,  et~al., 2013, \mn@doi [\apjl] {10.1088/2041-8205/768/1/L20},
  \href {https://ui.adsabs.harvard.edu/abs/2013ApJ...768L..20Q} {768, L20}

\bibitem[\protect\citeauthoryear{{Reichardt} et~al.,}{{Reichardt}
  et~al.}{2013}]{2013ApJ...763..127R}
{Reichardt} C.~L.,  et~al., 2013, \mn@doi [\apj] {10.1088/0004-637X/763/2/127},
  \href {http://adsabs.harvard.edu/abs/2013ApJ...763..127R} {763, 127}

\bibitem[\protect\citeauthoryear{{Riess}, {Press}  \& {Kirshner}}{{Riess}
  et~al.}{1996}]{1996ApJ...473...88R}
{Riess} A.~G.,  {Press} W.~H.,   {Kirshner} R.~P.,  1996, \mn@doi [\apj]
  {10.1086/178129}, \href {http://adsabs.harvard.edu/abs/1996ApJ...473...88R}
  {473, 88}

\bibitem[\protect\citeauthoryear{{Riess} et~al.,}{{Riess}
  et~al.}{1998}]{1998AJ....116.1009R}
{Riess} A.~G.,  et~al., 1998, \mn@doi [\aj] {10.1086/300499}, \href
  {http://adsabs.harvard.edu/abs/1998AJ....116.1009R} {116, 1009}

\bibitem[\protect\citeauthoryear{{Riess} et~al.,}{{Riess}
  et~al.}{2004}]{2004ApJ...600L.163R}
{Riess} A.~G.,  et~al., 2004, \mn@doi [\apjl] {10.1086/378311}, \href
  {http://adsabs.harvard.edu/abs/2004ApJ...600L.163R} {600, L163}

\bibitem[\protect\citeauthoryear{{Rigault} et~al.,}{{Rigault}
  et~al.}{2013}]{2013A&A...560A..66R}
{Rigault} M.,  et~al., 2013, \mn@doi [\aap] {10.1051/0004-6361/201322104},
  \href {http://adsabs.harvard.edu/abs/2013A\%26A...560A..66R} {560, A66}

\bibitem[\protect\citeauthoryear{{Rigault} et~al.,}{{Rigault}
  et~al.}{2018}]{2018arXiv180603849R}
{Rigault} M.,  et~al., 2018, 
 \href {http://adsabs.harvard.edu/abs/2018arXiv180603849R} {arXiv:1806.03849}

\bibitem[\protect\citeauthoryear{{Rodney} et~al.,}{{Rodney}
  et~al.}{2015}]{2015ApJ...811...70R}
{Rodney} S.~A.,  et~al., 2015, \mn@doi [\apj] {10.1088/0004-637X/811/1/70},
  \href {http://adsabs.harvard.edu/abs/2015ApJ...811...70R} {811, 70}

\bibitem[\protect\citeauthoryear{{Roman} et~al.,}{{Roman}
  et~al.}{2018}]{2018A&A...615A..68R}
{Roman} M.,  et~al., 2018, \mn@doi [\aap] {10.1051/0004-6361/201731425}, \href
  {http://adsabs.harvard.edu/abs/2018A\%26A...615A..68R} {615, A68}

\bibitem[\protect\citeauthoryear{{Rosen} et~al.,}{{Rosen}
  et~al.}{2016}]{2016A&A...590A...1R}
{Rosen} S.~R.,  et~al., 2016, \mn@doi [\aap] {10.1051/0004-6361/201526416},
  \href {http://adsabs.harvard.edu/abs/2016A\%26A...590A...1R} {590, A1}

\bibitem[\protect\citeauthoryear{{Rubin} et~al.,}{{Rubin}
  et~al.}{2018}]{2017arXiv170704606R}
{Rubin} D.,  et~al., 2018, \mn@doi [\apj] {10.3847/1538-4357/aad565}, \href
  {http://adsabs.harvard.edu/abs/2018ApJ...866...65R} {866, 65}

\bibitem[\protect\citeauthoryear{{Sako} et~al.,}{{Sako}
  et~al.}{2018}]{2018PASP..130f4002S}
{Sako} M.,  et~al., 2018, \mn@doi [\pasp] {10.1088/1538-3873/aab4e0}, \href
  {http://adsabs.harvard.edu/abs/2018PASP..130f4002S} {130, 064002}

\bibitem[\protect\citeauthoryear{{Santos} et~al.,}{{Santos}
  et~al.}{2011}]{2011A&A...531L..15S}
{Santos} J.~S.,  et~al., 2011, \mn@doi [\aap] {10.1051/0004-6361/201117190},
  \href {http://cdsads.u-strasbg.fr/abs/2011A\%26A...531L..15S} {531, L15}

\bibitem[\protect\citeauthoryear{{Santos} et~al.,}{{Santos}
  et~al.}{2015}]{2015MNRAS.447L..65S}
{Santos} J.~S.,  et~al., 2015, \mn@doi [\mnras] {10.1093/mnrasl/slu180}, \href
  {http://cdsads.u-strasbg.fr/abs/2015MNRAS.447L..65S} {447, L65}

\bibitem[\protect\citeauthoryear{{Schlafly} \& {Finkbeiner}}{{Schlafly} \&
  {Finkbeiner}}{2011}]{2011ApJ...737..103S}
{Schlafly} E.~F.,  {Finkbeiner} D.~P.,  2011, \mn@doi [\apj]
  {10.1088/0004-637X/737/2/103}, \href
  {http://adsabs.harvard.edu/abs/2011ApJ...737..103S} {737, 103}

\bibitem[\protect\citeauthoryear{{Scolnic} et~al.,}{{Scolnic}
  et~al.}{2018}]{2018ApJ...859..101S}
{Scolnic} D.~M.,  et~al., 2018, \mn@doi [\apj] {10.3847/1538-4357/aab9bb},
  \href {http://adsabs.harvard.edu/abs/2018ApJ...859..101S} {859, 101}

\bibitem[\protect\citeauthoryear{{Silverman} et~al.,}{{Silverman}
  et~al.}{2012}]{2012MNRAS.425.1789S}
{Silverman} J.~M.,  et~al., 2012, \mn@doi [\mnras]
  {10.1111/j.1365-2966.2012.21270.x}, \href
  {http://adsabs.harvard.edu/abs/2012MNRAS.425.1789S} {425, 1789}

\bibitem[\protect\citeauthoryear{{Smette} et~al.,}{{Smette}
  et~al.}{2015}]{2015A&A...576A..77S}
{Smette} A.,  et~al., 2015, \mn@doi [\aap] {10.1051/0004-6361/201423932}, \href
  {http://adsabs.harvard.edu/abs/2015A\%26A...576A..77S} {576, A77}

\bibitem[\protect\citeauthoryear{{Stalder} et~al.,}{{Stalder}
  et~al.}{2013}]{2013ApJ...763...93S}
{Stalder} B.,  et~al., 2013, \mn@doi [\apj] {10.1088/0004-637X/763/2/93}, \href
  {http://adsabs.harvard.edu/abs/2013ApJ...763...93S} {763, 93}

\bibitem[\protect\citeauthoryear{{Stritzinger} et~al.,}{{Stritzinger}
  et~al.}{2011}]{2011AJ....142..156S}
{Stritzinger} M.~D.,  et~al., 2011, \mn@doi [\aj]
  {10.1088/0004-6256/142/5/156}, \href
  {http://adsabs.harvard.edu/abs/2011AJ....142..156S} {142, 156}

\bibitem[\protect\citeauthoryear{{Sullivan} et~al.,}{{Sullivan}
  et~al.}{2006}]{2006ApJ...648..868S}
{Sullivan} M.,  et~al., 2006, \mn@doi [\apj] {10.1086/506137}, \href
  {http://adsabs.harvard.edu/abs/2006ApJ...648..868S} {648, 868}

\bibitem[\protect\citeauthoryear{{Sullivan} et~al.,}{{Sullivan}
  et~al.}{2010}]{2010MNRAS.406..782S}
{Sullivan} M.,  et~al., 2010, \mn@doi [\mnras]
  {10.1111/j.1365-2966.2010.16731.x}, \href
  {http://adsabs.harvard.edu/abs/2010MNRAS.406..782S} {406, 782}

\bibitem[\protect\citeauthoryear{{Suzuki} et~al.,}{{Suzuki}
  et~al.}{2012}]{2012ApJ...746...85S}
{Suzuki} N.,  et~al., 2012, \mn@doi [\apj] {10.1088/0004-637X/746/1/85}, \href
  {https://ui.adsabs.harvard.edu/abs/2012ApJ...746...85S} {746, 85}

\bibitem[\protect\citeauthoryear{{Tody}}{{Tody}}{1986}]{1986SPIE..627..733T}
{Tody} D.,  1986, in {Crawford} D.~L.,  ed.,  Society of Photo-Optical
  Instrumentation Engineers (SPIE) Conference Series Vol. 627, Instrumentation
  in astronomy VI. p.~733

\bibitem[\protect\citeauthoryear{{Totani}, {Morokuma}, {Oda}, {Doi}  \&
  {Yasuda}}{{Totani} et~al.}{2008}]{2008PASJ...60.1327T}
{Totani} T.,  {Morokuma} T.,  {Oda} T.,  {Doi} M.,   {Yasuda} N.,  2008,
  \mn@doi [\pasj] {10.1093/pasj/60.6.1327}, \href
  {https://ui.adsabs.harvard.edu/abs/2008PASJ...60.1327T} {60, 1327}

\bibitem[\protect\citeauthoryear{{Tripp}}{{Tripp}}{1998}]{1998A&A...331..815T}
{Tripp} R.,  1998, \aap, \href
  {http://adsabs.harvard.edu/abs/1998A\%26A...331..815T} {331, 815}

\bibitem[\protect\citeauthoryear{{Vernet} et~al.,}{{Vernet}
  et~al.}{2011}]{2011A&A...536A.105V}
{Vernet} J.,  et~al., 2011, \mn@doi [\aap] {10.1051/0004-6361/201117752}, \href
  {http://adsabs.harvard.edu/abs/2011A\%26A...536A.105V} {536, A105}

\bibitem[\protect\citeauthoryear{{Webb} et~al.,}{{Webb}
  et~al.}{2015}]{2015ApJ...809..173W}
{Webb} T.,  et~al., 2015, \mn@doi [\apj] {10.1088/0004-637X/809/2/173}, \href
  {http://adsabs.harvard.edu/abs/2015ApJ...809..173W} {809, 173}

\bibitem[\protect\citeauthoryear{{Wilson} et~al.,}{{Wilson}
  et~al.}{2009}]{2009ApJ...698.1943W}
{Wilson} G.,  et~al., 2009, \mn@doi [\apj] {10.1088/0004-637X/698/2/1943},
  \href {http://adsabs.harvard.edu/abs/2009ApJ...698.1943W} {698, 1943}

\bibitem[\protect\citeauthoryear{{Wright} et~al.,}{{Wright}
  et~al.}{2016}]{2016MNRAS.460..765W}
{Wright} A.~H.,  et~al., 2016, \mn@doi [\mnras] {10.1093/mnras/stw832}, \href
  {http://adsabs.harvard.edu/abs/2016MNRAS.460..765W} {460, 765}

\bibitem[\protect\citeauthoryear{{Wu} et~al.,}{{Wu}
  et~al.}{2018}]{2018ApJ...852...96W}
{Wu} J.,  et~al., 2018, \mn@doi [\apj] {10.3847/1538-4357/aa9ff3}, \href
  {http://adsabs.harvard.edu/abs/2018ApJ...852...96W} {852, 96}

\makeatother
\end{thebibliography}




\appendix

\begin{landscape}
\section{VLT observation log}
\begin{table}
	\centering
	\caption{See Change transients/host-galaxies observed with the VLT. Note that the exposure time details here refer to the total exposure targeted at an object, whether the data were used or not. The positions here are the calculated position of the transients themselves, rather than the host galaxies}
	\label{tab:hostobs}
	\begin{tabular}{lllclc} 
\hline
	Transient &Position (J2000) & FORS2 Obs. dates &FORS2 Exp. time (h) & X-Shooter Obs. dates & X-Shooter Exp. time (h)\\
\hline
	SCP15A01 &$02^{\mathrm{h}}05^{\mathrm{m}}46^{\mathrm{s}}\!.345$ $-58^{\circ}28^{\prime}54^{\prime\prime}\!\!.94$ & 2016 Sep 7 -- 2016 Sep 20 & 2.00  & 2015 Feb 13 -- 2015 Feb 18 &5.00\\
	SCP15A02 &$02^{\mathrm{h}}05^{\mathrm{m}}45^{\mathrm{s}}\!.662$ $-58^{\circ}28^{\prime}16^{\prime\prime}\!\!.86$ & 2015 Oct 4 -- 2015 Nov 10 &2.33 &-- &--\\
	SCP15A03 &$02^{\mathrm{h}}05^{\mathrm{m}}44^{\mathrm{s}}\!.383$ $-58^{\circ}30^{\prime}01^{\prime\prime}\!\!.90$ & 2015 Oct 4 -- 2016 Sep 20 &4.33 &-- &--\\
	SCP15A04 &$02^{\mathrm{h}}05^{\mathrm{m}}51^{\mathrm{s}}\!.130$ $-58^{\circ}29^{\prime}27^{\prime\prime}\!\!.38$ & 2015 Oct 4 -- 2016 Sep 20 &4.33 &2016 Aug 5 -- 2016 Aug 8 &4.00\\
	SCP15A05 &$02^{\mathrm{h}}05^{\mathrm{m}}48^{\mathrm{s}}\!.119$ $-58^{\circ}29^{\prime}26^{\prime\prime}\!\!.69$ & 2015 Oct 4 -- 2015 Nov 10 &2.33 &2015 Aug 26 -- 2015 Nov 15 &4.00\\
	SCP15A06 &$02^{\mathrm{h}}05^{\mathrm{m}}43^{\mathrm{s}}\!.091$ $-58^{\circ}29^{\prime}36^{\prime\prime}\!\!.73$ & 2015 Oct 4 -- 2015 Nov 10 &2.33 &-- &--\\
	SCP15A07 &$02^{\mathrm{h}}05^{\mathrm{m}}49^{\mathrm{s}}\!.295$ $-58^{\circ}29^{\prime}41^{\prime\prime}\!\!.02$ & 2016 Sep 7 &0.67 &-- &--\\
	SCP16A01 &$02^{\mathrm{h}}05^{\mathrm{m}}44^{\mathrm{s}}\!.320$ $-58^{\circ}29^{\prime}06^{\prime\prime}\!\!.96$ & 2016 Sep 7 &0.67 &-- &--\\
	SCP16A02 &$02^{\mathrm{h}}05^{\mathrm{m}}45^{\mathrm{s}}\!.844$ $-58^{\circ}27^{\prime}50^{\prime\prime}\!\!.00$ & 2016 Sep 7 -- 2016 Sep 13 & 2.00 &-- &--\\
	SCP16A03 &$02^{\mathrm{h}}05^{\mathrm{m}}52^{\mathrm{s}}\!.390$ $-58^{\circ}29^{\prime}40^{\prime\prime}\!\!.28$ & 2016 Sep 13 &0.67 &-- &--\\
	SCP16A04 &$02^{\mathrm{h}}05^{\mathrm{m}}51^{\mathrm{s}}\!.086$ $-58^{\circ}29^{\prime}26^{\prime\prime}\!\!.91$ &\multicolumn{4}{c}{{\it - - - - - - - - - - - - - - - - - - - - - - - - - - - Same host as SCP15A04 - - - - - - - - - - - - - - - - - - - - - - - - - -}}\\
	SCP14C01 &$10^{\mathrm{h}}14^{\mathrm{m}}04^{\mathrm{s}}\!.056$ $+00^{\circ}38^{\prime}11^{\prime\prime}\!\!.61$ & 2015 Jan 20 &2.71 &-- &--\\
	SCP15C01 &$10^{\mathrm{h}}14^{\mathrm{m}}05^{\mathrm{s}}\!.174$ $+00^{\circ}37^{\prime}16^{\prime\prime}\!\!.39$ & 2017 Dec 19 -- 2018 Feb 11 & 4.00 &-- &--\\
    SCP15C02 &$10^{\mathrm{h}}14^{\mathrm{m}}07^{\mathrm{s}}\!.619$ $+00^{\circ}38^{\prime}08^{\prime\prime}\!\!.44$ & 2017 Dec 19 -- 2018 Feb 11 & 4.00 &-- &--\\
	SCP15C03 &$10^{\mathrm{h}}14^{\mathrm{m}}07^{\mathrm{s}}\!.237$ $+00^{\circ}37^{\prime}29^{\prime\prime}\!\!.42$ & 2017 Dec 19 -- 2018 Feb 11 & 4.00 &-- &--\\
    SCP15C04 &$10^{\mathrm{h}}14^{\mathrm{m}}06^{\mathrm{s}}\!.530$ $+00^{\circ}38^{\prime}47^{\prime\prime}\!\!.08$ & 2017 Dec 19 -- 2018 Feb 11 & 4.00 &-- &--\\
	SCP16C01 &$10^{\mathrm{h}}14^{\mathrm{m}}06^{\mathrm{s}}\!.748$ $+00^{\circ}39^{\prime}03^{\prime\prime}\!\!.98$ & 2017 Dec 19 -- 2018 Feb 11 & 2.00 &-- &--\\
	SCP16C02 &$10^{\mathrm{h}}14^{\mathrm{m}}08^{\mathrm{s}}\!.349$ $+00^{\circ}38^{\prime}52^{\prime\prime}\!\!.37$ & 2017 Dec 19 -- 2018 Feb 11 & 2.00 &-- &--\\
	SCP16C03 &$10^{\mathrm{h}}14^{\mathrm{m}}06^{\mathrm{s}}\!.374$ $+00^{\circ}38^{\prime}25^{\prime\prime}\!\!.36$ & -- & -- & 2016 Mar 4 -- 2016 Mar 5 & 5.00\\
	SCP15D01 &$21^{\mathrm{h}}06^{\mathrm{m}}11^{\mathrm{s}}\!.769$ $-58^{\circ}45^{\prime}20^{\prime\prime}\!\!.26$ & 2016 Jun 30 -- 2016 Jul 2 & 3.33 &-- &--\\
	SCP15D02 &$21^{\mathrm{h}}06^{\mathrm{m}}07^{\mathrm{s}}\!.313$ $-58^{\circ}44^{\prime}02^{\prime\prime}\!\!.73$ & 2016 Jun 30 -- 2016 Oct 1 & 6.67 &2015 Sep 18 -- 2016 Jun 30 &4.00\\
	SCP15D03 &$21^{\mathrm{h}}06^{\mathrm{m}}02^{\mathrm{s}}\!.299$ $-58^{\circ}44^{\prime}54^{\prime\prime}\!\!.03$ & 2016 Jun 30 -- 2016 Oct 1 & 6.67 &-- &--\\
	SCP15D04 &$21^{\mathrm{h}}06^{\mathrm{m}}00^{\mathrm{s}}\!.343$ $-58^{\circ}45^{\prime}48^{\prime\prime}\!\!.51$ & 2016 Jun 30 -- 2016 Oct 1  & 6.67 &-- &--\\
	SCP16D01 &$21^{\mathrm{h}}06^{\mathrm{m}}07^{\mathrm{s}}\!.157$ $-58^{\circ}45^{\prime}11^{\prime\prime}\!\!.35$ & 2016 Jun 30 -- 2016 Jul 2 & 3.33 &-- &--\\
	SCP16D02 &$21^{\mathrm{h}}06^{\mathrm{m}}03^{\mathrm{s}}\!.439$ $-58^{\circ}44^{\prime}16^{\prime\prime}\!\!.01$ & 2016 Sep 4 -- 2016 Oct 1 & 3.33 &-- &--\\
	SCP16D03 &$21^{\mathrm{h}}05^{\mathrm{m}}57^{\mathrm{s}}\!.151$ $-58^{\circ}45^{\prime}12^{\prime\prime}\!\!.28$ & 2016 Sep 4 -- 2016 Oct 1 & 3.33 &-- &--\\
	SCP15E01 &$20^{\mathrm{h}}41^{\mathrm{m}}04^{\mathrm{s}}\!.766$ $-44^{\circ}52^{\prime}01^{\prime\prime}\!\!.26$ & 2016 Jun 17 -- 2017 Apr 27 &2.00 &-- &--\\
	SCP15E02 &$20^{\mathrm{h}}40^{\mathrm{m}}58^{\mathrm{s}}\!.767$ $-44^{\circ}51^{\prime}02^{\prime\prime}\!\!.55$ & 2016 Oct 30 -- 2017 Apr 27 &1.33 &-- &--\\
	SCP15E03 &$20^{\mathrm{h}}41^{\mathrm{m}}02^{\mathrm{s}}\!.926$ $-44^{\circ}51^{\prime}31^{\prime\prime}\!\!.59$ & 2016 Oct 30 -- 2017 Apr 27 &1.33 &-- &--\\
	SCP15E04 &$20^{\mathrm{h}}40^{\mathrm{m}}59^{\mathrm{s}}\!.697$ $-44^{\circ}52^{\prime}05^{\prime\prime}\!\!.05$ & 2017 Apr 27 -- 2018 Apr 10 &8.00 &-- &--\\
	SCP15E05 &$20^{\mathrm{h}}41^{\mathrm{m}}01^{\mathrm{s}}\!.123$ $-44^{\circ}51^{\prime}12^{\prime\prime}\!\!.21$ & 2016 Oct 30 -- 2018 Apr 10 &7.33 &-- &--\\
	SCP15E06 &$20^{\mathrm{h}}40^{\mathrm{m}}57^{\mathrm{s}}\!.739$ $-44^{\circ}51^{\prime}06^{\prime\prime}\!\!.52$ & 2016 Oct 30 -- 2018 Apr 10 &9.33 &-- &--\\
	SCP15E07 &$20^{\mathrm{h}}41^{\mathrm{m}}03^{\mathrm{s}}\!.306$ $-44^{\circ}51^{\prime}39^{\prime\prime}\!\!.47$ & 2017 Apr 27 -- 2017 Jun 3 &4.00 & 2016 Sep 21 -- 2016 Sep 30 &4.00\\
	SCP15E08 &$20^{\mathrm{h}}41^{\mathrm{m}}03^{\mathrm{s}}\!.775$ $-44^{\circ}51^{\prime}28^{\prime\prime}\!\!.39$ & 2017 Jun 2 -- 2018 Apr 10 &6.00 &-- &--\\
	SCP16E02 &$20^{\mathrm{h}}41^{\mathrm{m}}05^{\mathrm{s}}\!.729$ $-44^{\circ}51^{\prime}15^{\prime\prime}\!\!.90$ & 2017 Apr 27 -- 2017 Jun 3 &8.00 &-- &--\\
	SCP15G01 &$00^{\mathrm{h}}44^{\mathrm{m}}02^{\mathrm{s}}\!.946$ $-20^{\circ}33^{\prime}49^{\prime\prime}\!\!.71$ & -- &-- & 2015 Aug 24 -- 2015 Dec 8 &5.00\\
	SCP16H01 &$03^{\mathrm{h}}30^{\mathrm{m}}55^{\mathrm{s}}\!.991$ $-28^{\circ}42^{\prime}46^{\prime\prime}\!\!.56$ & -- &-- & 2016 Feb 29 -- 2016 Mar 4 & 4.00\\
	SCP15I02 &$02^{\mathrm{h}}24^{\mathrm{m}}27^{\mathrm{s}}\!.593$ $-03^{\circ}22^{\prime}57^{\prime\prime}\!\!.85$ & -- &-- & 2016 Aug 31 -- 2017 Jan 1 &9.00\\
\hline
\end{tabular}
\\
\end{table}
\end{landscape}

\section{Zoomed-in finding charts (online material)} \label{appendix}
In Figure~\ref{app1} we present a zoomed-in finding chart of each SN candidate, indicating its position within its host galaxy. 

\begin{figure*}
    \centering
    \begin{subfigure}[t]{\textwidth}
\includegraphics[width=0.5\columnwidth]{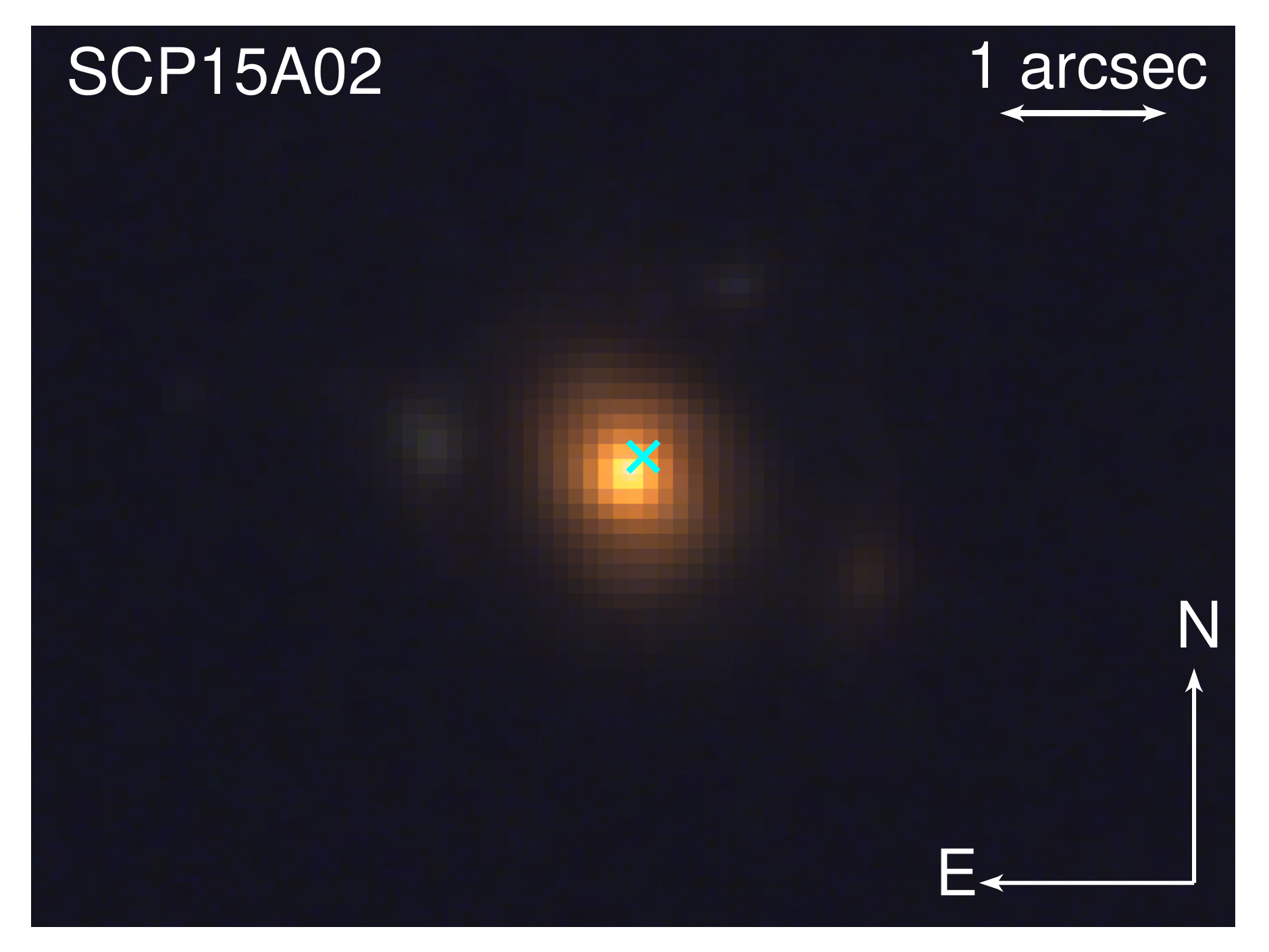}
\includegraphics[width=0.5\columnwidth]{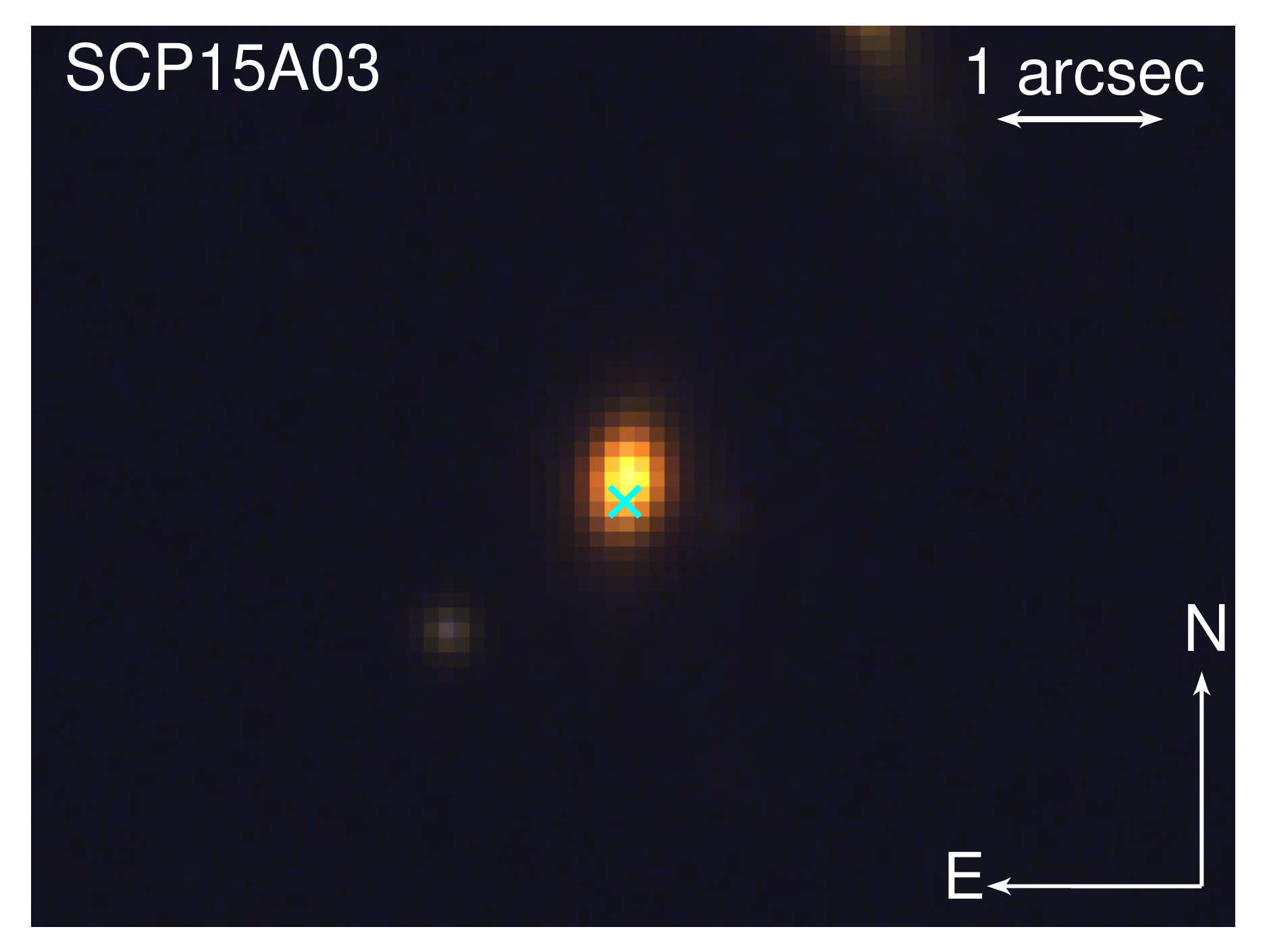}
\includegraphics[width=0.5\columnwidth]{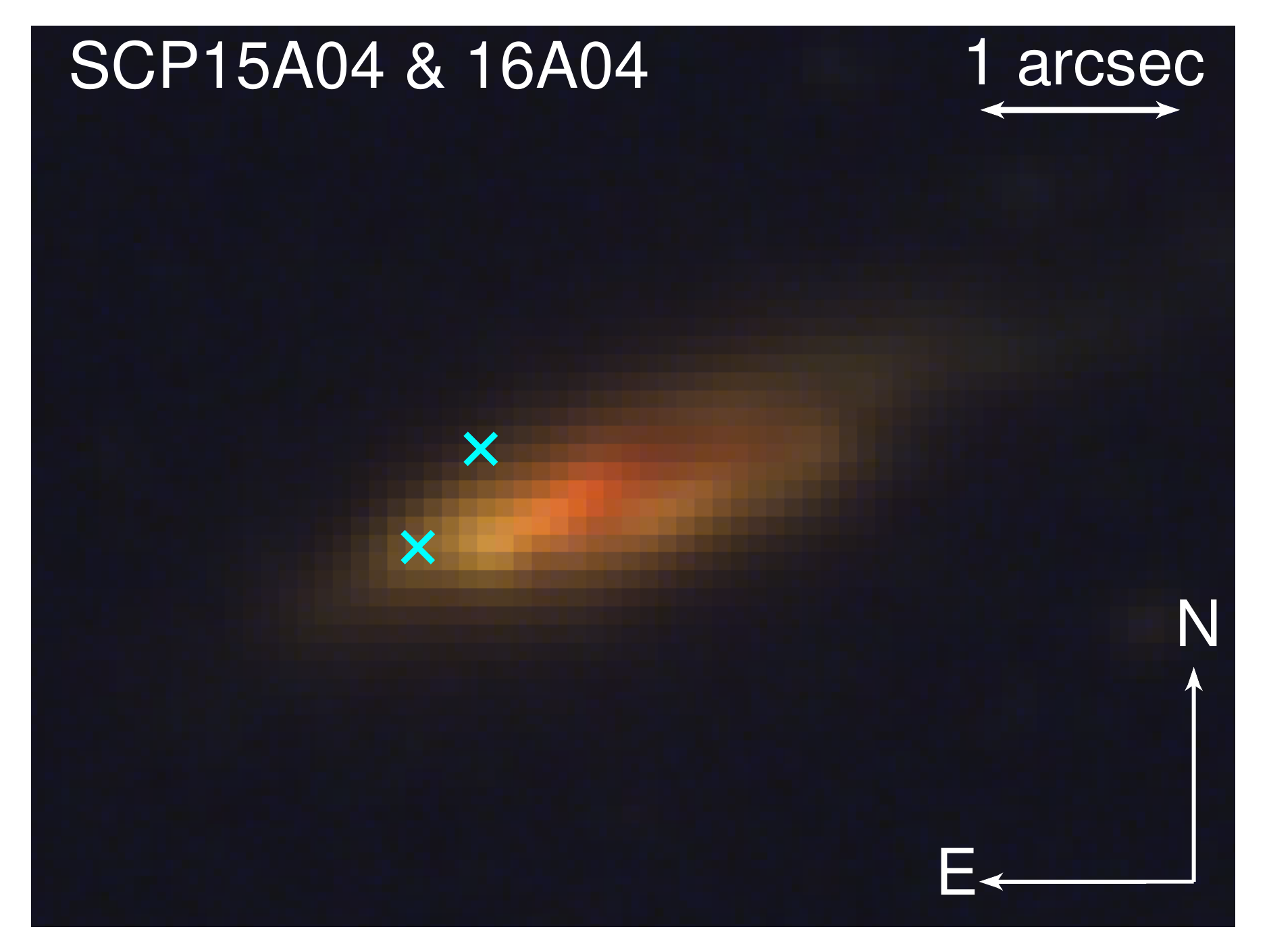}
\includegraphics[width=0.5\columnwidth]{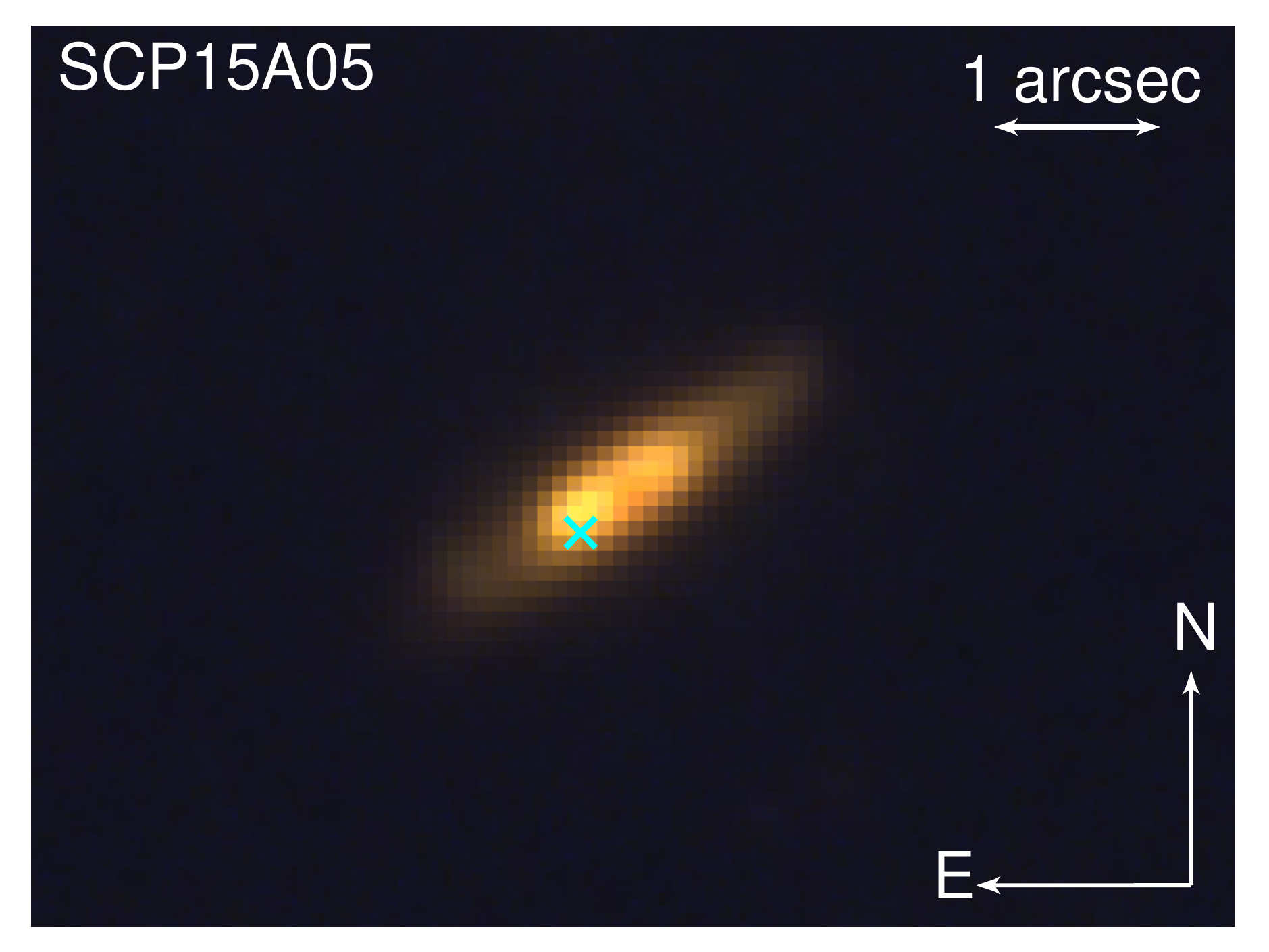}
\includegraphics[width=0.5\columnwidth]{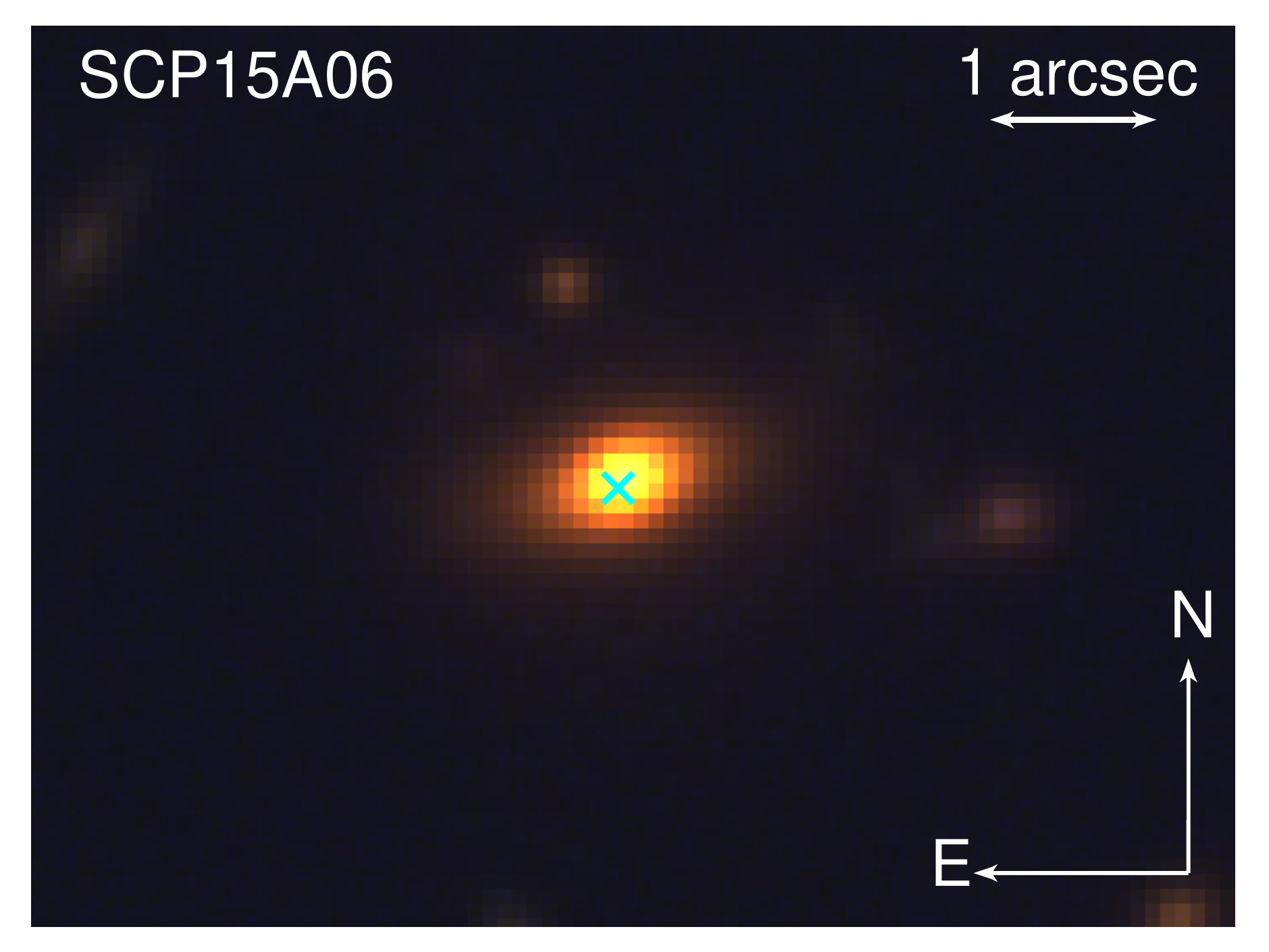}
\includegraphics[width=0.5\columnwidth]{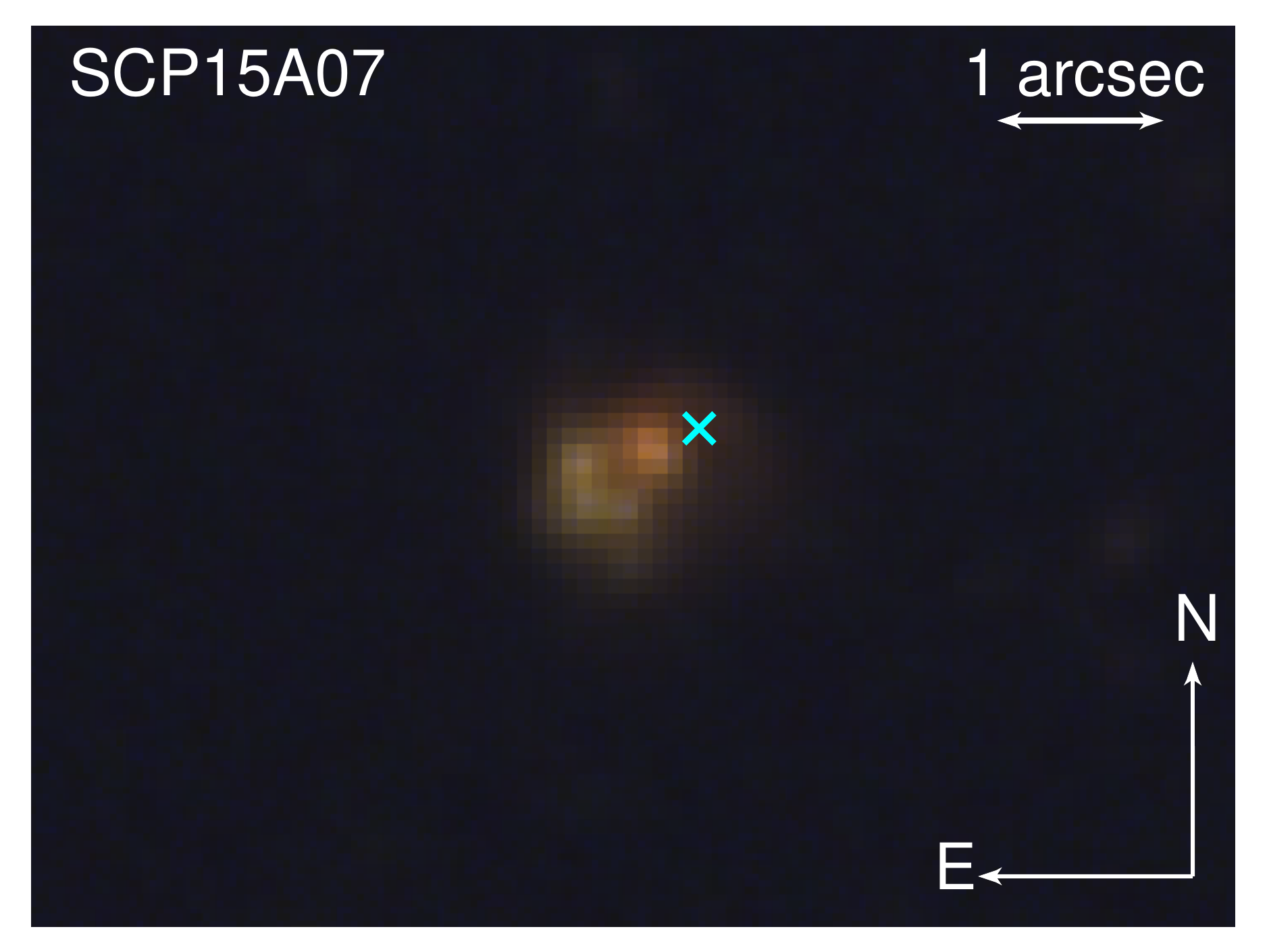}
    \end{subfigure}
\caption{Stacked colour-composite \textit{HST} \textit{F814W}, \textit{F105W} and \textit{F140W} See Change finding charts indicating the position of each SN within its host. SN position indicated by a cyan `$\times$'. In the case of SCP15A04 and SCP16A04, the upper $\times$ is indicating SCP16A04, with the lower being SCP15A04.}\label{app1}
\end{figure*}
\begin{figure*}\ContinuedFloat
    \begin{subfigure}[t]{\textwidth}
\includegraphics[width=0.5\columnwidth]{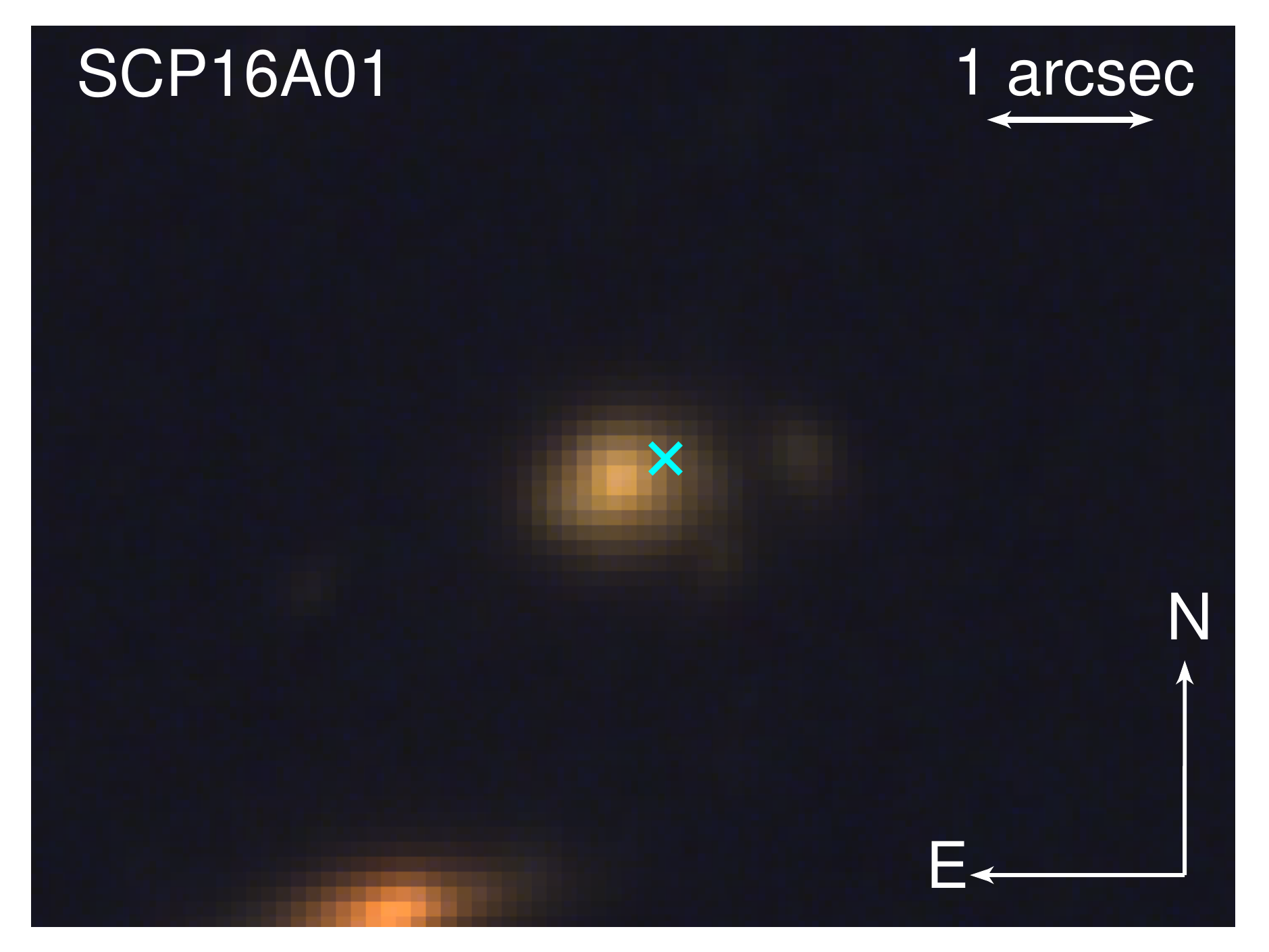}
\includegraphics[width=0.5\columnwidth]{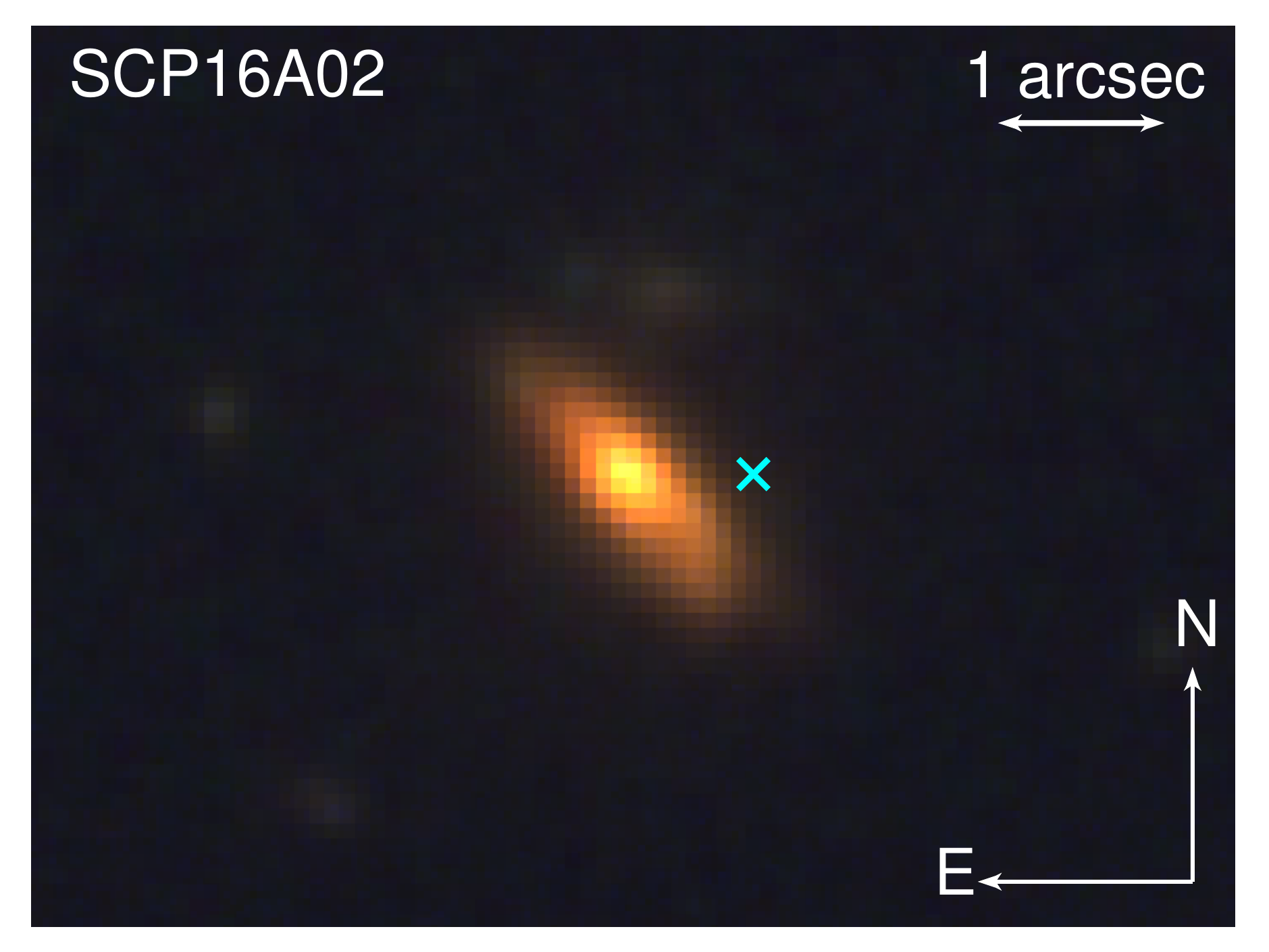}
\includegraphics[width=0.5\columnwidth]{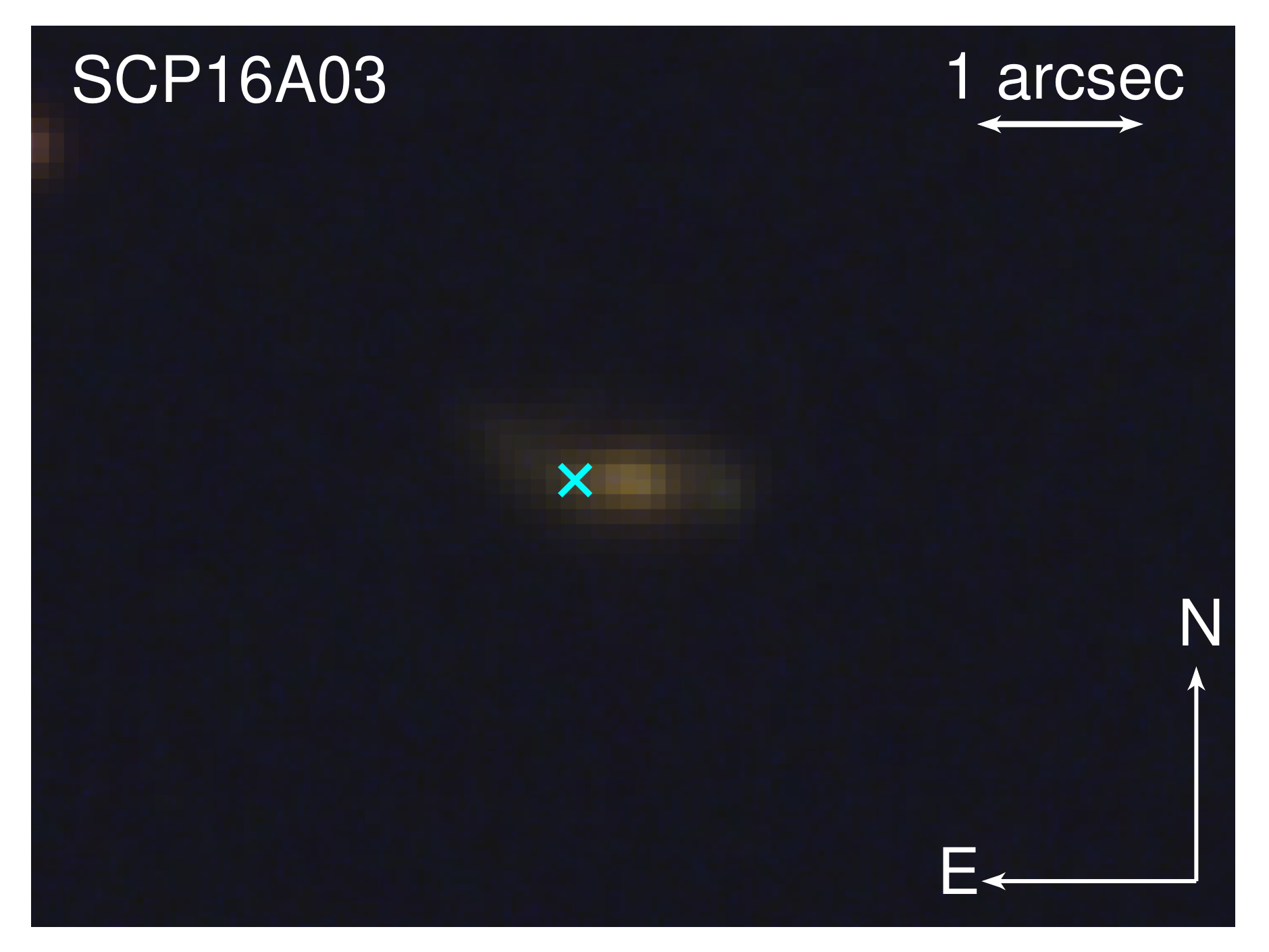}
\includegraphics[width=0.5\columnwidth]{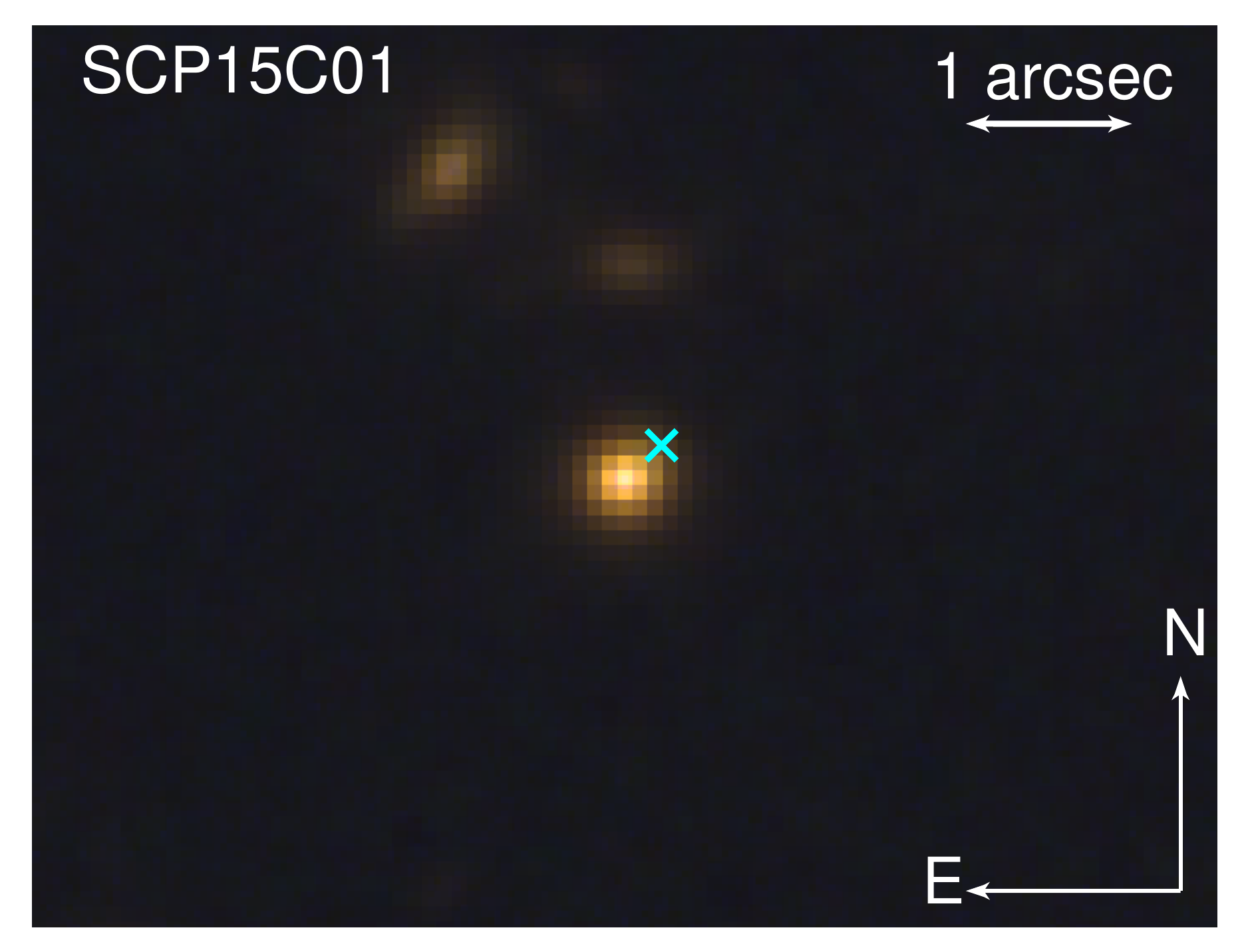}
\includegraphics[width=0.5\columnwidth]{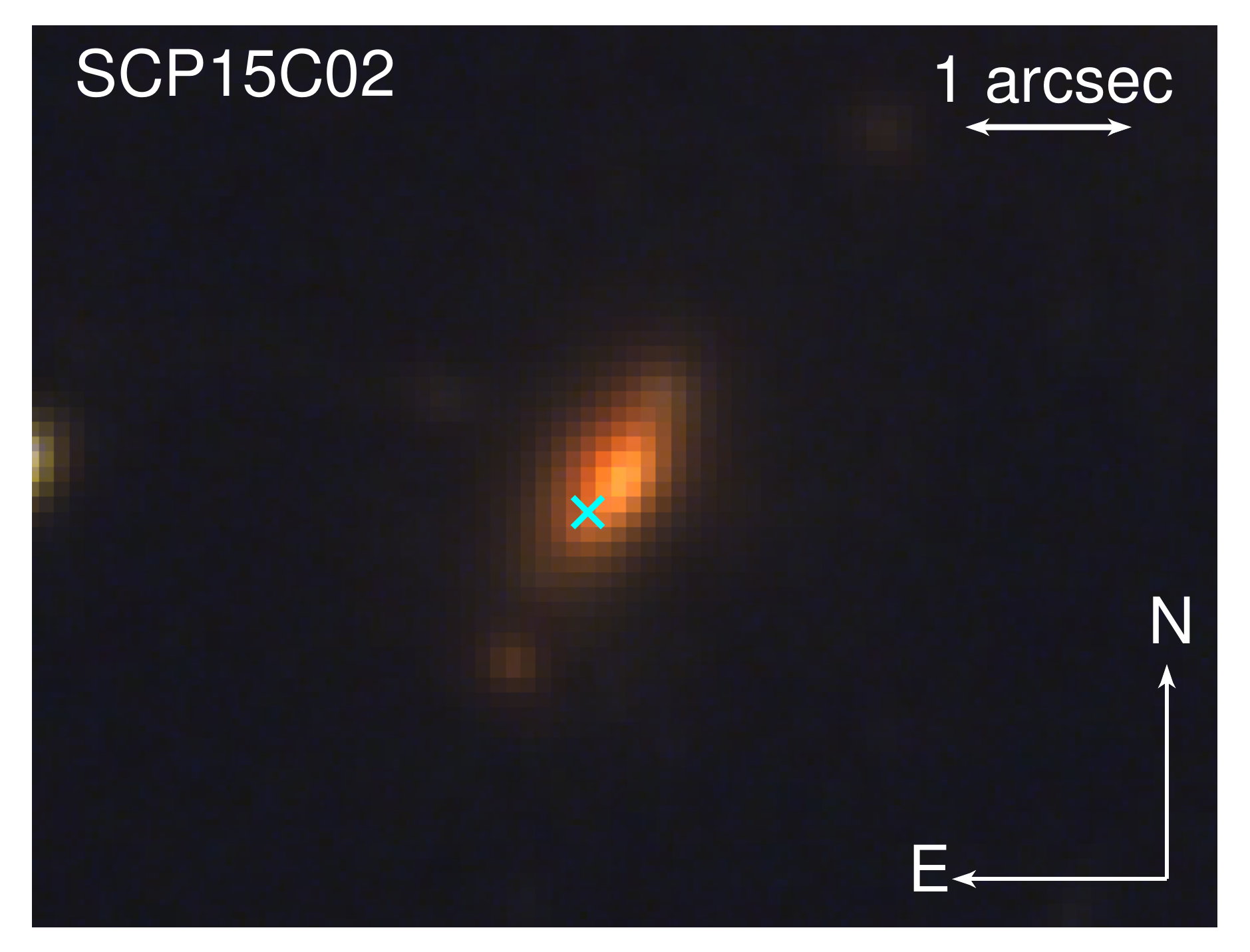}
\includegraphics[width=0.5\columnwidth]{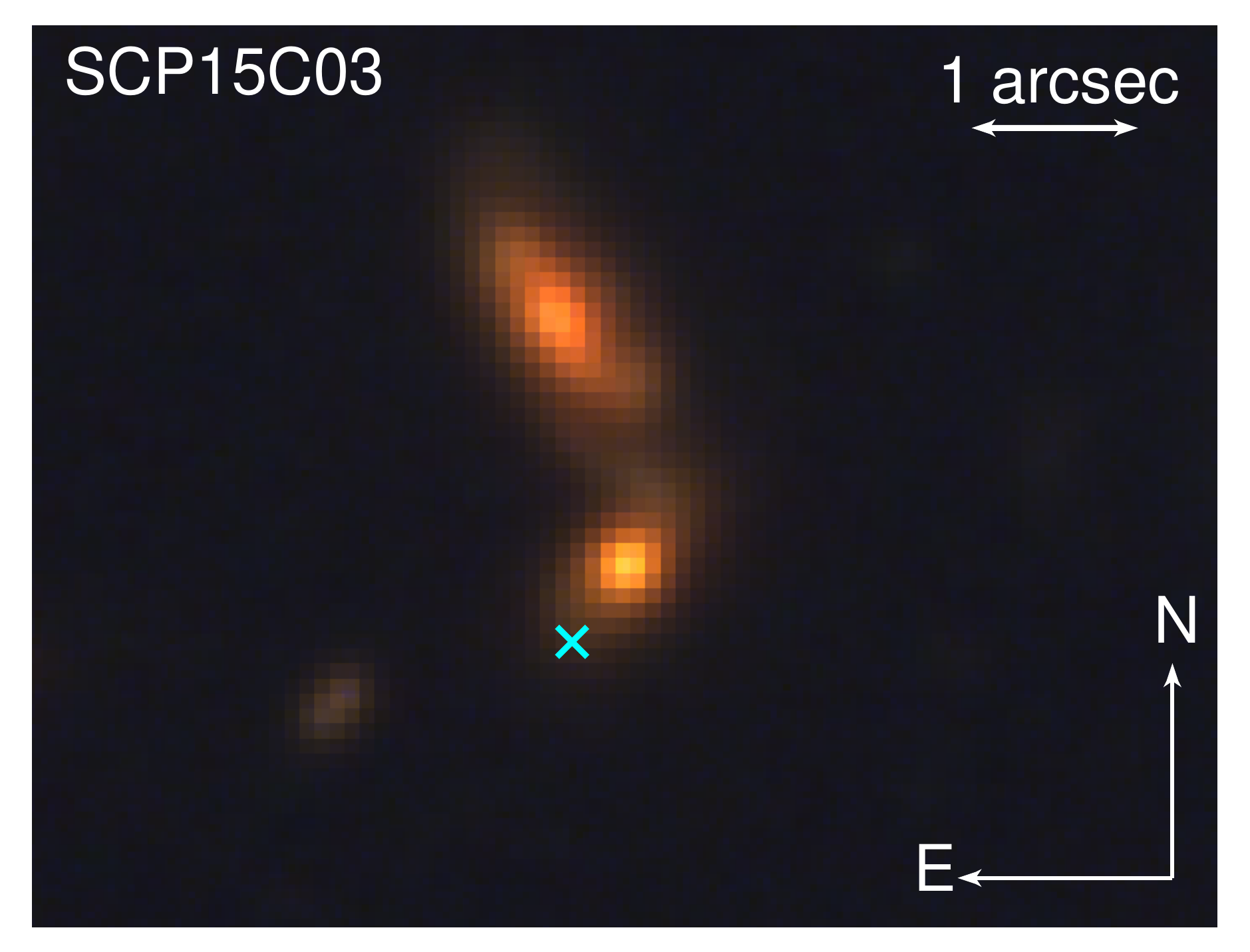}
    \end{subfigure}
\caption{\textbf{Continued.} Stacked colour-composite \textit{HST} \textit{F814W}, \textit{F105W} and \textit{F140W} See Change finding charts indicating the position of each SN within its host.}
\end{figure*}
\begin{figure*}\ContinuedFloat
    \begin{subfigure}[t]{\textwidth}
\includegraphics[width=0.5\columnwidth]{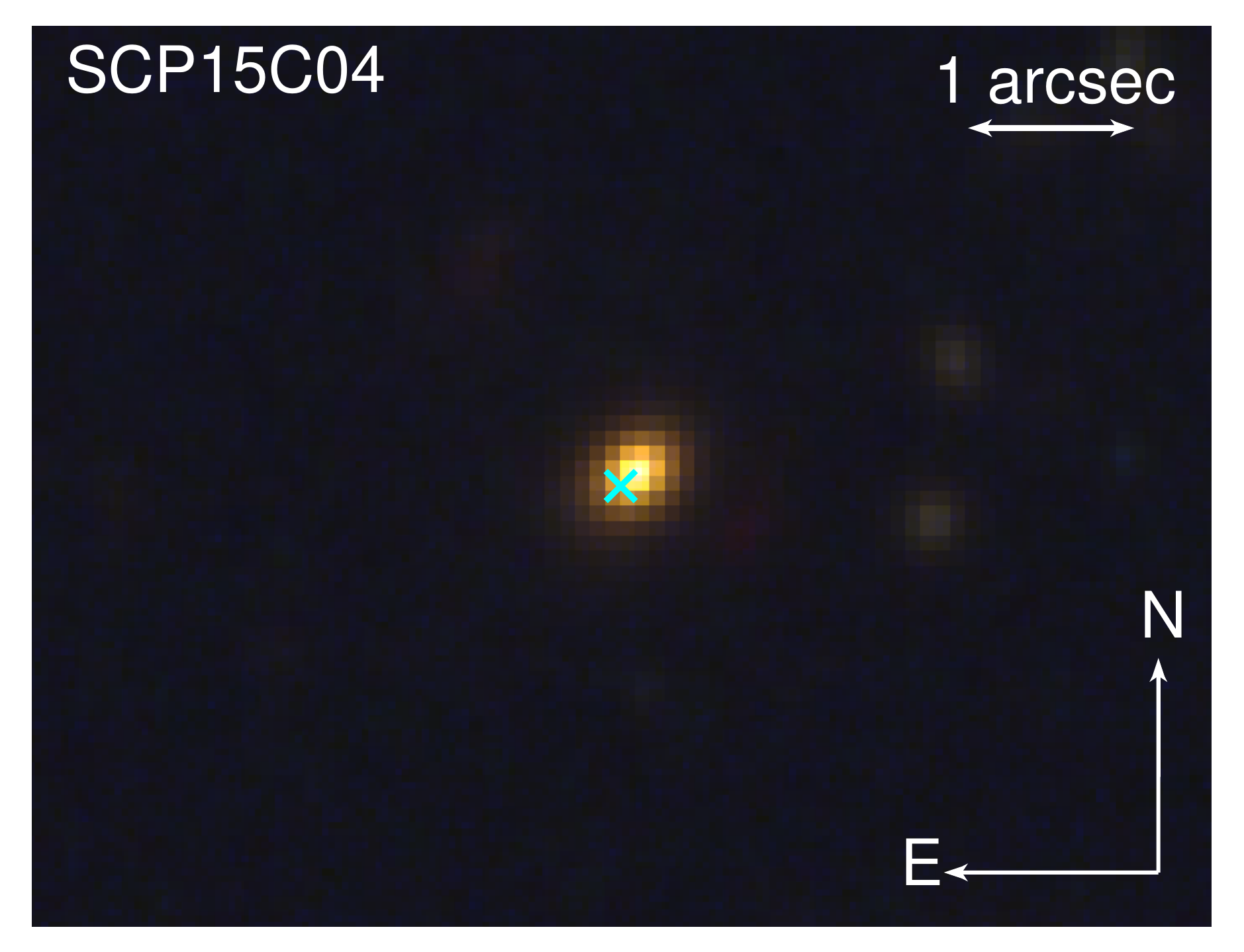}
\includegraphics[width=0.5\columnwidth]{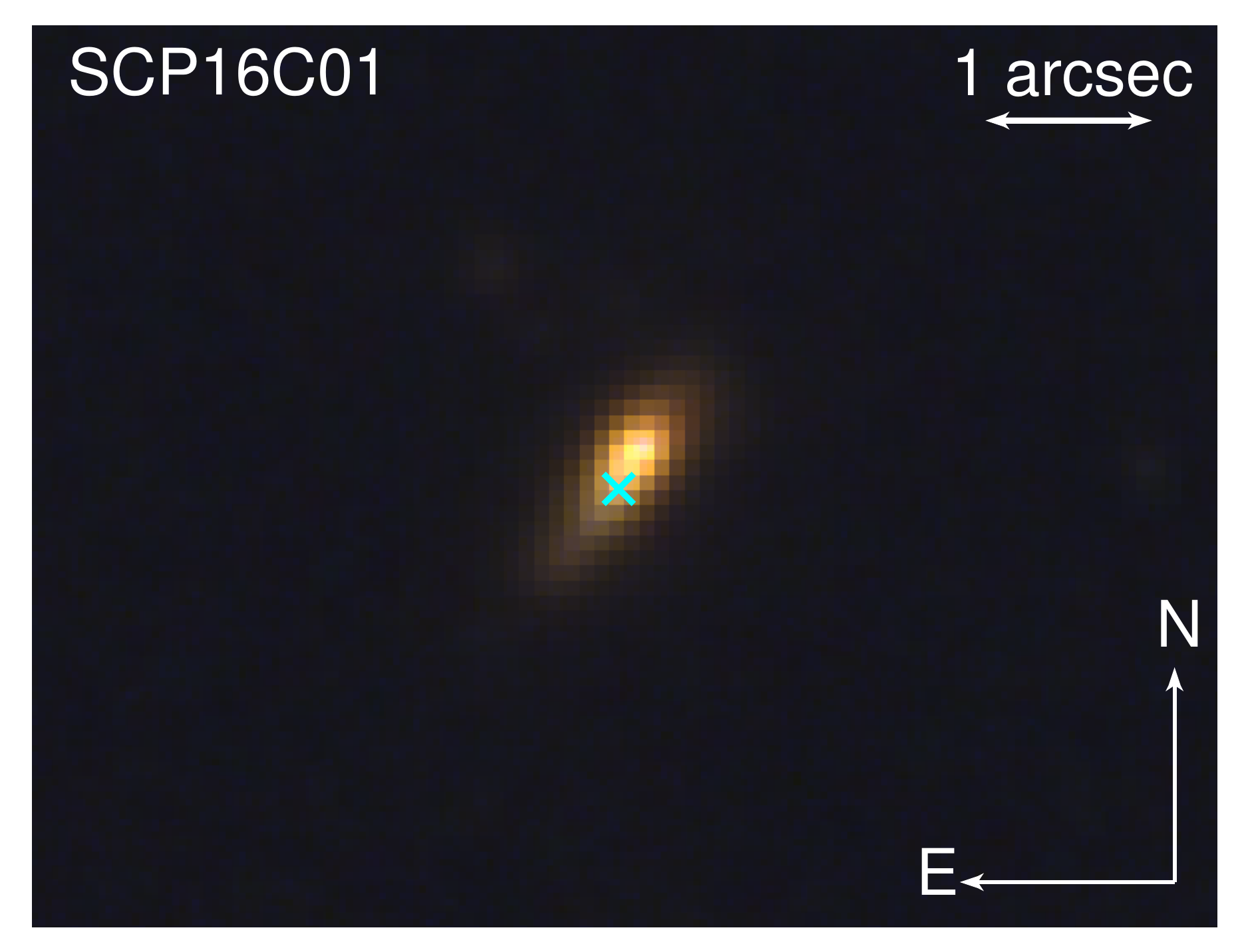}
\includegraphics[width=0.5\columnwidth]{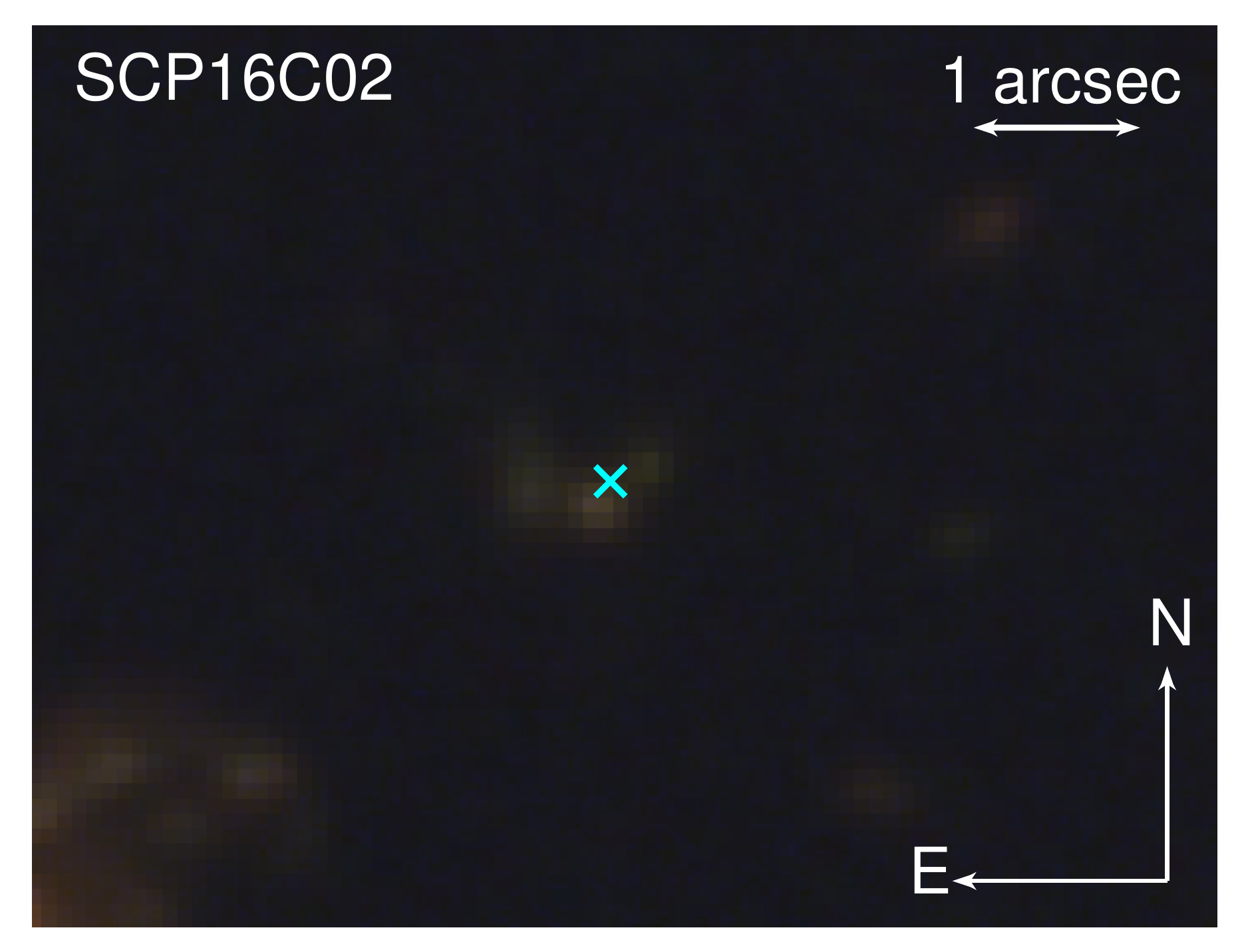}
\includegraphics[width=0.5\columnwidth]{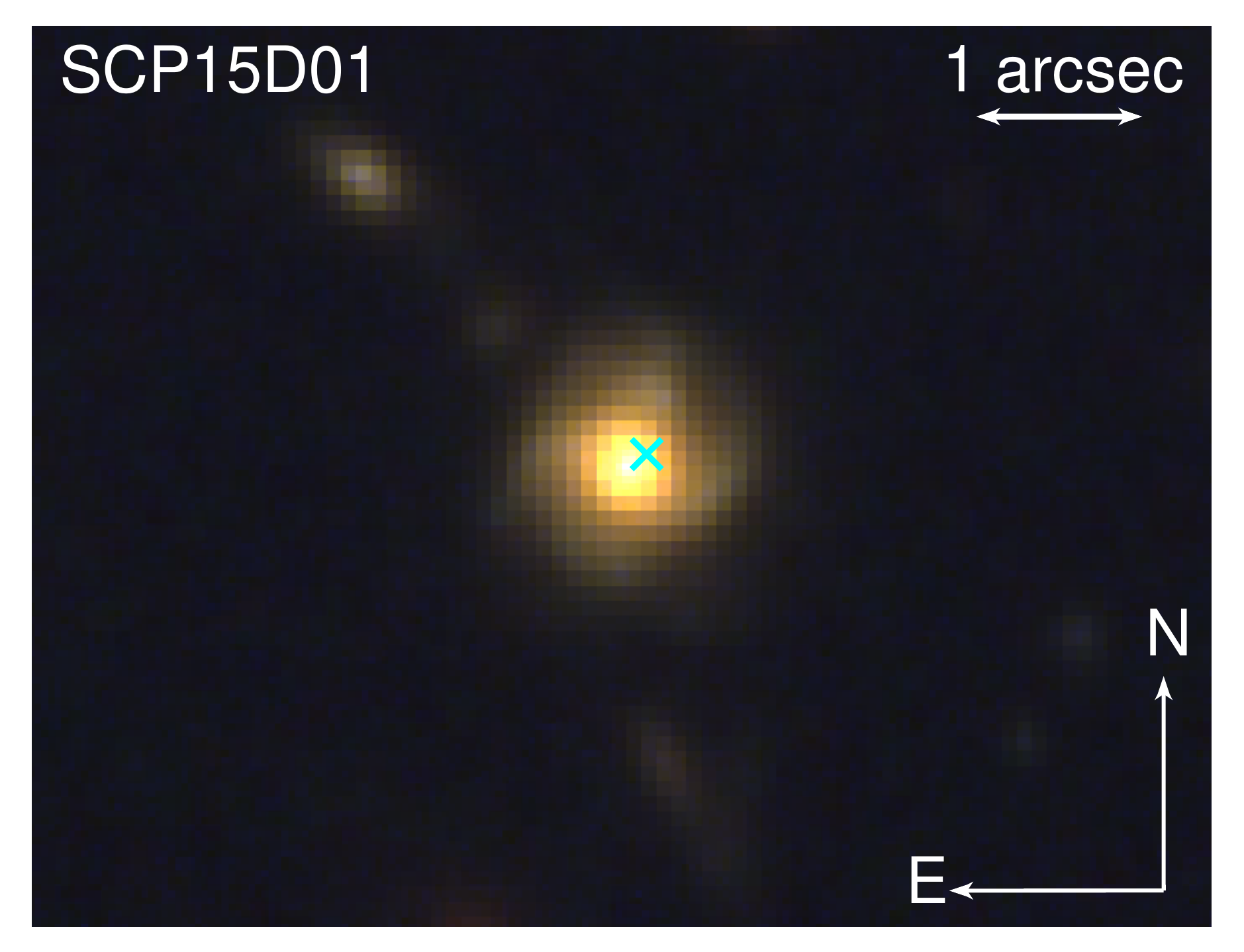}
\includegraphics[width=0.5\columnwidth]{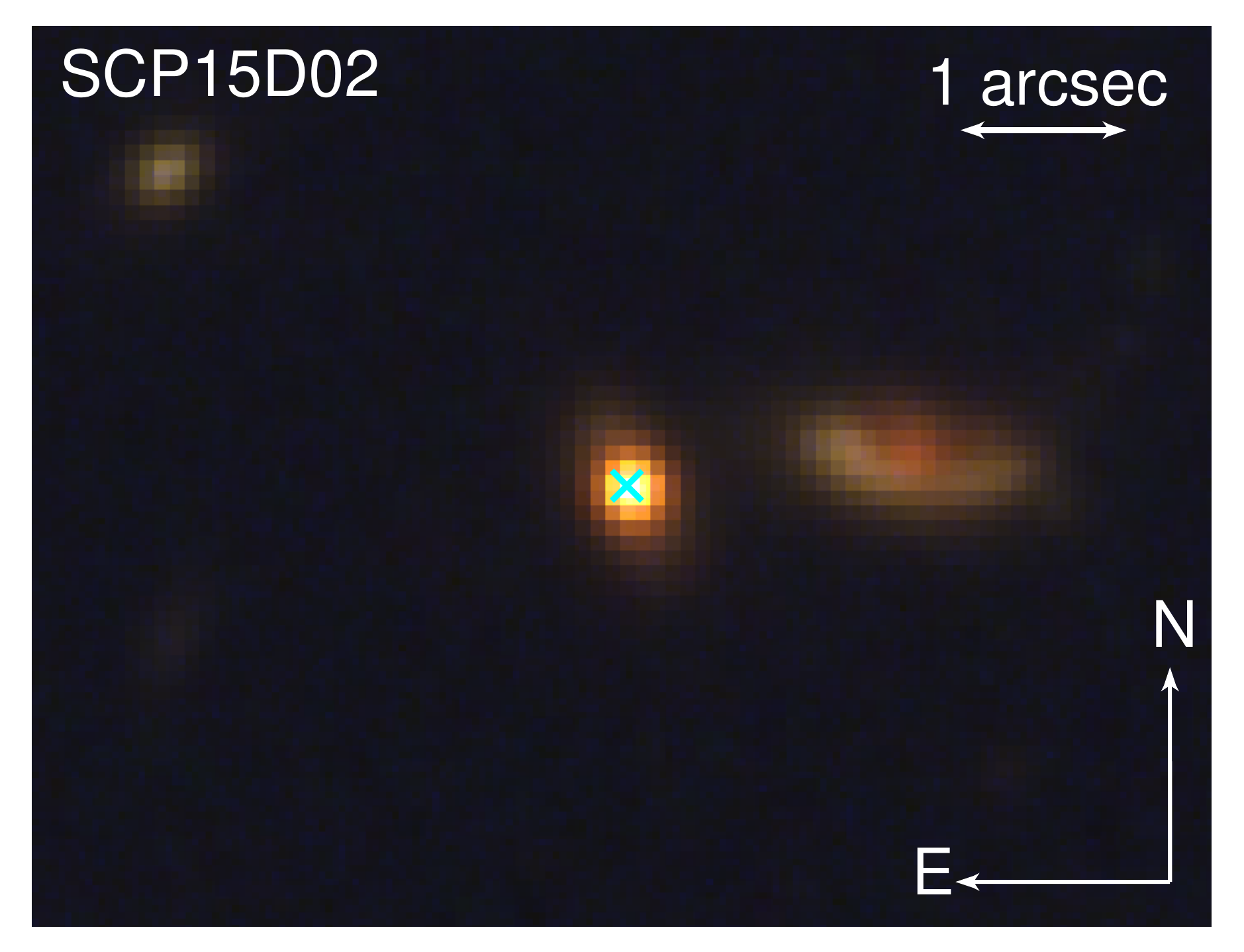}
\includegraphics[width=0.5\columnwidth]{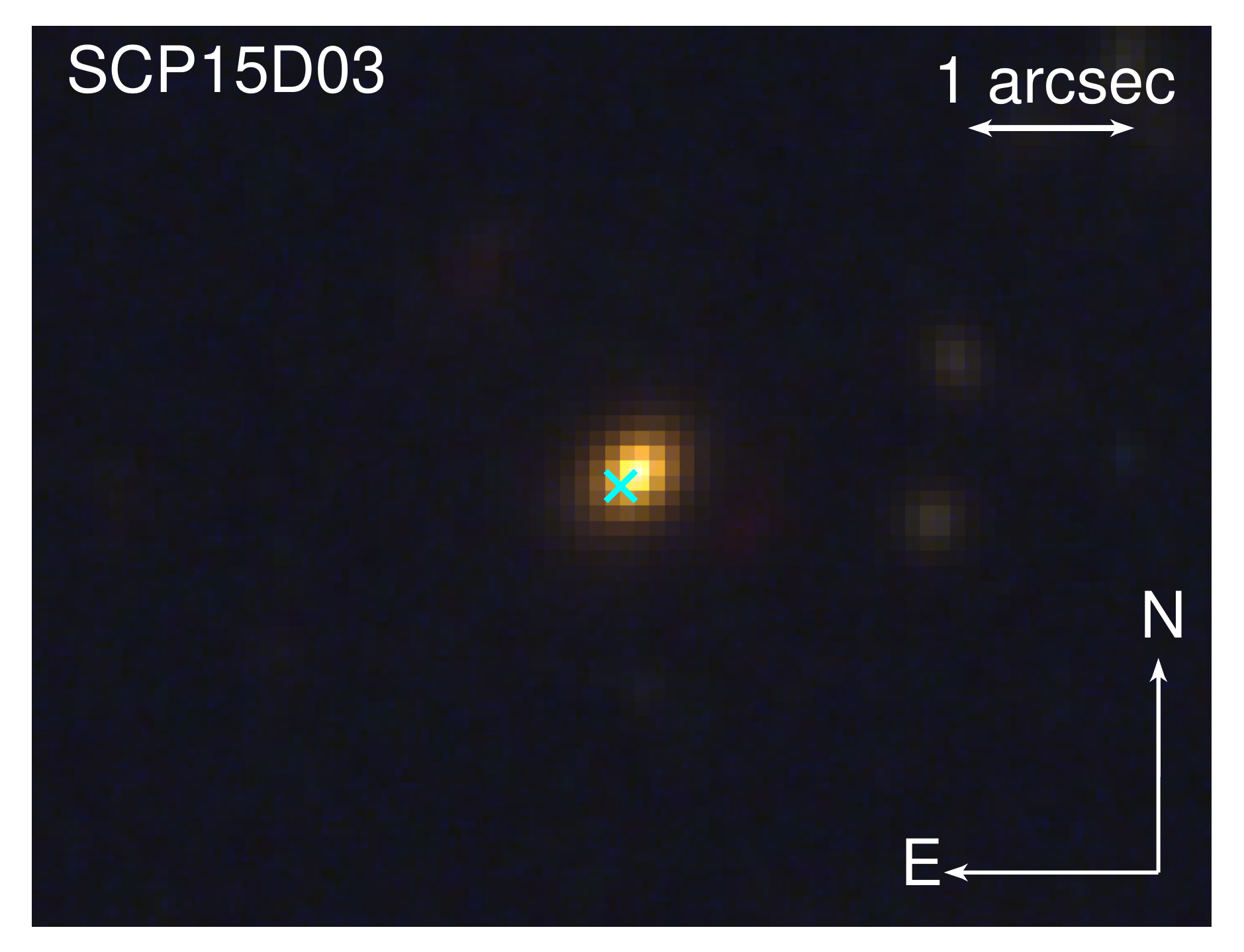}
    \end{subfigure}
\caption{\textbf{Continued.} Stacked colour-composite \textit{HST} \textit{F814W}, \textit{F105W} and \textit{F140W} See Change finding charts indicating the position of each SN within its host.}
\end{figure*}
\begin{figure*}\ContinuedFloat
    \begin{subfigure}[t]{\textwidth}
\includegraphics[width=0.5\columnwidth]{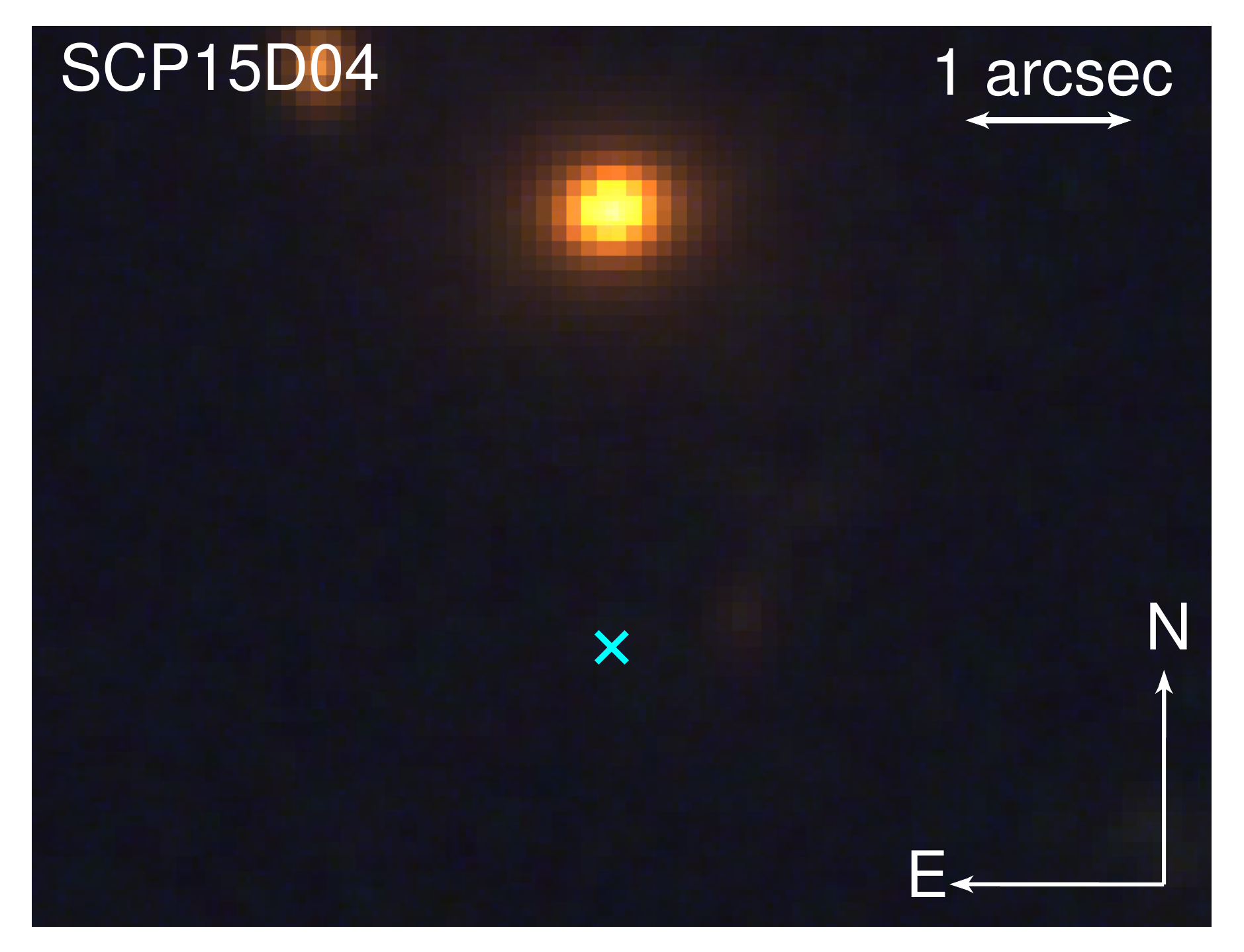}
\includegraphics[width=0.5\columnwidth]{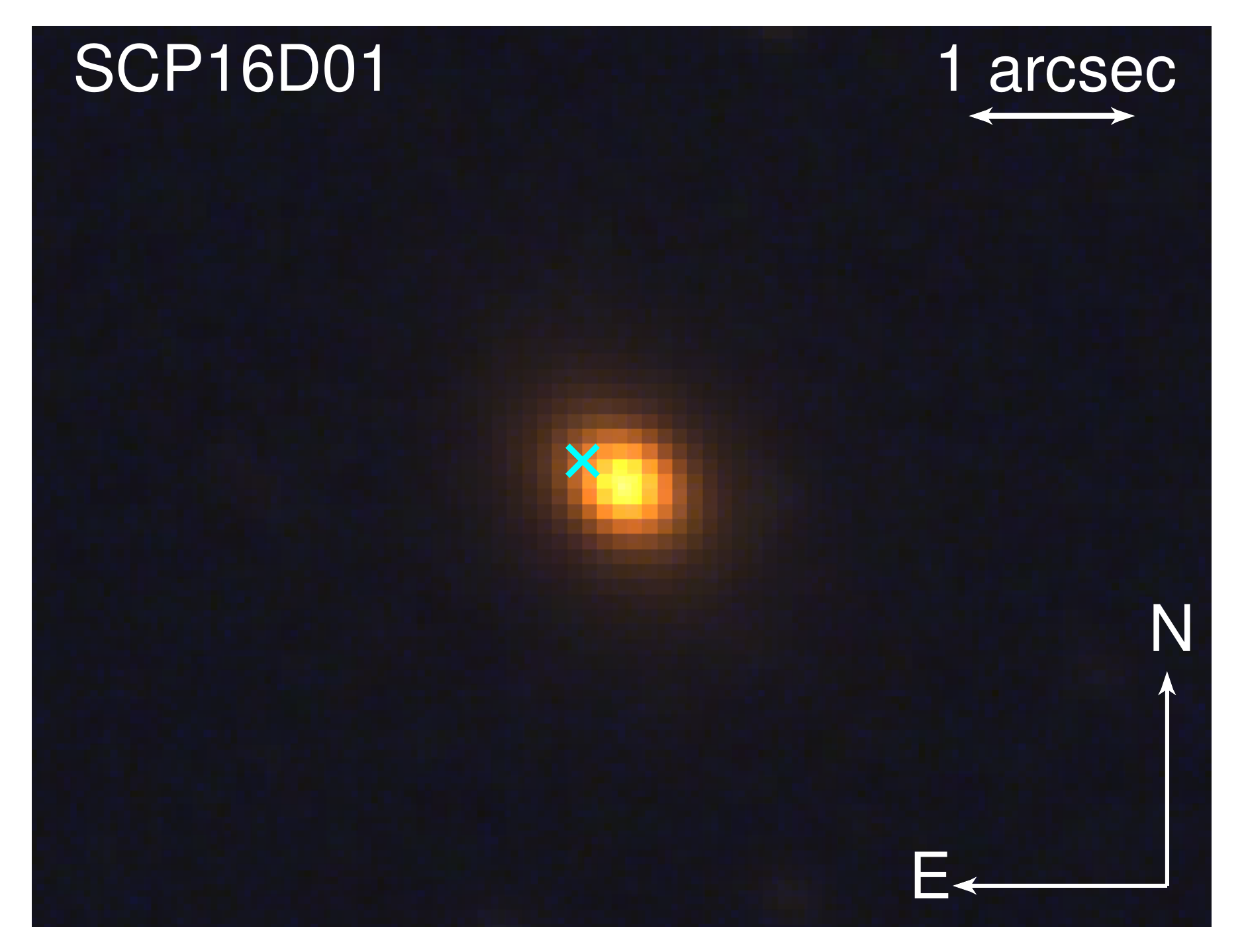}
\includegraphics[width=0.5\columnwidth]{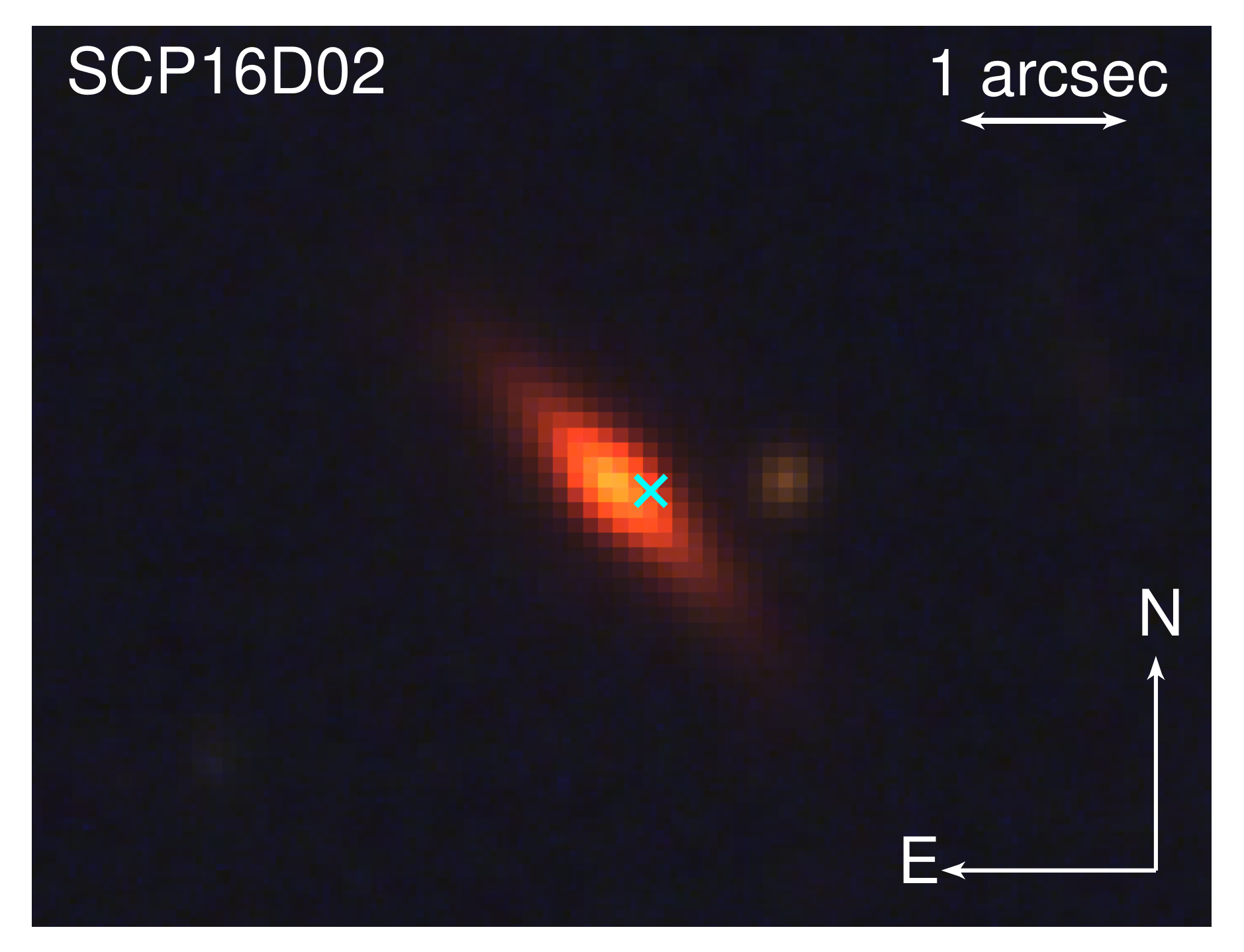}
\includegraphics[width=0.5\columnwidth]{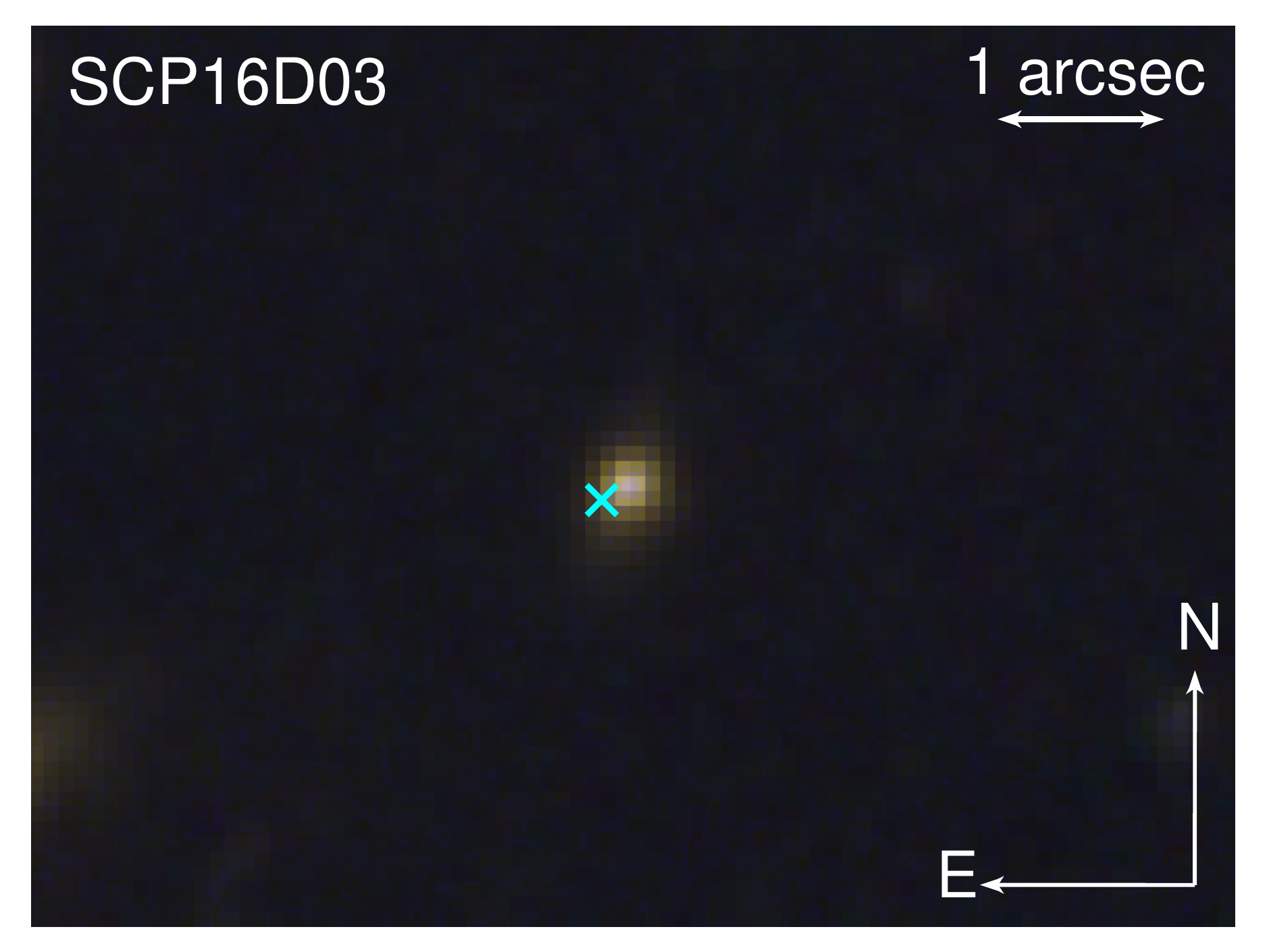}
\includegraphics[width=0.5\columnwidth]{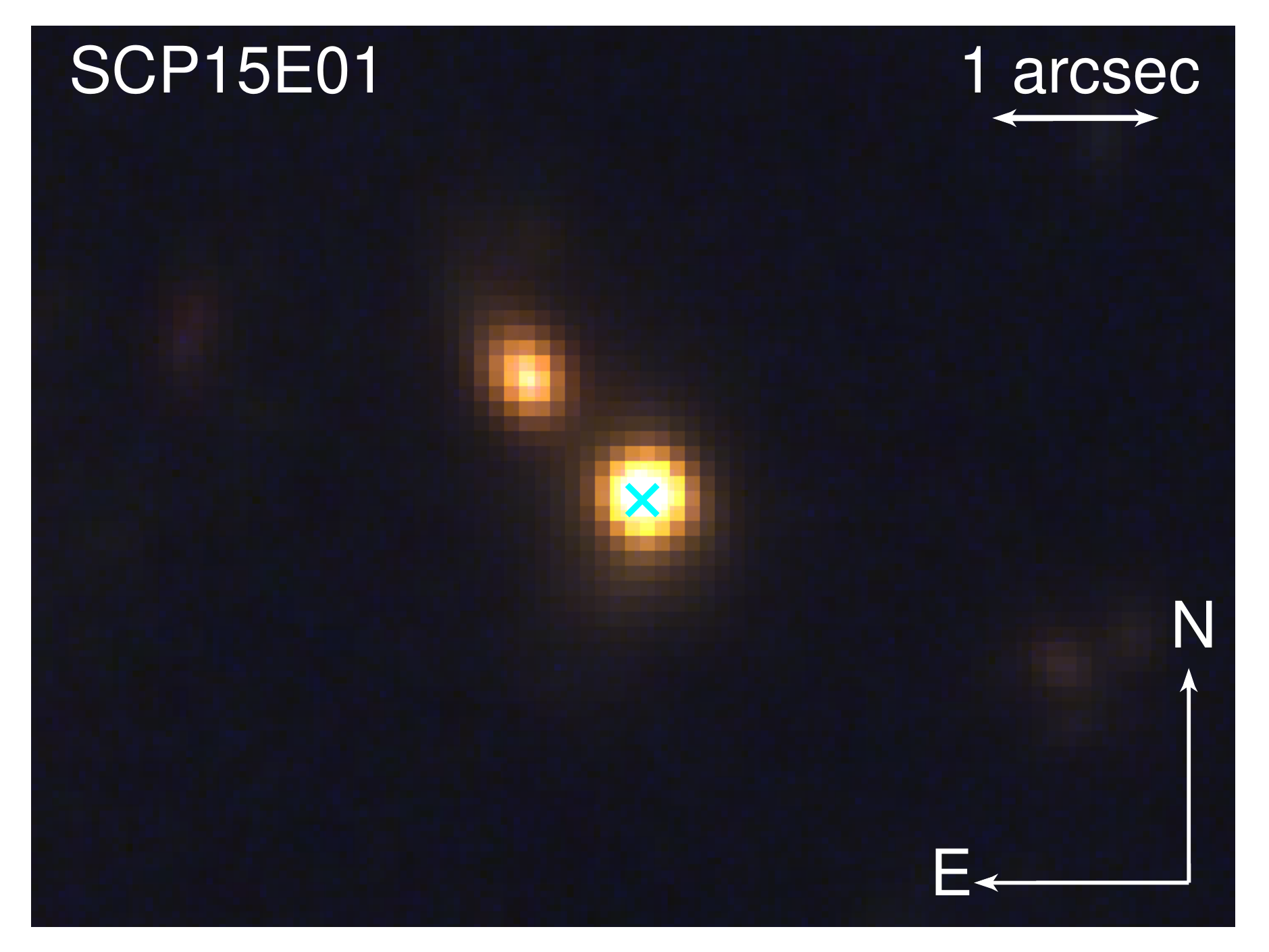}
\includegraphics[width=0.5\columnwidth]{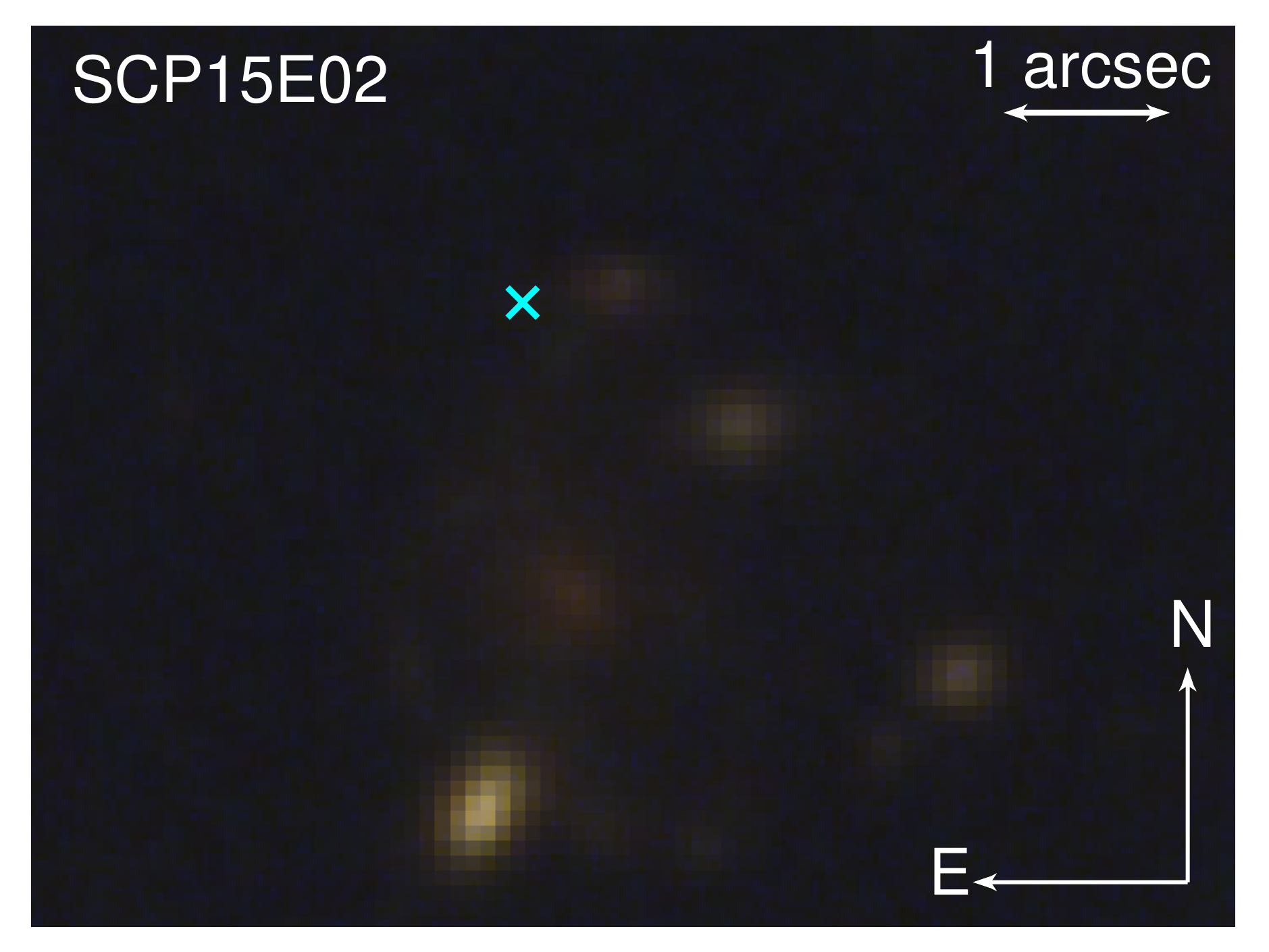}
    \end{subfigure}
\caption{\textbf{Continued.} Stacked colour-composite \textit{HST} \textit{F814W}, \textit{F105W} and \textit{F140W} See Change finding charts indicating the position of each SN within its host.}
\end{figure*}
\begin{figure*}\ContinuedFloat
    \begin{subfigure}[t]{\textwidth}
\includegraphics[width=0.5\columnwidth]{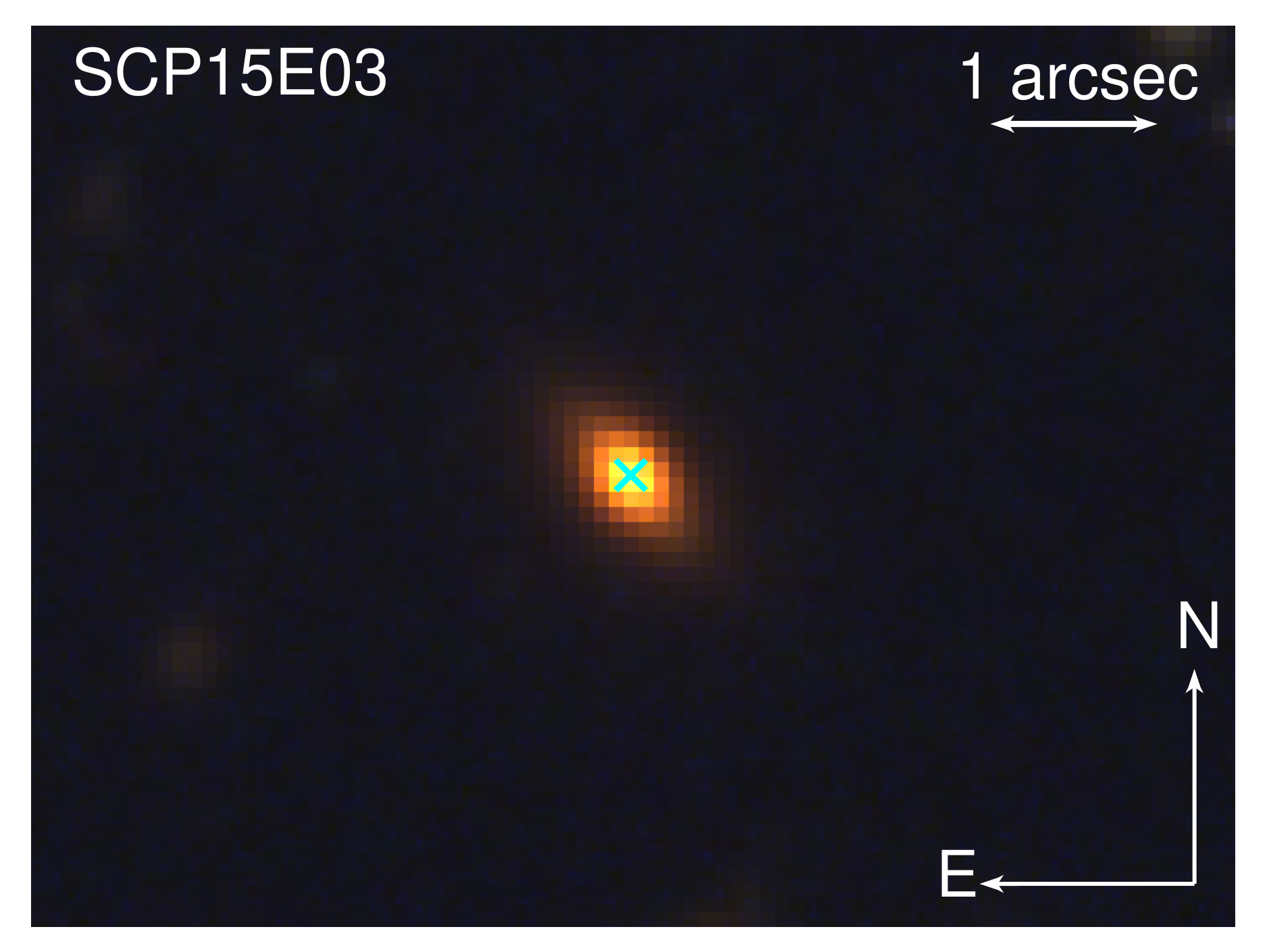}
\includegraphics[width=0.5\columnwidth]{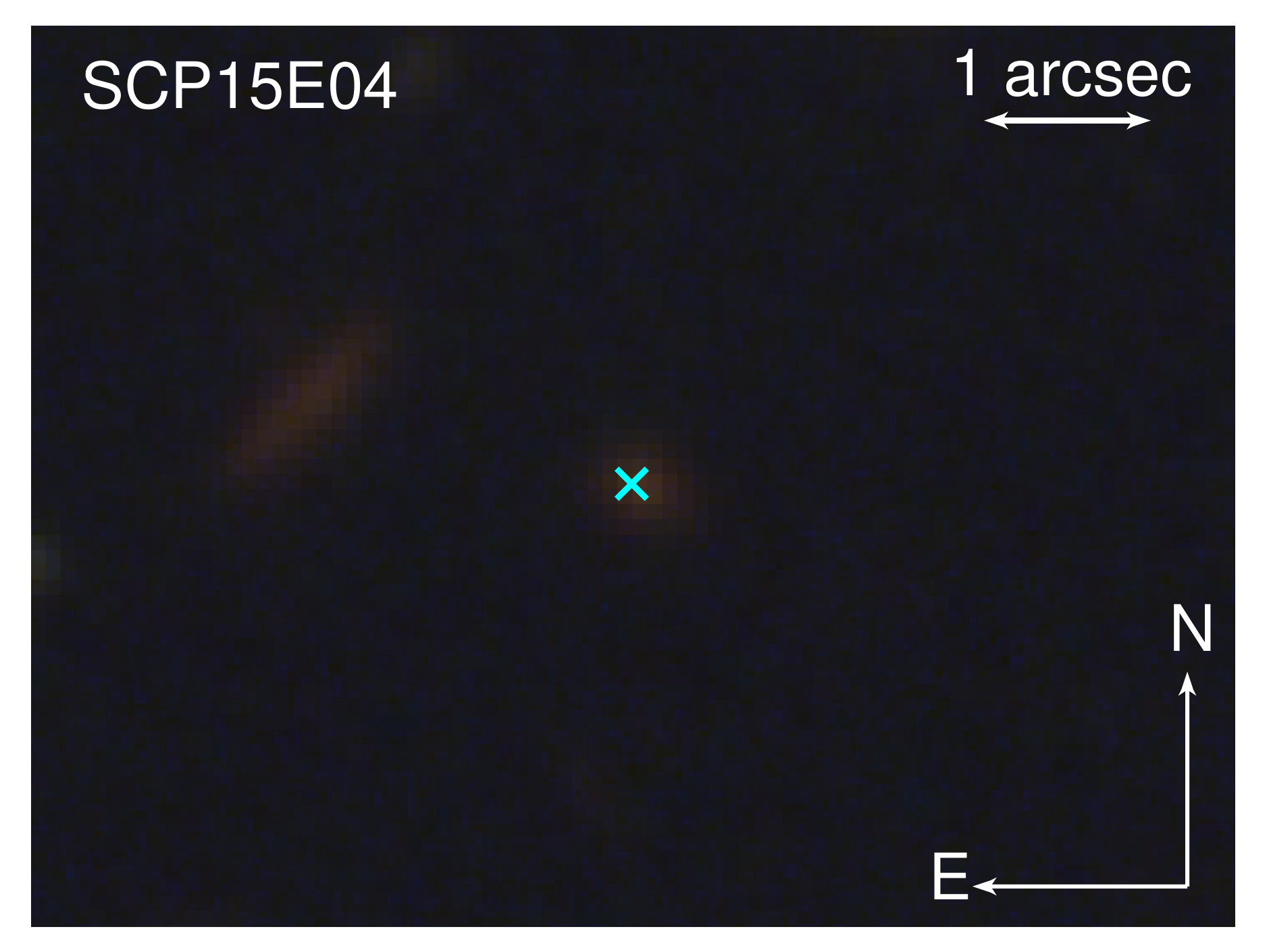}
\includegraphics[width=0.5\columnwidth]{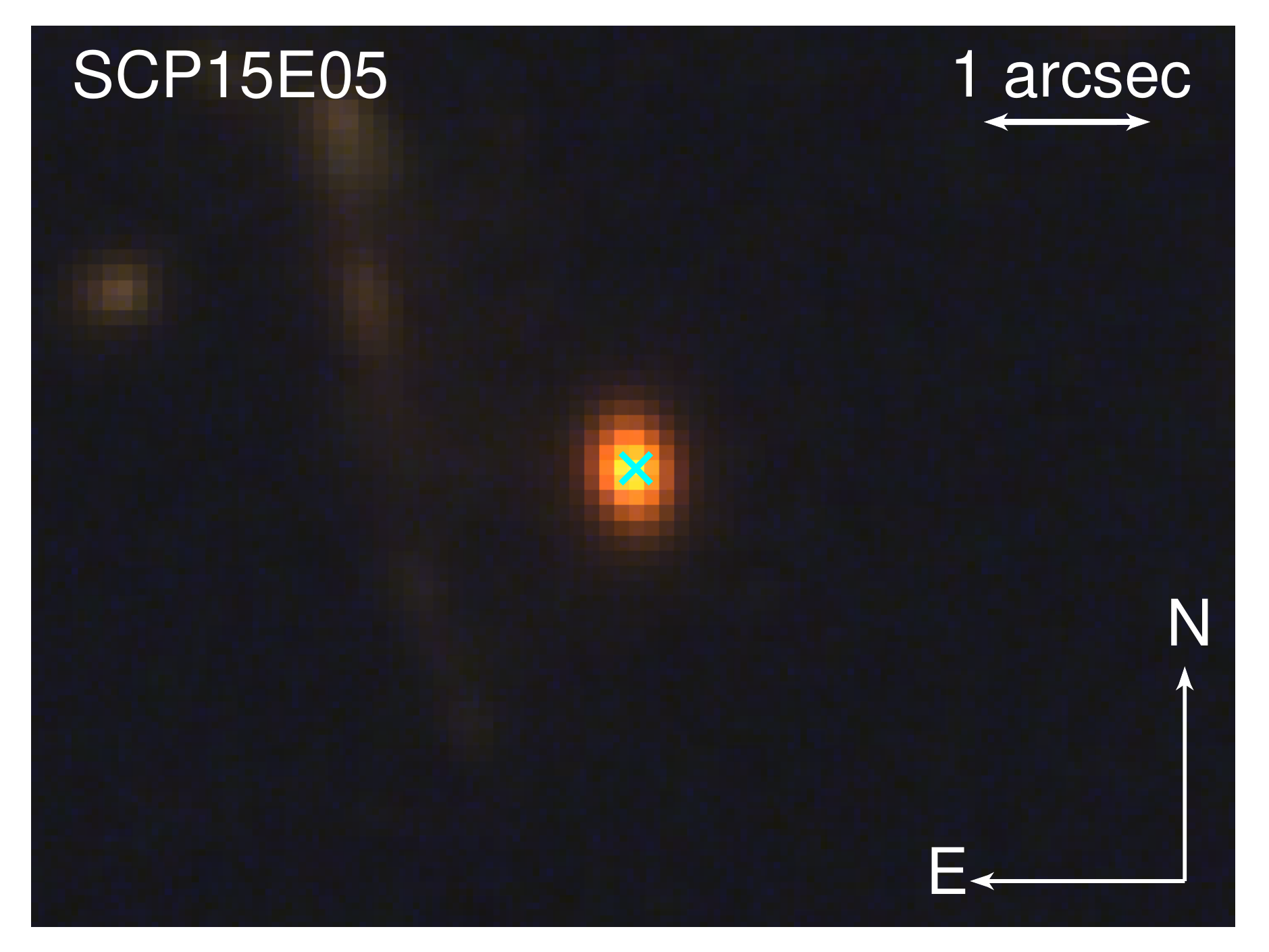}
\includegraphics[width=0.5\columnwidth]{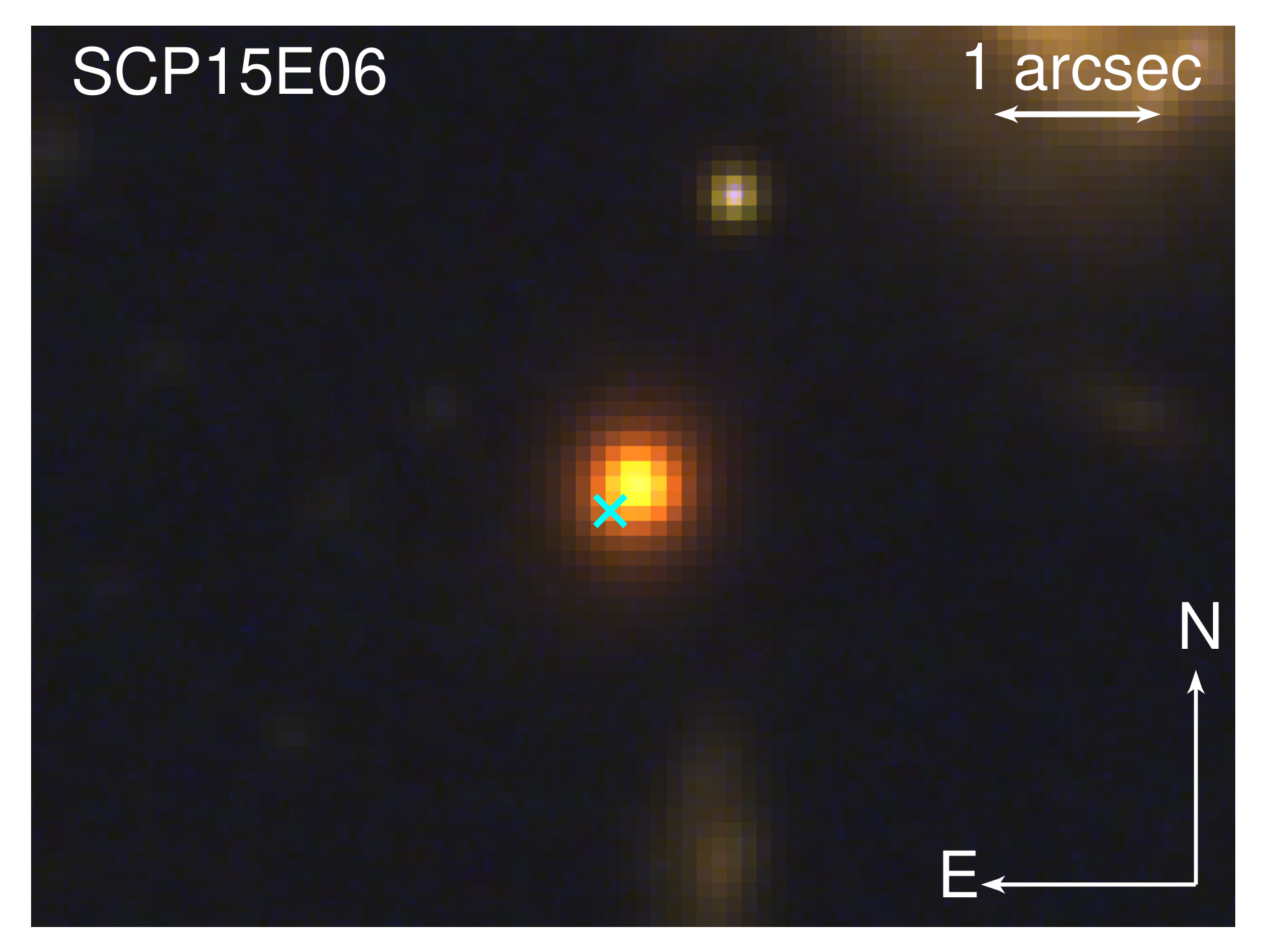}
\includegraphics[width=0.5\columnwidth]{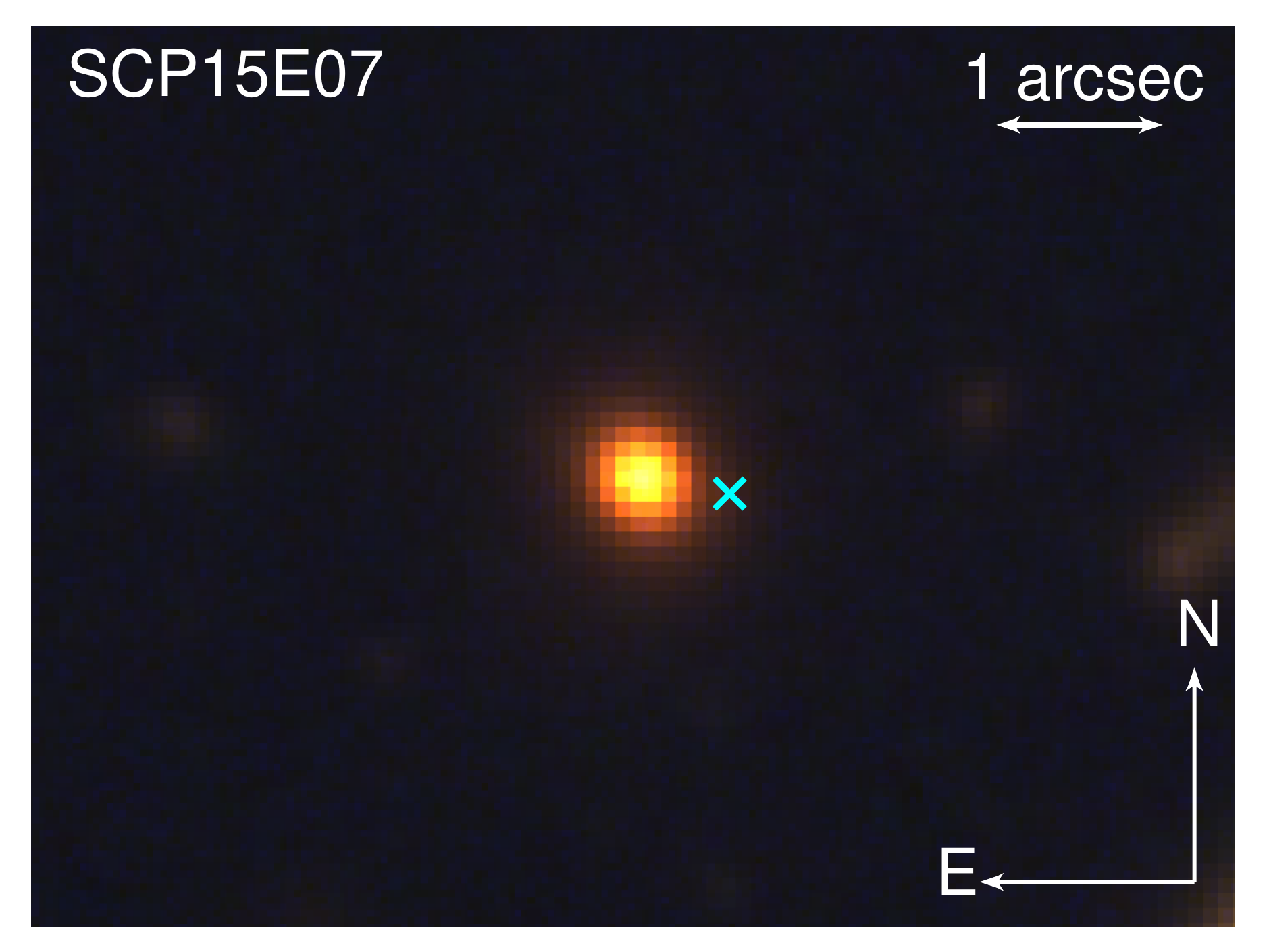}
\includegraphics[width=0.5\columnwidth]{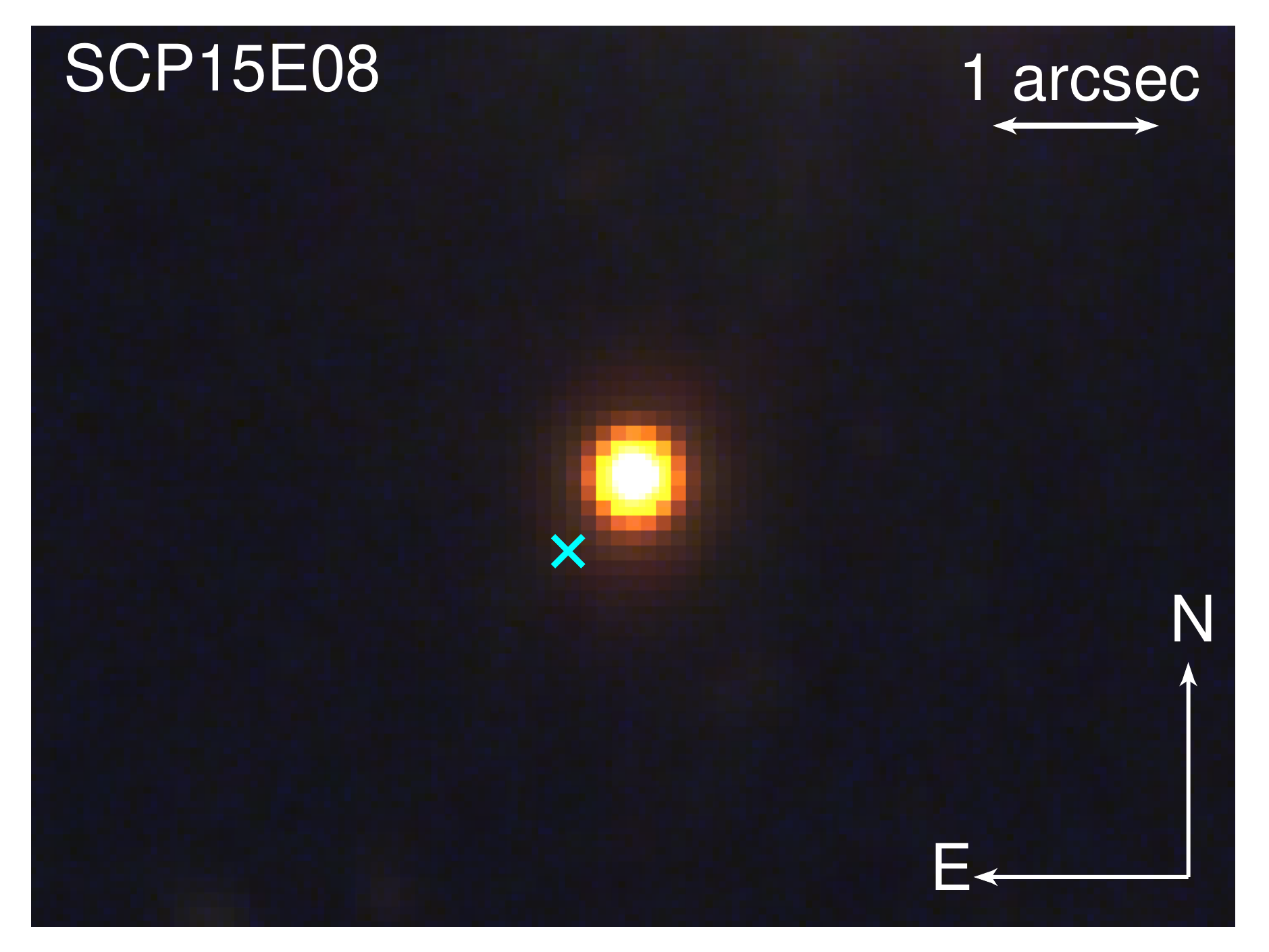}
    \end{subfigure}
\caption{\textbf{Continued.} Stacked colour-composite \textit{HST} \textit{F814W}, \textit{F105W} and \textit{F140W} See Change finding charts indicating the position of each SN within its host.}
\end{figure*}
\begin{figure*}\ContinuedFloat
    \begin{subfigure}[t]{\textwidth}
\includegraphics[width=0.5\columnwidth]{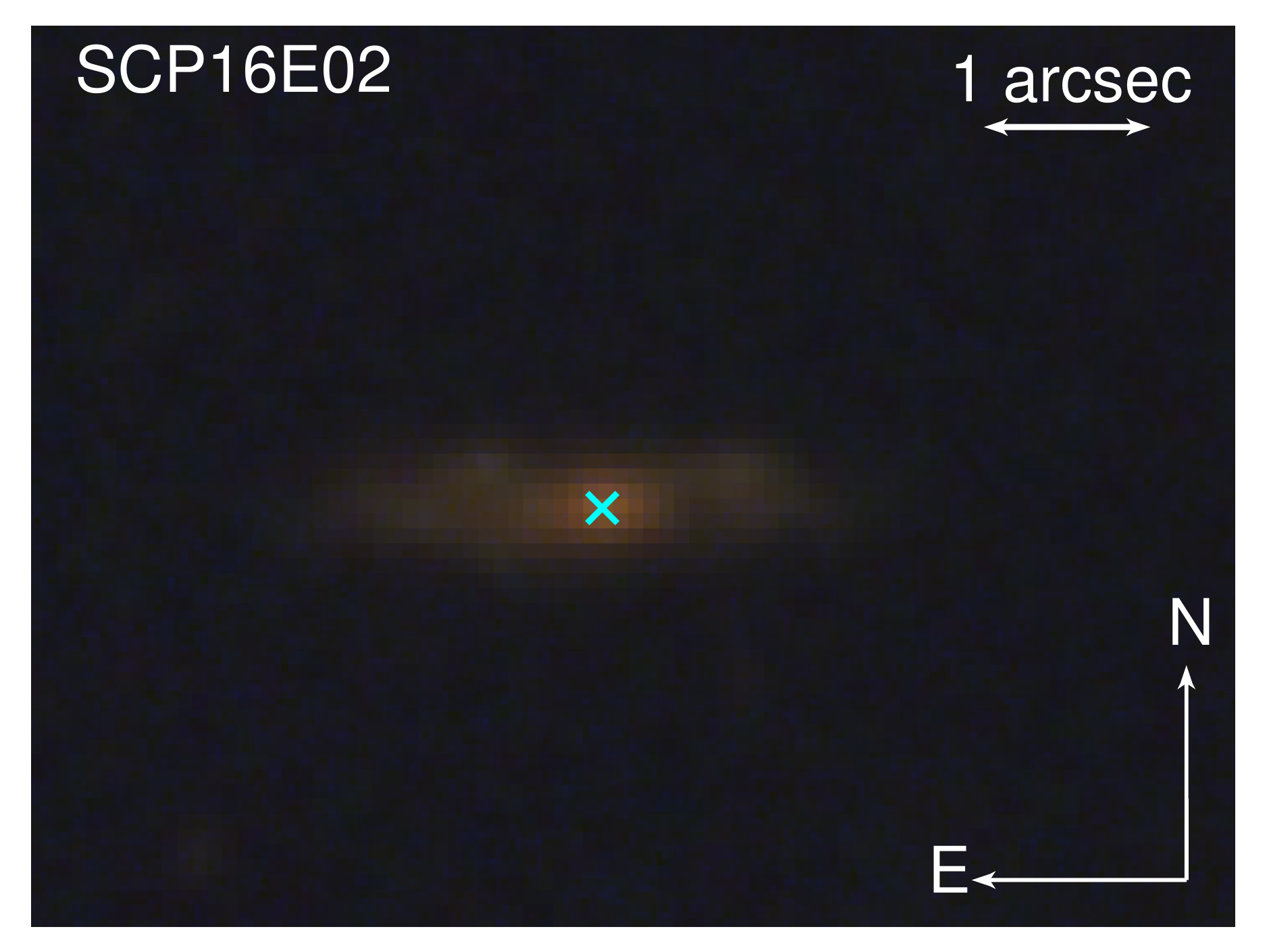}
\includegraphics[width=0.5\columnwidth]{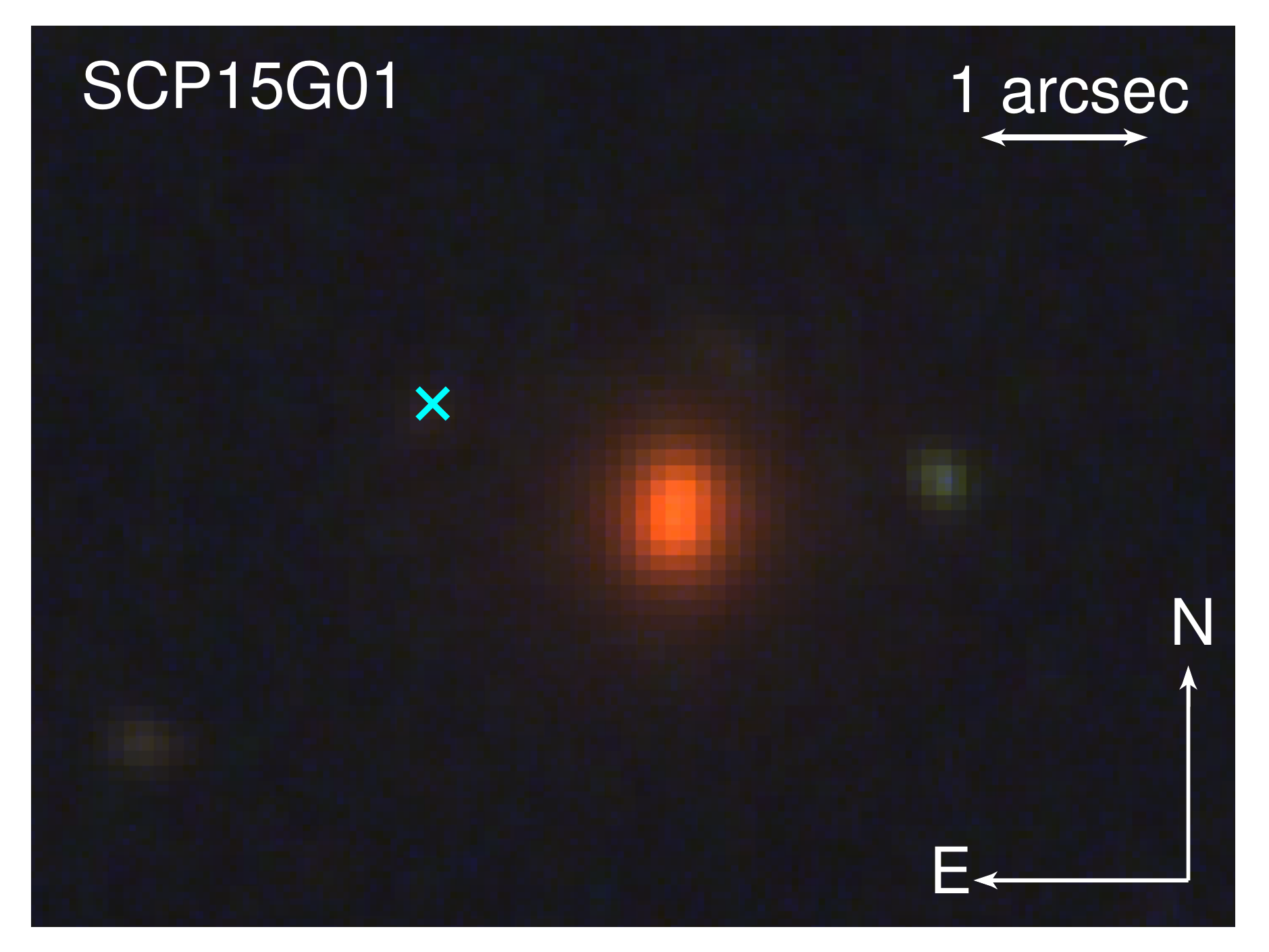}
\includegraphics[width=0.5\columnwidth]{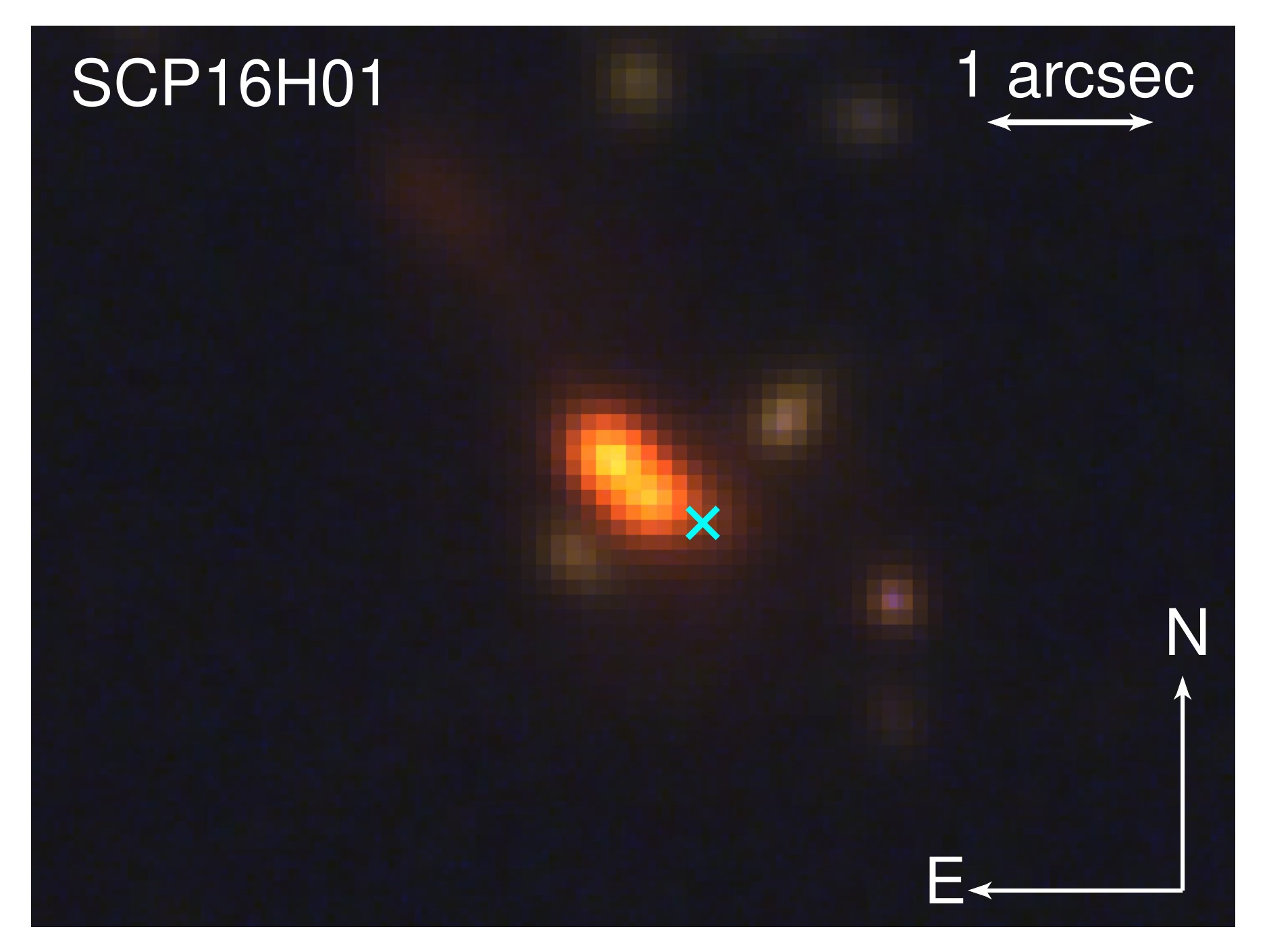}
\includegraphics[width=0.5\columnwidth]{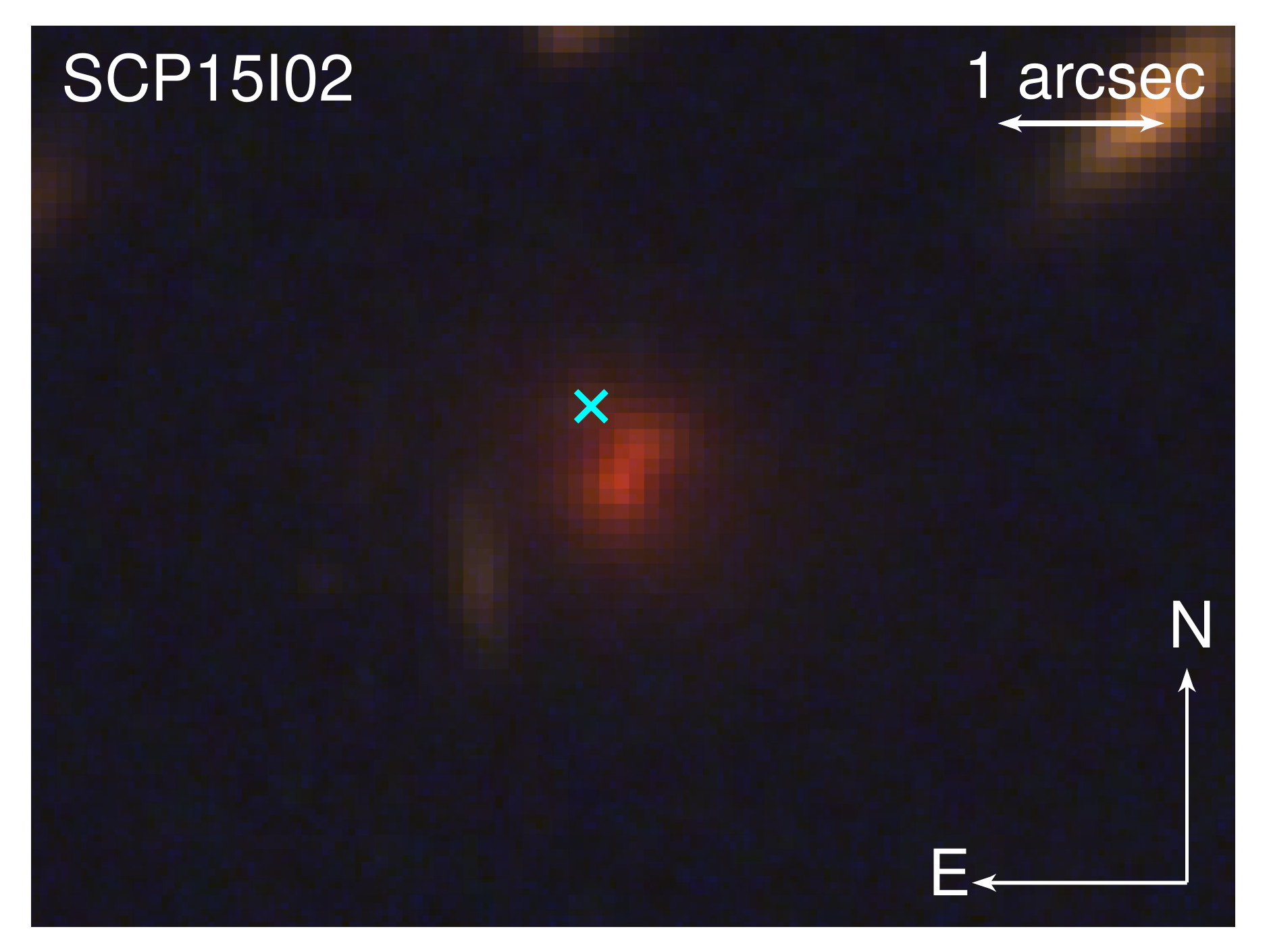}
    \end{subfigure}
\caption{\textbf{Continued.} Stacked colour-composite \textit{HST} \textit{F814W}, \textit{F105W} and \textit{F140W} See Change finding charts indicating the position of each SN within its host.}
\end{figure*}


\bsp	
\label{lastpage}
\end{document}